\documentclass[prd,12pt,reprint, twocolumn,noshowpacs,nofootinbib,amsmath,amssymb,superscriptaddress,preprintnumbers]{revtex4-1}

\usepackage{graphicx,color}
\usepackage{caption}
\usepackage{amssymb,amsmath,mathrsfs}
\usepackage{cancel}
\usepackage{subfig}
\usepackage{hyperref}
\usepackage{verbatim,slashed}
\usepackage{xspace}
\newcommand{\bp}{{\bf p}}
\newcommand{\bk}{{\bf k}}
\newcommand{\bq}{{\bf q}}

\newcommand{\beq}{\begin{equation}}
\newcommand{\eeq}{\end{equation}}
\newcommand{\be}{\begin{eqnarray}}
\newcommand{\ee}{\end{eqnarray}}

\newcommand{\bi}{\begin{itemize}}
\newcommand{\ei}{\end{itemize}}
\newcommand{\benum}{\begin{enumerate}}
\newcommand{\eenum}{\end{enumerate}}

\newcommand{\Msplit}{\Delta M_{21}^2}
\newcommand{\YB}{Y_B} 
\newcommand{\YBmL}{Y_{B-L,\,\mathrm{SM}}} 
\newcommand{\rhotot}{\rho^{\chi+\overline{\chi}}} 
\newcommand{\Yonetot}{Y_1^{\chi+\overline{\chi}}} 
\newcommand{\Ytwotot}{Y_2^{\chi+\overline{\chi}}} 
\newcommand{\Yeqtot}{Y_{\text{eq}}^{\chi+\overline{\chi}}} 
\newcommand{\Ya}{\delta Y^\alpha} 
\newcommand{\Ychi}{\delta Y^{\chi}} 
\newcommand{\Mchitwo}{M_2} 
\newcommand{\Mchione}{M_1} 

\def\lsim{\mathrel{\rlap{\lower4pt\hbox{\hskip1pt$\sim$}}
    \raise1pt\hbox{$<$}}}
\def\gsim{\mathrel{\rlap{\lower4pt\hbox{\hskip1pt$\sim$}}
    \raise1pt\hbox{$>$}}}

\begin{document}
\title{Baryogenesis and Dark Matter from Freeze-In}
\author{Brian Shuve}
\affiliation{Harvey Mudd College, 301 Platt Blvd., Claremont, CA 91711, USA}
\affiliation{University of California, 900 University Ave., Riverside,
  CA 92521, USA}
\author{David Tucker-Smith}
\affiliation{Department of Physics, Williams College, Williamstown, MA 01267, USA}

\date{\today}

\begin{abstract}
We propose a simple  model in which the baryon asymmetry and dark matter are created  via the decays and inverse decays of QCD-triplet scalars, at least one of which must be in the TeV mass range.  Singlet fermions produced in these decays constitute the dark matter.   The singlets never reach equilibrium, and their coherent production, propagation, and annihilation generates a baryon asymmetry.  We find that that the out-of-equilibrium condition and the dark matter density constraint typically require the lightest scalar to be long-lived, giving good prospects for detection or exclusion in current and upcoming colliders. In generalizing the leptogenesis mechanism of Akhmedov, Rubakov and Smirnov, our model  expands the phenomenological possibilities for low-scale baryogenesis.

\end{abstract}
\maketitle

\section{Introduction}
One of the most important  questions in particle physics is the origin of the baryon asymmetry. While the Standard Model (SM) contains $CP$ violation that distinguishes between rates of particle and antiparticle interactions, it is widely accepted that the degree of $CP$ violation in the SM is insufficient to explain the magnitude of the observed asymmetry (see, for example, Ref.~\cite{Cline:2006ts}). Furthermore, the generation and preservation of an asymmetry requires a departure from equilibrium that is not realized in the SM:~with the observed Higgs boson mass, the SM predicts a second-order electroweak phase transition, which is insufficient to generate a baryon asymmetry \cite{Shaposhnikov:1986jp,Shaposhnikov:1987tw,Bochkarev:1987wf,Kajantie:1996mn}.

Resolving the origin of the baryon asymmetry necessitates the existence of new particles and interactions beyond the SM. 
Various theoretical scenarios for baryogenesis exist, including but not limited to electroweak baryogenesis \cite{Kuzmin:1985mm,Shaposhnikov:1986jp,Shaposhnikov:1987tw,Cohen:1993nk}, leptogenesis \cite{Fukugita:1986hr} (inspired by the see-saw mechanism for neutrino mass generation \cite{Minkowski:1977sc,Mohapatra:1979ia,GellMann:1980vs,Yanagida:1980xy,Schechter:1980gr,Schechter:1981cv}), and realizations within grand-unified models \cite{Yoshimura:1978ex,Ignatiev:1978uf}. Some models of baryogenesis are challenging to test, whether because the relevant mechanism operates at high scales that are not kinematically accessible to current or future experiments, or because satisfying the out-of-equilibrium condition for baryogenesis predicts small couplings for the new particles relative to other SM couplings. There are, however,   baryogenesis scenarios that are testable in their minimal incarnations:~electroweak baryogenesis, whose dynamics are necessarily constrained to lie around the weak scale and which accommodates large couplings of beyond-SM states to the Higgs in order to give rise to a first-order phase transition; and freeze-in leptogenesis, also known as the Akhmedov-Rubakov-Smirnov (ARS) mechanism or leptogenesis via neutrino oscillations \cite{Akhmedov:1998qx,Asaka:2005pn}.

In this paper, we study a  new class of models inspired by  ARS leptogenesis.  
We consider a framework in which light, gauge-singlet Majorana fermions $\chi_I$ interact feebly through a  Yukawa coupling 
\be
F^i_{\alpha I} {\overline \psi}_\alpha \chi_I \Phi_i + \text{h.c.},
\label{eqn:interaction}
\ee 
where $\psi_\alpha$ are SM fermions and $\Phi_i$ are new scalars with the same gauge quantum numbers as $\psi_\alpha$.  Here we focus on scenarios where the scalars carry quantum chromodynamics (QCD) charge, with $\psi = Q_L, u_R$, or $d_R$.  Collider searches then constrain the masses of the QCD-triplet scalars $\Phi_i$ to be at or above the TeV scale.   We impose a $Z_2$ symmetry under which only $\chi_I$ and $\Phi_i$ are odd, making  the  $\chi$ particles dark matter (DM) candidates.  

For appropriate parameter choices, these ingredients are sufficient to generate a baryon asymmetry.   The relevant dynamics are somewhat involved, but that should not obscure the simplicity of the model setup.
 Decays of $\Phi$ particles produce coherent superpositions of $\chi$ mass eigenstates, whose subsequent time evolution and scattering can produce an overall $\Phi$ asymmetry.  The net  $(B-L)_\Phi$ and hypercharge $\mathcal{Y}_\Phi$ stored in the $\Phi$ sector are balanced by opposite charges $(B-L)_\text{SM}$ and $\mathcal{Y}_\text{SM}$ stored in SM particles.  At temperatures above the electroweak scale, rapid sphaleron and SM-Yukawa-induced processes re-distribute the $(B-L)_\text{SM}$ asymmetry among baryons and leptons,
whereas $B_\Phi$, the  baryon number in $\Phi$, is left unchanged.  Because the resultant $B_\text{SM}$ differs in magnitude from  $B_\Phi$, a net baryon asymmetry survives after the $\Phi$ particles decay and disappear, provided  $\Phi$ particles survive until the time of sphaleron decoupling.

 Certain essential phenomenological considerations parallel the ARS case.  
To satisfy the out-of-equilibrium Sakharov condition and generate an asymmetry \cite{Sakharov:1967dj}, the Yukawa couplings must satisfy $|F_{\alpha I}| \lesssim 10^{-7}$. This is a model of freeze-in baryogenesis because the $\chi_I$ do not come into equilibrium while the asymmetry is being generated.  The baryon asymmetry is enhanced for $\chi_I$ mass splittings of order 10 keV, so that $\chi_I$ oscillations have time to develop before sphaleron decoupling, but are not so rapid that the asymmetry generation saturates at early times and gives a smaller asymmetry.  However, we find that our model predictions are qualitatively distinct from ARS, giving rise to significant enhancements in the baryon asymmetry in parts of parameter space as well as new phenomenological probes.

In place of the right-handed neutrinos (RHNs) of ARS are new cosmologically stable neutral states $\chi_I$ that we  identify as the DM; these singlet $\chi_I$ states have negligible mixing with SM fermions.   It is the oscillations of the DM particles themselves that are responsible for baryogenesis, and our model  generally favors  DM states with a non-degenerate mass spectrum.   This is unlike viable ARS models, where the dynamics of DM is unrelated to the generation of the baryon asymmetry via RHN oscillations, and which typically require highly degenerate $\chi$ masses.

The beyond-SM (BSM) QCD-charged scalars $\Phi_i$ can qualitatively alter the baryon asymmetry calculation.   We pay particular attention to the possibility of having more than one scalar, which tends to dramatically enhance the baryon asymmetry.  In the two-scalar case, the different channels for $\chi$ production and annihilation lead to an asymmetry at $\mathcal{O}(F^4)$,
 rather than at $\mathcal{O}(F^6)$ as in standard ARS leptogenesis.

Successful baryogenesis requires a $B-L$ asymmetry to be stored in the $\Phi$ sector until sphaleron decoupling.  When combined with the DM abundance constraint, we find that this favors the mass of the lightest $\Phi$ particle to be not far above the TeV scale and its lifetime to be comparable to or larger than the Hubble time at electroweak-scale temperatures, corresponding to values of $c\tau \gsim 1$ cm.  Consequently, the model can be probed by the Large Hadron Collider (LHC), and much of the parameter space predicts long-lived particle signatures.  

The properties of the heavier $\Phi$ scalar(s) are much less constrained.  In the two-scalar case, it is viable to have $M_{\Phi_2} \gg M_{\Phi_1}$, and in this ``decoupled-$\Phi_2$'' regime the baryon asymmetry and DM abundance depend on the properties of  $\Phi_2$ only through the characteristics of the coherent background of $\chi_I$ particles left behind after the $\Phi_2$ particles have entirely decayed/annihilated away.  It is worth emphasizing that this coherent background can be $CP$-symmetric ``initially,'' that is, just after the $\Phi_2$ particles have disappeared.  The $CP$ violation arises from time-evolution phases, 
in tandem with phases encoded in the coherent $\chi$ background when expressed in the $\Phi_1$ interaction eigenbasis.

More generally, the asymmetry in the decoupled-$\Phi_2$ regime is independent of the origin of the coherent background of DM particles. It could be left behind by the decays of a heavy particle with different quantum numbers than $\Phi$, for example the inflaton.

Our baryon asymmetry and DM results for the 
two-$\Phi$ model in the decoupled-$\Phi_2$ regime are summarized in Figures ~\ref{fig:max_ranges_various_angles}, \ref{fig:Mchi_Mphi_plot}, and \ref{fig:scans_nonzero_mchi1_rhothird}, which show the preference for sub-MeV $\chi$ masses, and for the lighter $\Phi$ particle to be in the few-TeV range and long-lived for collider purposes.

Meanwhile, the analysis of the {\em single}-scalar scenario (in which the same BSM particle is involved in $\chi$ production and $\chi$ annihilation) is dramatically impacted by the fact that the SM states participating directly in the asymmetry generation are quarks rather than leptons.  Unlike the situation for the leptonic case, where different flavors of leptons can have different chemical potentials, quark flavor mixing drives the quark chemical potentials towards a universal value, thereby suppressing one possible source of asymmetry (here we work in the approximation of flavor-universal quark chemical potentials  and save a more careful study for future work). 

 On the other hand, the large top Yukawa coupling opens up the possibility that flavor dependence in the thermal masses of the active fermions plays a role in generating the baryon asymmetry at $\mathcal{O}(F^4 y_t^2)$.
In fact, we find that top-Yukawa effects make  the single-$\Phi$, two-$\chi$ model  viable for obtaining the observed baryon asymmetry and DM abundance, although it is more constrained than the two-scalar scenario (see Fig.~\ref{fig:top_assisted_single_scalar}), and most of the parameter space will be tested by searches for heavy scalars at the Large Hadron Collider. 

 In the absence of SM-Yukawa effects, we need three or more $\chi$ particles to get an $\mathcal{O} (F^6)$ contribution to the asymmetry.  In this case, we find that the model's ability to 
simultaneously satisfy the baryon asymmetry and dark-matter abundance constraints is marginal at best (see Fig.~\ref{fig:singlescalar_DM_baryo}). 
The observed baryon asymmetry  can still be realized if one imagines a different explanation for the DM and a decay mechanism for massive $\chi$ particles, and in this scenario the prospects for $\Phi$ discovery at colliders are quite promising (see Figs.~\ref{fig:singlescalar_comparechimass_benchmark} and \ref{fig:singlescalar_maxMPhi}).  

Our study is organized as follows:~in Sec.~\ref{sec:mechanism}, we provide a qualitative review of the mechanism of freeze-in baryogenesis, and we perform an analytic calculation of the asymmetry in the weak-washout limit for a representative model that illustrates the parametric dependence of the asymmetry on the model parameters. In Sec.~\ref{sec:twoscalar}, we  investigate the model where the singlets responsible for baryogenesis couple to SM quarks and two new QCD-charged scalars, providing a comprehensive study of the parameters giving rise to successful baryogenesis and DM. In Sec.~\ref{sec:single_scalar}, we study  the more constrained model with only a single new scalar.  Finally, we discuss these models' experimental signatures and prospects for discovery in Sec.~\ref{sec:signals}.

We now discuss  connections between our work and earlier studies of ARS leptogenesis and freeze-in DM. In minimal extensions of the SM, the only renormalizable coupling of singlet fermions $\chi$ is via the neutrino portal, giving rise to the ARS mechanism.
In the context of ARS leptogenesis, the $\chi$ particles  
are RHNs that lie in the GeV mass range.  Because the RHNs are low in mass relative to the electroweak scale, they can potentially be  produced in intensity frontier and other collider experiments \cite{Gorbunov:2007ak}. ARS leptogenesis therefore provides a well-motivated, testable mechanism for both the generation of neutrino masses as well as baryogenesis. Consequently, this mechanism for baryogenesis and its discovery prospects have been well studied in the literature \cite{Kersten:2007vk,Shaposhnikov:2008pf,Canetti:2010aw,Asaka:2011wq,Canetti:2012kh,Canetti:2012vf,Drewes:2012ma,Garbrecht:2013urw,Shuve:2014zua,Canetti:2014dka,Garbrecht:2014bfa,Hernandez:2015wna,Abada:2015rta,Drewes:2015iva,Hambye:2016sby,Hernandez:2016kel,Drewes:2016gmt,Asaka:2016zib,Drewes:2016lqo,Drewes:2016jae,Chun:2017spz,Ghiglieri:2017gjz,Eijima:2017cxr,Eijima:2017anv,Asaka:2017rdj,Abada:2017ieq,Ghiglieri:2018wbs,Eijima:2018qke}. Indeed, the first dedicated searches for GeV-scale RHNs have now been performed at the ATLAS, CMS, and LHCb experiments at the Large Hadron Collider \cite{Aaij:2014aba,Shuve:2016muy,Sirunyan:2018mtv,Aad:2019kiz}. Baryogenesis from freeze-in is also possible in models without oscillations, in which the relevant $CP$-violating phases for baryogenesis originate from the interference of tree and loop diagrams in scattering processes \cite{Hall:2010jx,Hook:2011tk,Unwin:2014poa}.

Models of freeze-in are inherently sensitive to other couplings of the sterile states to the SM.
This is well-known in the case of freeze-in models of DM, where the largest coupling of the hidden particle tends to dominate its cosmology and phenomenology \cite{McDonald:2001vt,Choi:2005vq,Kusenko:2006rh,Petraki:2007gq,Hall:2009bx} (for a recent review, see Ref.~\cite{Bernal:2017kxu}). 
While there have been a few studies of freeze-in baryogenesis where there exist new fields beyond the minimal ARS model \cite{Shuve:2014zua,Khoze:2016zfi,Caputo:2018zky,Baumholzer:2018sfb}, there has not to our knowledge been a comprehensive attempt to study of the parametric regimes and signatures associated with non-minimal scenarios.

\section{Mechanism of Freeze-In Baryogenesis}
\label{sec:mechanism}

\subsection{Qualitative Overview of Freeze-In Baryogenesis} 
\label{sec:qualitative}
The mechanism of baryogenesis via singlet oscillations, which is most studied as a mechanism for low-scale leptogenesis 
\cite{Akhmedov:1998qx,Asaka:2005pn}, generates an asymmetry through the out-of-equilibrium \emph{production} of singlets and their subsequent annihilation; this differs from conventional leptogenesis, which generates an asymmetry through the singlets' decay. 
We now review the  mechanism of baryogenesis via singlet oscillations,
highlighting certain aspects of the parametric dependence of the asymmetry.

In this section we focus on  the minimal case, 
with exactly two massive Majorana singlets, $\chi_I$.  We couple $\chi_I$  to a SM  field, $\psi_\alpha$,  and a set of scalars, $\Phi_i$, which have the same SM gauge quantum numbers as $\psi_\alpha$:
\be\label{eq:ARS_model}
\mathcal{L} &\supset & - \frac{M_I}{2}\overline \chi_I^{\rm c}\chi_I - ( F^i_{\alpha I}\,\overline \psi_\alpha \chi_I \Phi_i + \mathrm{h.c.}).
\ee 
The standard Type-I see-saw mechanism is realized if the $\chi_I$ fields are the RH neutrinos, $\psi_\alpha$ are the left-handed lepton doublets, and there is a single $\Phi$ which is the SM Higgs field. However, different SM fermions $\psi_\alpha$ and scalars $\Phi_i$ can realize baryogenesis as well; in that case, the $\Phi_i$ must be new scalars. We have expressed the Yukawa couplings, $F_{\alpha I}$, in the basis where the $\chi_I$ Majorana masses are diagonal. 

As we will soon see, baryogenesis favors a low mass scale for the $\chi_I$, and in fact we are  mainly interested in scenarios with $M_I \lesssim$ MeV.   The $\chi$ masses are essential for inducing $\chi$ oscillations, but we can otherwise neglect them throughout the baryon-asymmetry calculation, and we take all $\chi$ interactions to respect the $U(1)_{\chi - \Phi}$ symmetry realized in the massless-$\chi$ limit.  Note that we label the singlet states so that   $\Phi$ decays produce $\overline\chi$ particles and $\Phi^{*}$ decays produce $ \chi$ particles\footnote{The corresponding helicity assignments depend on the  identity of the active fermions. For $\psi = Q_L$, the $\chi_I$ have positive helicity and the $\overline{\chi}_I$ have negative helicity;  for $\psi = u_R$ the opposite is true; \it{etc}.}.

\begin{figure}
        \includegraphics[width=3in]{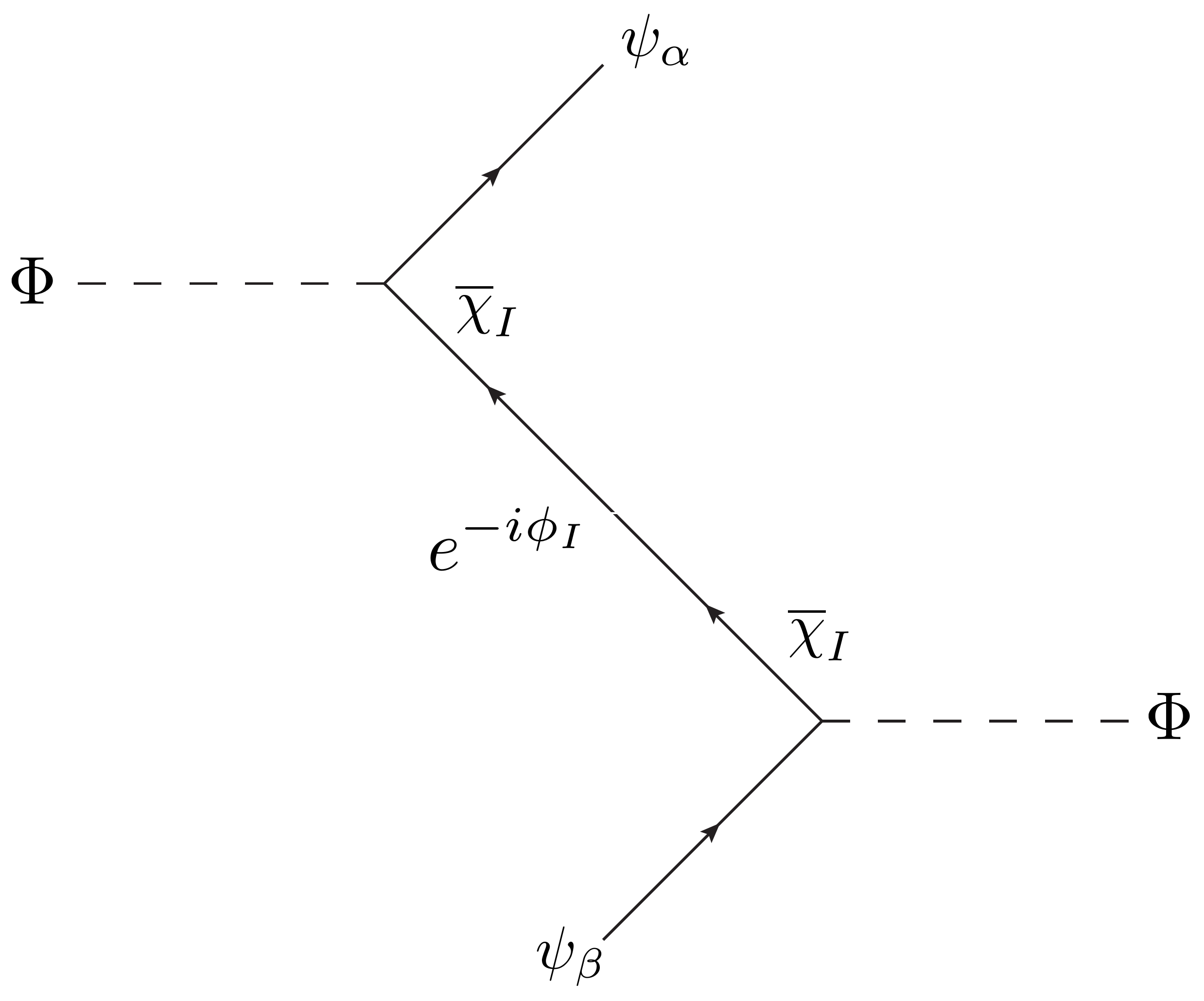}
   \caption{
Feynman diagram illustrating the process of $\chi_I$ production, propagation, and annihilation. First, the scalar $\Phi$ decays into the SM quark $\psi_\alpha$ and $\overline\chi_I$; the $\overline\chi_I$ field propagates and then annihilates with another SM field $\psi_\beta$ to reconstitute $\Phi$. The net reaction is $\psi_\beta \Phi \rightarrow \psi_\alpha \Phi$. In general, since  $\chi_I$ is out of equilibrium, the full process is a coherent sum over intermediate $\chi_I$ states. 
    }
    \label{fig:feynman_diagram}
\end{figure}

Consider first the case with a single scalar, $\Phi$. In order for the singlets not to come into equilibrium, their Yukawa couplings must be very small, $|F_{\alpha I}|\ll1$. As in most freeze-in scenarios, we assume that at initial times $n_{\chi_I}=0$. 
The decay $\Phi\rightarrow \psi_\alpha \overline\chi$ produces an interaction-basis state of $\chi$ fields which is  a coherent superposition of $\chi_I$ mass eigenstates. Because the $\chi$ scattering is out of equilibrium, each interaction-basis particle propagates coherently
\footnote{The mass splittings among $\chi_I$ states that we consider are sufficiently small that there is negligible decoherence in the production process.},
with the mass eigenstates $\chi_I$ acquiring  phases $e^{-i\phi_I}$, where $\phi_I=\int E_I\,dt$. At some later time, the $\chi$ fields annihilate with a potentially different SM fermion flavor $\psi_\beta$ into $\Phi$ via the inverse-decay process $\psi_\beta \overline\chi\rightarrow \Phi$. The net process is $\psi_\beta \Phi \rightarrow \psi_\alpha \Phi$, with coherent contributions from the propagation of both $\chi_I$ particles; see Fig.~\ref{fig:feynman_diagram}. The matrix element for this process is proportional to
\be
\mathcal{M}(\psi_\beta \Phi\rightarrow \psi_\alpha \Phi) \propto F_{\alpha 1}F_{\beta 1}^* e^{-i\phi_1} + F_{\alpha 2}F_{\beta 2}^* e^{-i\phi_2}.
\ee
The matrix element for the $CP$-conjugate process has $F\rightarrow F^*$, with propagation phases unchanged. The result is a $CP$-violating asymmetry 
\be
\Delta\Gamma_{\alpha\beta}
&\equiv& |\mathcal{M}(\psi_\beta \Phi\rightarrow \psi_\alpha \Phi)|^2-|\mathcal{M}(\overline \psi_\beta \Phi^*\rightarrow \overline \psi_\alpha \Phi^*)|^2\\
 &\propto& \mathrm{Im}\left(F_{\alpha 1}^* F_{\beta 2}^* F_{\beta 1} F_{\alpha 2}\right) \sin(\phi_2-\phi_1)\\
&=& \mathrm{Im}\left(F_{\alpha 1}^* F_{\beta 2}^* F_{\beta 1} F_{\alpha 2}\right)\sin\left[\int_{t_d}^{t_{id}}\!dt\,(E_2-E_1)\right].\quad
\label{eq:asym_qual}
\ee
This factor appears in the contribution to the rate of asymmetry generation at inverse-decay time $t_{id}$, due to $\chi$ particles produced with a particular momentum at decay time $t_d$.

\paragraph{Generation of Non-Zero Asymmetry.} 
Eq.~\eqref{eq:asym_qual}, which applies in the single-$\Phi$ case, 
implies that  $\mathcal{O}(F^4)$ asymmetries can arise within individual $\psi_\alpha$ flavors.  At this order, however, we do not get an an overall asymmetry in SM particles relative to antiparticles.  Summing over active flavors gives
\be
\sum_{\alpha,\beta}\,\Delta\Gamma_{\alpha\beta} \propto \mathrm{Im}\left[(F^\dagger F)_{12}(F^\dagger F)_{21}\right]=0,
\ee
confirming that the total $\psi$ asymmetry vanishes at $\mathcal{O}(F^4)$.

 An overall asymmetry is possible at higher order in $F$; for example, if the $\psi$ are leptons and the flavor asymmetries in $e$, $\mu$, and $\tau$ are destroyed at different rates, then a total asymmetry results at $\mathcal{O}(F^6)$  \cite{Asaka:2005pn}. If the $\psi$ are quarks,  the flavors are brought into equilibrium with one another through their couplings to the SM Higgs, and there are no asymmetries in individual flavors even at $\mathcal{O}(F^4)$; nevertheless, a source for the total asymmetry still results at $\mathcal{O}(F^6)$ if the number of $\chi_I$ is greater than or equal to three \cite{Abada:2018oly}. We discuss this further in Sec.~\ref{sec:ars_compare}.

To show that  the total $\psi$ asymmetry vanishes at $\mathcal{O}(F^4)$ in the single-scalar case, we assumed that the relevant interaction rates depend on the active fermion flavor only though the couplings $F_{\alpha I}$.  Thermal mass effects can invalidate this assumption.  These effects are  of higher order in some coupling, but they can be important for the top quark in particular, as we show in Sec.~\ref{sec:top_thermal_mass}.

Now we turn to the case with two scalars, $\Phi_1$ and $\Phi_2$. Remarkably, in this situation a total $\psi$ asymmetry results at $\mathcal{O}(F^4)$! The reason is that we now have two sets of Yukawa couplings, $F^1_{\alpha I}$ and $F^2_{\alpha I}$, and since $\mathrm{Im}\left[({F^1}^\dagger F^1)_{12}({F^2}^\dagger {F^2})_{21}\right] \neq 0$, an asymmetry can be generated.
In fact, the underlying mechanism works even if only a single active flavor of $\psi$ has couplings to $\chi_I$ and $\Phi_i$.  

We perform a direct calculation of the asymmetry in Sec.~\ref{sec:calculation} and show that it is indeed non-zero, but here we provide a qualitative understanding of how the mechanism works.  Take  $M_{\Phi_2}\gg M_{\Phi_1}$ for concreteness, 
and consider
two possible net processes initiated by $\Phi_2$ decay: (1) the feebly interacting ${\overline \chi}$ state produced in the decay  may not participate in any subsequent scattering, so that the net process is $\Phi_2 \rightarrow \psi \overline\chi$, or (2) the ${\overline \chi}$ state may later scatter to produce a $\Phi_1$ particle, so that the net process may be summarized as $\Phi_2 \rightarrow \Phi_1$ (with one $\psi$ emitted into the plasma and one $\psi$  absorbed from it).  The first net process induces equal and opposite changes to the total $\Phi$ and $\psi$ abundances, while the second has no effect on these total abundances.

The point is that $CP$ violation can lead to a difference between the fraction of decaying $\Phi_2$ particles that participate in $\Phi_2 \rightarrow \psi \overline\chi$ (as opposed to $\Phi_2 \rightarrow \Phi_1$) and the fraction of decaying $\Phi_2^*$ particles that participate in $\Phi_2^* \rightarrow \overline \psi \chi$ (as opposed to $\Phi_2^* \rightarrow \Phi_1^*$).  Consequently, a $\psi$ asymmetry arises at $\mathcal{O}(F^4)$.

One may worry that this asymmetry is cancelled when one includes processes initiated by $\Phi_1$ decay, with the roles of $\Phi_1$ and $\Phi_2$ reversed.  
However, because the number of $\Phi_2$ particles produced at lower temperatures is Boltzmann suppressed, the net process $\Phi_1 \rightarrow \Phi_2$ typically completes at high temperatures $T \sim M_{\Phi_2}$.  In effect, the processes  $\Phi_1 \rightarrow \Phi_2$ and $\Phi_2 \rightarrow \Phi_1$ are active during different cosmological time periods, and so the numbers of $\Phi$ particles involved in them are unrelated.

Moreover, it is possible that the timescale for $\chi$ oscillations to have an effect is comparable to the Hubble time at $T \sim M_{\Phi_1}$, but much longer than the Hubble time at $T \sim M_{\Phi_2}$.  In this case, $CP$ violation is negligible in $\Phi_1 \rightarrow \Phi_2$, which therefore cannot cancel an asymmetry produced by  $\Phi_2 \rightarrow \Phi_1$.

Note that it is crucial that the decay and the inverse decay happen at \emph{different} times, and that Hubble expansion changes the particle kinematics during the interval between those times.  This gives rise to the necessary differences in the thermal suppressions that apply at the decay and inverse-decay times.  If the universe were not expanding and instead in a state of exact equilibrium, then any Boltzmann suppressions would affect $\Phi_1 \rightarrow \Phi_2$ and $\Phi_2 \rightarrow \Phi_1$  processes equally, leading to a vanishing asymmetry. Indeed, we have checked that in the limit of zero Hubble expansion, or alternatively when $M_{\Phi_1}=M_{\Phi_2}$ such that both fields have the same thermal distribution, the net asymmetry is exactly zero at $\mathcal{O}(F^4)$.

\paragraph{$\chi$ Oscillations and Hubble Expansion.}
Having outlined the conditions for obtaining a non-vanishing total asymmetry in SM states, we now discuss the parametrics of the requisite $\chi$ oscillations.   
Taking the $\chi$ production time to be much earlier than the inverse-decay time $t$, we can approximate the oscillation factor in Eq.~(\ref{eq:asym_qual}) as
\be
\sin\left[\int_{0}^{t}\!dt'\,(E_2-E_1)\right]
& \simeq &
\sin\left[\int_0^t\,dt'\,\frac{\Msplit}{2p(t')}\right] \label{eq:asymmetry_rate_qual}\\
& \simeq &
\sin\left[\frac{\Delta M_{21}^2}{6p(t)H(t)}\right],
\ee
where $H(t)$ is the Hubble expansion rate at time $t$, $\Delta M_{21}^2 = M_2^2-M_1^2$, and we have assumed that the $\chi$ fields are highly relativistic, $E_i \approx p + M_i^2 / 2p$. 

Because both $p$ and $H$ decrease with  Hubble expansion, the phase factor oscillates at a frequency that increases with time. 
Assuming that $p\sim T$, we thus find that at times for which $\Delta M_{21}^2/T \ll H$, the argument of the sine function is very small and so is the asymmetry production rate. Conversely, for $\Delta M_{21}^2/T \gg H$, the asymmetry production rate undergoes rapid oscillations that time-average to zero. The bulk of the asymmetry is thus created when $\Delta M_{21}^2/T\sim H$, which corresponds to $T_{\rm osc} \sim (\Delta M_{21}^2 M_{\rm Pl})^{1/3}$. We are then led to expect the dependence
\begin{equation}
\mathrm{asymmetry} \propto (\Delta M_{21}^2)^{-2/3},
\label{eqn:rough_scale}
\end{equation}
or something roughly similar,
based on the assumption that the final asymmetry should scale approximately as $1/H(T_{\rm osc})$.

This estimate ignores the fact that asymmetry generation is suppressed below some temperature $T_{\rm cutoff}$.   If $\Delta M_{21}^2$ is too small, the oscillations develop too late to produce a significant asymmetry.    Regardless of the identity of $\psi_\alpha$, a non-zero final baryon asymmetry requires that the $\psi$ asymmetry is processed by sphaleron transitions that are only active in the unbroken electroweak phase at $T\gsim T_\text{ew}$.  So $T_{\rm cutoff}$ is in general at least as high as  $T_\text{ew}$.  If the $\psi_\alpha$ are quarks, then we need the oscillations to begin even earlier, because collider constraints require the masses of QCD-charged $\Phi_i$ to be $\gtrsim1$ TeV.   For $T_{\rm osc}\ll M_{\Phi_1}$, $\Phi_1$ inverse decays are highly suppressed at what would otherwise have been the time of maximal asymmetry generation.   In a model with heavy BSM scalars, then, we have $T_{\rm cutoff} \sim M_{\Phi_1}$.

The general point is that to avoid suppression of the asymmetry, we need $T_{\rm osc}\gtrsim T_{\rm cutoff}$, with the asymmetry maximized when the two temperatures are comparable.   According to the very rough estimates given above (which we will refine in our subsequent calculations), this corresponds to a $\chi$ mass-squared splitting of $\Delta M_{21}^2 \sim \mathrm{keV}^2$ for $T_{\rm cutoff} = T_\text{ew}$. Thus, freeze-in baryogenesis naturally favors light $\chi$ fields, $M_\chi\sim$  keV. However, an asymmetry can still be obtained for heavier $\chi$ fields, provided they are highly degenerate.  In the ARS scenario, $M_\chi\sim$ GeV whereas the mass splitting $M_2-M_1$ is often eV or smaller (although there are exceptional cases with non-degenerate spectra \cite{Drewes:2012ma}). In any case, the preference for small singlet masses makes the $\chi$ states kinematically accessible at laboratory experiments.

For $T_\text{osc} \gg T_\text{ew}$, $\chi$ oscillations have time to become rapid before sphaleron decoupling, and in this regime one finds that the standard ARS model does indeed exhibit the scaling of Eq.~(\ref{eqn:rough_scale}); see Eq.~(\ref{eq:ARS_final}) below.   
More generally, the exact dependence on 
$\Delta M_{21}^2$ in the rapid-oscillation regime is slightly model-dependent.  For example, in the decoupled-$\Phi_2$ regime, the two-scalar scenarios of Sec.~(\ref{sec:calculation}) lead to a  final asymmetry that scales as $ (\Delta M_{21}^2)^{-1}$ for $T_\text{osc} \gg M_{\Phi_1}$; see Eqs.~(\ref{eqn:YB_large_Mphi_alt}), (\ref{eqn:beta_tilde}),  and (\ref{eqn:aymp_beta}).

\paragraph{Survival of the Asymmetry.} 
In subsequent sections, we will focus on scenarios in which the only non-gauge interactions of $\Phi_i$ are those of Eq.~(\ref{eq:ARS_model}), and with $\langle \Phi_i \rangle = 0$.
Then all $\Phi_i$ interactions conserve both $B$ and $L$, with $\Phi_i$ carrying the same charges as $\psi_\alpha$.  
Thus, it seems that when the $\Phi_i$ decay (as they must), all asymmetries are destroyed!   This, however, ignores two important effects. The first is that SM spectator effects (including $B$- and $L$-violating sphaleron processes) distribute the $\psi_\alpha$ asymmetry into all SM fermions, causing the $B$ and $L$ stored in SM fermions to differ in magnitude from those stored in the $\Phi_i$.  The second is that the connection between  quark and lepton asymmetries is broken after the electroweak phase transition (when $B$ and $L$ become separately conserved). If one of the $\Phi_i$ states does not decay entirely until after the electroweak phase transition, then although its eventual decays cancel the $B-L$ asymmetry that had been stored in SM fermions, equal and opposite $B$ and $L$ asymmetries survive.

These arguments suggest that the lifetime of the lightest scalar $\Phi_i$ should be at least comparable to the inverse Hubble scale at $T_{\rm ew}$. By an interesting numerical coincidence, $H(T_{\rm ew})^{-1} \sim 1\,\,\mathrm{cm}$, and hence one of the $\Phi_i$ states is typically long lived on collider scales. Since the $\Phi_i$ carry SM gauge charges, they can be produced at colliders and give rise to long-lived particle signatures (see Sec.~\ref{sec:signals}).

\subsection{Calculation of the $\mathcal{O}(F^4)$ Asymmetry in  Two-Scalar Scenarios}
\label{sec:calculation}
In this section we perform a detailed calculation of the $\mathcal{O}(F^4)$ baryon asymmetry for the two-$\chi$,  two-$\Phi$ case: $I=1,2$ and $i=1,2$  in Eq.~(\ref{eq:ARS_model}).  Our final expression for the $\mathcal{O}(F^4)$ asymmetry, as expressed in Eqs.~\eqref{eqn:final_I_fn} and \eqref{eq:final_asymmetry}, is applicable for any choice of the SM fermions $\psi_\alpha$, including couplings to all three generations of active flavors.  One can think of these two-scalar models as a proxy for more general scenarios in which a coherent background of oscillating DM particles (of unspecified origin) participates in inverse decays of BSM particles carrying SM charges.  
The ``decoupled-$\Phi_2$'' results presented in Section \ref{sec:twoscalar} are particularly relevant from that perspective.  Those results also apply for arbitrary $\psi_\alpha$, but we do assume $M_{\Phi_1} \gg T_\text{ew}$ to arrive at the simplified final expressions of Eqs.~(\ref{eqn:YB_large_Mphi_alt}) and (\ref{eqn:large_Mphi_int_alt}).

Although the results of this section are broadly applicable,  we will refer to the $\psi_\alpha$ as quarks to align with our focus in later sections, which present numerical results for the particular case in which the $\psi_\alpha$ fields are right-handed up-type quarks, with $\Phi_1$ and $\Phi_2$ carrying charges 
$(\mathbf{3},\mathbf{1},2/3)$  under the $\mathrm{SU}(3)_{\rm c}\times\mathrm{SU}(2)_{\rm w}\times \mathrm{U}(1)_{\mathcal Y}$ SM gauge group.  We focus on  QCD-charged scenarios because the parameters for viable baryogenesis and the model phenomenology are very different from the well-studied ARS leptogenesis scenario where the $\psi_\alpha$ are leptons. One can conceive of various scenarios in which a multiplet of SM-singlet DM particles couples feebly to a SM field and BSM fields.  We leave study of some of these alternatives to future work.

To streamline our discussion and derivation, we couple only a single flavor of SM fermion, ``$Q$,''  to $\Phi_i$ and $\chi_I$, so that our Lagrangian becomes
\be\label{eq:our_Lagrangian}
\mathcal{L} &\supset&- \frac{M_I}{2} \bar\chi_I^{\rm c}\chi_I 
 - \left(F^1_{I} \,\bar Q\Phi_1\chi_I - F^2_{I}\,\bar Q \Phi_2\chi_I + \mathrm{h.c.} \right).
 \quad\quad
\ee
However, we will express our final results so that they apply equally well to the three-active-flavor case.  We impose a $Z_2$ symmetry  under which $\Phi_1$, $\Phi_2$, and $\chi_I$ have charge $-1$ and the SM fields all have charge $+1$. In this case, the $\chi_I$ states are also stable DM candidates, and the neutrino-portal coupling 
$\bar{L}_\alpha H \chi_I$ is forbidden.

We adopt a perturbative approach to the calculation of the baryon asymmetry, both for physical clarity and because the requirement of $\chi$ as a viable DM candidate mandates that we are in the weak-washout regime. 
Strong washout effects do become important when we study single-scalar models without the DM constraint in Sec.~(\ref{sec:no_DM}), and there we adopt a fully numerical treatment of the relevant system of quantum kinetic equations.

Our perturbative calculation uses Maxwell-Boltzmann statistics throughout.  It  also neglects thermal contributions to the $\Phi_i$ and $Q$ masses, along with the production and scattering of $\chi$ from $2\leftrightarrow2$ processes. These effects are most important at $T\gg M_{\Phi_i}$, while  decays and inverse decays predominantly occur at $T\sim M_{\Phi_i}$. Thus, neglecting thermal masses is not expected to have a huge effect. The principal exception is if $\chi$ oscillations occur at $T\gg M_{\Phi_i}$, in which case the result presented here will underestimate the production and scattering rates.  

With these simplifications, our final expressions for the baryon asymmetry appear below in Eqs.~\eqref{eqn:final_I_fn} and \eqref{eq:final_asymmetry}.  We show in Appendix \ref{sec:thermal_mass_appendix} that generalizing these expressions to include thermal masses has only a modest quantitative impact.   We further find in Appendix \ref{sec:kinetic_two_scalar} that results based on Eqs.~\eqref{eqn:final_I_fn} and \eqref{eq:final_asymmetry} match rather well with what we get by numerically solving the  quantum kinetic equations for the $\chi$ and ${\overline \chi}$ density matrices, taking into account thermal masses,  quantum statistics, and back-reaction/washout effects.

The calculation proceeds in four steps:~first, a coherently propagating population of $\chi_I$ states is produced from the decay of the heavier scalar $\Phi_2$. Second, some part of this population subsequently re-scatters into $\Phi_1$. Third, we account for the phases from the coherent propagation and compute the difference in rates between $\bar\chi Q\rightarrow \Phi_1$ and $\chi \bar Q\rightarrow \Phi_1^*$, which leads to a baryon asymmetry\footnote{In our full expression for the baryon asymmetry, we also track the asymmetry resulting from the opposite process where $\chi$ particles are initially produced from $\Phi_1$ decay and re-scatter into $\Phi_2$. However, we focus  on only one of these processes for now, as it is straightforward to obtain the other by interchanging $\Phi_1\leftrightarrow\Phi_2$.}. Fourth, we evolve the asymmetry down to the electroweak phase transition temperature, $T_{\rm ew}$, to determine the size of the ultimate baryon asymmetry.

\subsubsection*{Step 1:~$\chi$ Production}
The important $\bar\chi$ ($\chi$) production mode is from $\Phi_2\rightarrow Q\bar\chi$ ($\Phi_2^*\rightarrow \bar Q\chi$); the $\chi$ population from $\Phi_1$ decay can be found by simply interchanging $F^2\leftrightarrow F^1$ in this calculation. 
We wish to calculate the spectrum of $\bar\chi$ particles present at a dimensionless time $z\equiv T_{\rm ew}/T$ corresponding to temperature $T$, since the $\chi$ momentum affects the oscillation phase according to Eq.~\eqref{eq:asymmetry_rate_qual}; 
we must consider contributions from  $\Phi_2$ decays that occur at any $z_2\equiv T_{\rm ew}/T_{\Phi_2\,\rm decay} < z$.
 The energy of a $\bar\chi$ particle produced at time $z_2$ is not preserved by the Hubble expansion; however, since $\chi$ is relativistic throughout the asymmetry generation process, $E_{\bar\chi}\approx p_{\bar\chi}$, and the co-moving energy $y\equiv E_{\bar\chi}/T$ is constant with respect to time.

The differential $\overline \chi$ production in time $dt_2$ due to decays of $\Phi_2$ particles having momenta in some window
$d^3{\bf p}_{\Phi_2}$ is
\be\label{eq:decay_rate}
dY^{\bar\chi} &=& \frac{1}{s(z_2)}\frac{M_{\Phi_2}}{E_{\Phi_2}}\,\Gamma_{\Phi_2}\,\frac{g_\Phi}{(2\pi)^3}f^{\rm eq}_{\Phi_2}(E_{\Phi_2})\,d^3{\bf p}_{\Phi_2}\,dt_2,
\ee
where $Y^{\bar\chi}\equiv n^{\bar\chi}/s$ is the co-moving number density of $\bar\chi$ particles,  $s$ is the entropy density, $g_\Phi$ is the number of $\Phi_2$ degrees of freedom ($g_\Phi = 6$ if $Q=Q_{\rm L}$ and $g_\Phi = 3$ for $Q=u_{\rm R}$ or $Q=d_{\rm R}$), and we have included a time dilation factor to account for the fact that the plasma-frame decay rate of $\Phi_2$ is slower than its rest-frame decay rate. 
We may express  $d^3{\bf p}_{\Phi_2}$ as $E_{\Phi_2}|{\bf p}_{\Phi_2}|dE_{\Phi_2} \,d\phi \,d\cos\theta$, where $\cos\theta$
is the angle between the $\Phi_2$ momentum in the plasma frame and the $\bar\chi$ momentum in the $\Phi_2$ rest frame.
Assuming Maxwell-Boltzmann statistics for $\Phi_2$ and integrating over $\phi$, we then have
\be
dY^{\bar\chi} &=& \frac{M_{\Phi_2}}{s(z_2)}\,\Gamma_{\Phi_2}\,\frac{g_\Phi}{4\pi^2}e^{-E_{\Phi_2} z_2/T_{\rm ew}}|{\bf p}_{\Phi_2}|\,d\cos\theta\, dE_{\Phi_2} \;dt_2.\nonumber
\ee
Neglecting the thermal mass for  $Q$, we find $y= z_2(E_{\Phi_2}+|{\bf p}_{\Phi_2}|  \cos\theta)/(2T_{\rm ew})$, which restricts $E_{\Phi_2}\ge E_{\bar\chi} + M_{\Phi_2}^2/4E_{\bar\chi}$.  Finally, we  change variables from $(\cos\theta, t_2)$ to $(y,z_2)$, and we integrate over $E_{\Phi_2}$ to obtain
\begin{multline}
\label{eq:chi_abundance}
dY^{\bar\chi}=\frac{45g_\Phi}{4\pi^4g_{*}}\,\frac{M_{\Phi_2}\Gamma_{\Phi_2}M_0}{T_{\rm ew}^3}e^{-y}\\
\times e^{-M_{\Phi_2}^2 z_2^2/(4T_{\rm ew}^2 y)} z_2^2 \,dy\,dz_2,
\end{multline}
where $M_0\approx M_{\rm Pl}/(1.66\sqrt{g_*})\approx7\times10^{17}$ GeV is defined so that the Hubble expansion rate is $H=T^2/M_0$, and $g_*$ is the effective number of relativistic degrees of freedom. We thus have an expression for the (co-moving) number density of $\bar\chi$ particles with a particular co-moving energy $y$ produced at time $z_2$. 

In addition to being relevant for the baryon asymmetry, this abundance of $\chi+\bar\chi$ particles can also account for the observed abundance of DM. Assuming that the Yukawa coupling is sufficiently small that $\chi$ is not brought into equilibrium, Eq.~\eqref{eq:chi_abundance} gives the leading result for the $\chi+\bar\chi$ abundance in perturbation theory.  Integrating over all $y$ and $z_2$ gives
the summed abundance of all $\chi$ and ${\overline \chi}$ particles produced by $\Phi_i^{(*)}$ decays $(i = 1,2)$:
\be
Y^{\chi+\bar\chi}_{i}
 = \frac{135 g_\Phi}{4 \pi^3 g_*}   \left( \frac{T_{\rm ew}}{M_{\Phi_i}}\right)^{2}  
\left( 
 \frac{\Gamma_{\Phi_i}}{H_{\rm ew}}
 \right)
,
\label{eqn:chi_Y_from_Phi_decay}
\ee
where $H_{\rm ew}$ is the Hubble expansion rate at sphaleron decoupling.

\subsubsection*{Step 2:~Inverse Decay}
In Eq.~\eqref{eq:chi_abundance}, we have calculated the abundance of $\bar\chi$ with co-moving energy $y$ and production time $z_2$. This abundance then leads to the inverse decay process $\bar\chi Q\rightarrow \Phi_1$, which changes the abundance of the field $\Phi_1$. Note that there is also a process $\bar\chi Q\rightarrow \Phi_2$, but according to the arguments of Sec.~\ref{sec:qualitative}, this cannot lead to an asymmetry at $\mathcal{O}(F^4)$. We also emphasize that \emph{any} primordial process that populates a coherent superposition of $\chi_1$ and $\chi_2$ states can lead to an asymmetry from Steps 2--4 outlined here, independent of their origin.

We calculate the number of $\bar\chi Q\rightarrow \Phi_1$ inverse decays that occur between times $z_1$ and $z_1+dz_1$, where as usual, $z_1\equiv T_{\rm ew} / T_1$. The Boltzmann equations specify that the inverse decay rate is the same as the decay rate, but with the substitution of the distribution functions as $f_{\Phi_1}^{\rm eq}(E_{\Phi_1})\rightarrow f_{\bar\chi} f_Q^{\rm eq}(E_Q)$. The limits of the phase-space integrals are otherwise unchanged. We already know the decay rate as a function of $f_{\Phi_1}(E_{\Phi_1})$:~we simply take Eq.~\eqref{eq:decay_rate}, substitute $\Phi_2\rightarrow \Phi_1$, and make the substitution $f_{\Phi_1}^{\rm eq}(E_{\Phi_1})\rightarrow f_{\bar \chi} f_Q^{\rm eq}(E_Q)$. Furthermore, conservation of energy dictates that $E_Q = E_{\Phi_1}-E_{\bar\chi} = E_{\Phi_1}-yT$, and thus assuming Boltzmann statistics,
\be
f^{\rm eq}_Q(E_Q) &=& e^y f^{\rm eq}_{\Phi_1}(E_{\Phi_1}).
\ee
 Substituting and integrating over $E_{\Phi_1}$ gives a similar result to before:
\be
dY^{\Phi_1} &=& \frac{45g_\Phi}{4\pi^4g_{*}}\,\frac{M_{\Phi_1}\Gamma_{\Phi_1}M_0}{T_{\rm ew}^3}e^{-M_{\Phi_1}^2 z_1^2/(4T_{\rm ew}^2 y)} z_1^2 \,dy\,dz_1\,f_{\bar\chi}.\nonumber
\ee
Finally, we have that 
\be
\frac{dY^{\bar\chi}}{dy} &=& \frac{45}{4\pi^4 g_{*}}\,y^2f_{\bar\chi},
\ee
and so
\be\label{eq:phi_diff}
dY^{\Phi_1}&=&  \frac{g_\Phi M_{\Phi_1}\Gamma_{\Phi_1}M_0}{T_{\rm ew}^3}\\
&&\quad\quad e^{-M_{\Phi_1}^2 z_1/(4T_{\rm ew}^2 y)} \frac{z_1^2}{y^2} \,dz_1dY^{\bar\chi}.
\ee
 We can readily substitute the result from Eq.~\eqref{eq:chi_abundance}, or the distribution $f_{\bar\chi}$ from any other out-of-equilibrium $\bar\chi$ production process. To obtain the abundance of $\Phi_1^*$ from the $CP$-conjugate process, we replace $F\rightarrow F^*$.

\subsubsection*{Step 3:~Oscillations and Asymmetry}

The results of Steps 1 and 2 are valid for the single-$\chi$ case.  
We now modify those results to apply when linear combinations of $\chi$ mass eigenstates propagate 
coherently between the points of $\chi$ production and $\chi$ annihilation.

Our single-$\chi$ result has
\be
d Y^{\Phi_1} \propto \Gamma_{\Phi_1}\Gamma_{\Phi_2}\propto |F^1|^2 |F^2|^2,
\ee
where $F^1$ and $F^2$ are the couplings to $\Phi_1$ and $\Phi_2$, respectively.
Consistent with the arguments of Section \ref{sec:qualitative}, we replace 
\be
|F^1|^2 |F^2|^2 &\rightarrow& \left|F_1^2 {F_1^1}^*e^{-i\phi_1} + F_2^2{F_2^1}^*e^{-i\phi_2}\right|^2
\ee
for the two-$\chi$ case.  That is, we sum coherently over the production and inverse decay processes mediated by different $\chi_I$ mass eigenstates. The phases are calculated from the time of  $\Phi_2$ decay to the time of  $\Phi_1$ inverse decay, and the physical phase is the difference\footnote{Our notation is slightly confusing because there are two types of scalar $\Phi_{1,2}$, as well as  two types of fermion $\chi_{1,2}$. The phases $\phi_{1,2}$ and energies $E_{1,2}$ refer to the propagation of the $\chi_{1,2}$ mass eigenstates, while the times $t_{1,2}$ refer to decay/inverse-decay of $\Phi_{1,2}$. }
\be
\phi_2-\phi_1 &=& \int_{t_2}^{t_1} \,dt\left(E_2-E_1\right)\\
&\approx& \int_{t_2}^{t_1} \,dt\,\frac{\Delta M_{21}^2}{2yT}\\
&=& \frac{\Delta M_{21}^2 M_0}{6T_{\rm ew}^3}\,\frac{z_1^3-z_2^3}{y}.
\ee
The coherent oscillation phase thus depends on the co-moving energy of the propagating singlet ($y$), as well as the times of production ($z_2$) and scattering ($z_1$).

Combining this with the result of Eq.~\eqref{eq:phi_diff}, we find
\begin{multline}
dY^{\Phi_1}-dY^{\Phi_1^*} \propto 4\,\mathrm{Im}\left({F_1^1} {F_1^2}^* {F_2^2} {F_2^1}^*\right)\\
\times \sin\left(\frac{\Delta M_{21}^2M_0}{6T_{\rm ew}^3}\,\frac{z_1^3-z_2^3}{y}\right).
\end{multline}
Finally, an analogous calculation gives an asymmetry in $\Phi_2$ from the process $\Phi_1\rightarrow Q\bar\chi$, $\bar\chi Q\rightarrow \Phi_2$; the result is found by simply interchanging the $\Phi$ index $1\leftrightarrow2$ in all of our results so far.

For an asymmetry to be generated,  $\chi$ production, propagation, and annihilation must all be coherent processes \cite{Akhmedov:2017mcc}.  Given the relatively large $\Phi - \Phi^*$ annihilation rate to gluons, $\Gamma_{\Phi,\text{col}} \gsim 10$ GeV, the overall energy uncertainty in $\Phi$ decays and inverse decays is many orders of magnitude larger than the energy splitting between $\chi$ mass eigenstates, $\Delta E \sim \frac{\Delta M^2}{E}$.  Therefore, we do not expect coherence loss in $\chi$ production or annihilation to be an issue.  Propagation decoherence seems more likely to be important.  In the wave-packet picture, the group velocities of the constituent $\chi$ mass eigenstates differ by $\Delta v \sim \Delta M^2/E^2$.  Approximating the spatial spread in the $\chi$ wave packet to be $\sigma_x \sim \Gamma_{\Phi,\text{col}}^{-1}$, the requirement that the spatial separation between the two mass eigenstates remains less than $\sigma_x$ leads to a coherence time of $t_\text{coh} \sim E^2/( \Gamma_{\Phi,\text{col}} \Delta M^2)$.  This coherence time is longer than the time for oscillations to develop, $t_\text{osc} \sim E/\Delta M^2$, provided that $\Gamma_{\Phi,\text{col}} \lesssim E$ is satisfied.  The scale of the $\chi$ energy $E$ is set by the larger of $M_\Phi$ and the temperature $T$.  We therefore expect  $\Gamma_{\Phi,\text{col}}$ to be perturbatively smaller than $E$, which would imply that $\chi$ coherence survives long enough for oscillations to have an effect.  While these rough, qualitative considerations are reassuring, a more careful study of decoherence in this framework is certainly merited.

\subsubsection*{Step 4:~From $\Phi$ Asymmetry to Baryon Asymmetry}
The interactions of Eq.~(\ref{eq:our_Lagrangian}) conserve baryon number, with $B=1/3$ assigned to $\Phi_i$.  Taking only those interactions into account, we get equal and opposite baryon asymmetries in $\Phi$ and $Q$, and no final asymmetry survives once the $\Phi$ particles decay to $Q\bar\chi$. 

Spectator processes among the other SM quarks and leptons can, however, prevent this destruction from happening and directly connect the phenomenology of $\Phi_{1,2}$ to the baryon asymmetry. The \emph{rate} of asymmetry production in $\Phi$ is equal and opposite to the that in $Q$; however, the asymmetry in $Q$ is quickly distributed amongst all SM quark, lepton, and Higgs species via sphalerons and SM Yukawa interactions. By contrast, the $\Phi$ asymmetry is not distributed amongst any other particles. Thus, the baryon asymmetries stored in the SM and $\Phi$ sectors have different magnitudes when spectator processes are taken into account.

To make this quantitative, we  solve a  system of equations relating SM chemical potentials and abundances in equilibrium \cite{Harvey:1990qw}, with the  hypercharge and $B-L$ conservation equations modified to include the $\Phi_i$ abundances.  In this way we can relate asymmetries to the $B-L$ asymmetry in the SM sector:
\be
\delta Y^{\Phi_1}+\delta Y^{\Phi_2} &=& \mathcal{K}_\Phi \,\YBmL
\label{eqn:spec1}
\\
\YB &=& \mathcal{K}_B\,\YBmL.  
\label{eqn:spec2}
\ee
Here $\YB$ is the total baryon asymmetry, including that stored in $\Phi_{1,2}$, but we exclude the $\Phi$ asymmetries in calculating $\YBmL$.  Conservation of $B-L$ leads directly to $\mathcal{K}_\Phi = -3$, while the value of $\mathcal{K}_B$ depends on the gauge charges of $Q$:~we find $\mathcal{K}_B=-54/79$ for $Q=Q_{\rm L}$, $-63/79$ for $Q=u_{\rm R}$, and $-45/79$ for $Q=d_{\rm R}$\footnote{For the leptonic cases, we have $\mathcal{K}_\Phi=1$, along with $\mathcal{K}_B=25/79$ for $L_{\rm L}$ and  $\mathcal{K}_B = 22/79$ for $e_{\rm R}$.}.  

Taking the sphaleron-decoupling temperature to be $T_\text{ew} =  131.7$ GeV \cite{DOnofrio:2014rug}, we work in the approximation that electroweak-symmetric conditions apply for $z = T_\text{ew}/T  < 1$, transitioning abruptly to $B$ conservation for $z>1$.  The final baryon asymmetry is then  $\mathcal{K}_B/\mathcal{K}_\Phi (\delta Y^{\Phi_1}+\delta Y^{\Phi_2})_{z=1}$.  We estimate that this instantaneous-transition approximation introduces an error of at most $\sim 15\%$ in the final baryon asymmetry, based on the broken-phase equilibrium value for $\YB/\YBmL$.   We adopt a more careful treatment of sphaleron decoupling only for Sec.~(\ref{sec:no_DM}), in the context of a strong washout scenario that features potentially rapid variations in the $\Phi$ asymmetry at $z \sim 1$.   

The ultimate baryon asymmetry depends on the $\Phi_{1,2}$ asymmetries at sphaleron decoupling, which are in turn proportional to the fraction of $\Phi_{1,2}$ particles that survive until $z =1$. For each scalar, we therefore dress
the contribution to its asymmetry from inverse decays at $z_1$ with the survival factor
\be
S_{\Phi_i}(z_1) &=& \exp\left(-\int_{t_1}^{t_{\rm ew}}\,dt\,\langle\Gamma_{\Phi_i}\rangle\right) ,
\ee
 where $ \langle \Gamma_{\Phi_i} \rangle$ is the thermally averaged decay width.  
Neglecting thermal masses and adopting Maxwell-Boltzmann statistics, we get
\begin{eqnarray}
S_{\Phi_i} (z_1)& =&\exp
\left(
-\frac{\Gamma_{\Phi_i}}{H_\text{ew}}  
\int_{z_1}^{1} \! dz \;z\; \frac{\mathcal{K}_1\left( \frac{M_{\Phi_i }}{T_\text{ew}} z\right)}{\mathcal{K}_2 \left( \frac{M_{\Phi_i }}{T_\text{ew}} z\right)}
 \right),
 \label{eqn:survival}
\end{eqnarray}
where $\mathcal{K}_i$ are  modified Bessel functions of the second kind. Washout from $\Phi_i$ decay can also be taken into account by solving the quantum kinetic equations including both source and washout terms, as done in Appendix \ref{sec:kinetic_two_scalar}.

The survival of a substantial fraction of $\Phi_i$ scalars down to the electroweak scale suggests that $\Gamma_{\Phi_i}\lesssim H_{\rm ew}$ for at least one of the scalars. This leads to the conclusion that the decay length of one of the scalars satisfies $c\tau_\Phi\gtrsim1$ cm. This is interesting from a phenomenological perspective, since this is precisely the set of decay lengths that lead to long-lived particle signatures at colliders. The freeze-in baryogenesis mechanism therefore provides a very explicit link between the baryon asymmetry, the Hubble expansion rate at the electroweak phase transition time, and collider signatures. We explore collider signatures in more detail in Sec.~\ref{sec:signals}.

\subsubsection*{Final Result}
Putting together the results from the four steps of our calculation, we find the baryon asymmetry today equals the asymmetry at the time of the electroweak phase transition:
\be
\YB &=& \frac{45 g_\Phi^2}{256 g_*\pi^6}\,\frac{\mathcal{K}_B}{\mathcal{K}_\Phi}\,\frac{M_{\Phi_1}^2M_{\Phi_2}^2 M_{\rm 0}^2}{T_{\rm ew}^6} \nonumber\\
&&{}\times \mathrm{Im}\left({F_1^1} {F_1^2}^* {F_2^2} {F_2^1}^*\right)\left(
I_{12}-I_{21}\right),\\
I_{ij} &=& \int_0^\infty\,dy\,\frac{e^{-y}}{y^2}\int_0^1\,dz_1\,z_1^2\, S_{\Phi_i}(z_1)\,e^{-\alpha_i z_1^2/y} \nonumber\\
&&{} \int_0^{z_1}\,dz_2\,z_2^2\,e^{-\alpha_j z_2^2/y}\sin\left[\frac{\beta_{\rm osc}}{y}(z_1^3-z_2^3)\right],
\label{eqn:final_I_fn}
\ee
where $\alpha_i = (M_{\Phi_i}/2T_{\rm ew})^2$ and $\beta_{\rm osc}=M_0\Delta M_{21}^2/6T_{\rm ew}^3$. Again, we have neglected thermal masses throughout.

To make more transparent the connections between the baryon asymmetry and physical properties of the new states such as masses and decay widths, we re-parametrize the asymmetry as follows:
\be\label{eq:final_asymmetry}
\YB &=& \frac{45g_\Phi^2}{4\pi^4g_*}\,\frac{\mathcal{K}_B}{\mathcal{K}_\Phi}\,\mathcal{J}\left(\frac{M_{\Phi_1}}{T_{\rm ew}}\right)\left(\frac{M_{\Phi_2}}{T_{\rm ew}}\right)\nonumber\\
&&{}\times \left(\frac{\Gamma_{\Phi_1}}{H_{\rm ew}}\right)\left(\frac{\Gamma_{\Phi_2}}{H_{\rm ew}}\right)\left(
I_{12}-I_{21}\right),
\label{eqn:final_YB}
\ee
where we have used
\be
\Gamma_{\Phi_i} &=& \frac{\mathrm{Tr}\left[{F^i}^\dagger{F^i}\right]}{16\pi} M_{\Phi_i},
\label{eq:gam}
\ee
and defined the Jarlskog-like invariant $\mathcal{J}$  by
\be
4\,\mathrm{Im}\left({F_1^1} {F_1^2}^* {F_2^2} {F_2^1}^*\right) &=& \mathcal{J}\,\mathrm{Tr}\left[{F^1}^\dagger{F^1}\right]\mathrm{Tr}\left[{F^2}^\dagger {F^2}\right]. \nonumber
\ee
This  invariant can be parametrized in terms of six mixing angles,
\be\label{eq:jarlskog_invariant}
\mathcal{J} &=& \sin2\theta_1\sin2\theta_2\cos\rho_1\cos\rho_2\sin(\phi_2-\phi_1),
\ee
where
\be\label{eq:mixing_angles}
\cos\theta_i &=& \sqrt{\frac{({F^i}^\dagger {F^i})_{11}}{\mathrm{Tr}({F^i}^\dagger F^i)}},\\
\cos\rho_i &=& \frac{|({F^i}^\dagger F^i)_{12}|}{\sqrt{({F^i}^\dagger F^i)_{11}({F^i}^\dagger F^i)_{22}}},\\
\phi_i &=& \arg({F^i}^\dagger F^i)_{12}, \label{eq:phases}
\ee
with $0\le(\theta_i, \rho_i) \le \pi/2$.
Our derivation assumed couplings to a single quark flavor, 
in which case we should regard $F^1$ and $F^2$ as $\chi$-space row vectors in the above equations, consistent with the index placement in Eq.~(\ref{eq:ARS_model}).
However, Eqs.~(\ref{eqn:final_YB}-\ref{eq:phases}) apply equally well in the three-flavor case, $F^i_I \rightarrow F^i_{\alpha I}$.
The $\theta_i$ angles parametrize the relative strength of the coupling to $\chi_1$ vs.~$\chi_2$, 
while the $\rho_i$ angles parametrize the degree to which the couplings to  $\chi_1$ and $\chi_2$ are aligned in quark-flavor space; for a single quark flavor, $\cos\rho_i=1$. 
Finally, the $\phi_i$ give the relative phases.

Our final result is consistent with our arguments from Sec.~\ref{sec:qualitative}. In particular, we can specialize to the case of a single scalar by making $M_{\Phi_1}=M_{\Phi_2}$ and $F^1=F^2$ (and including only the $I_{12}$ term); in this case, the asymmetry vanishes at this order in perturbation theory, recovering the standard ARS result. In Sec.~\ref{sec:single_scalar}, we return to the single-scalar scenario, showing that 
asymmetries can arise at $\mathcal{O}(F^4 y_t^2)$ and at $\mathcal{O}(F^6)$
in the model where $\Phi$ couples to quarks, although even the $\mathcal{O}(F^6)$ asymmetry has a different parametric dependence than in ARS leptogenesis, due to the equilibration among quark flavors in the SM. We also note that, if we take $M_{\Phi_1}=M_{\Phi_2}$ and neglect washout effects,  we get $I_{12}-I_{21}=0$, in accordance with our arguments in Sec.~\ref{sec:qualitative} that the asymmetry should vanish in this limit. 

In the absence of  washout, our result is independent of a possible cross-quartic coupling,  $\lambda_{12} (\Phi_1^\dagger \Phi_2)^2 + \mathrm{h.c.}$ 
In that limit, $\Phi_1 \Phi_2^* \leftrightarrow \Phi_1^* \Phi_2$ scattering does not affect the final baryon asymmetry, which is determined by the total $\Phi$ asymmetry at sphaleron decoupling.  However, those scatterings can  impact decay-washout effects (which our perturbative result encodes in the $S_{\Phi_i}$ functions).   To arrive at Eq.~\eqref{eq:final_asymmetry}, we assumed  that the $\Phi_1/\Phi_1^{(*)}$ and  $\Phi_2/\Phi_2^{(*)}$ asymmetries evolve independently. In many cases of interest, for example  if the  $\chi$ oscillations necessary for  asymmetry generation begin at temperatures well below $M_{\Phi_2}$, this assumption is valid.  Moreover, when $\Phi_1 \Phi_2^* \leftrightarrow \Phi_1^* \Phi_2$ scattering does affect the final asymmetry it does not generally reduce it dramatically in viable parameter regions.  For simplicity, we therefore neglect $\Phi_1 \Phi_2^* \leftrightarrow \Phi_1^* \Phi_2$ scattering. 

We conclude by noting that a non-zero baryon asymmetry can still be obtained in the limit $M_{\Phi_2}\gg M_{\Phi_1}$. In this case, from the perspective of the low-energy effective theory, there exists a primordial coherent $\chi$ background that interacts \emph{once} to generate a baryon asymmetry. We therefore see that our two-scalar model readily generalizes to \emph{any} scenario where a non-thermal coherent $\chi$ ensemble is produced in the early universe. This could include, for example, production from inflaton decays, or from the decays of some other particle with different quantum numbers than $\Phi_1$. Thus, freeze-in baryogenesis can occur through any one of a large number of mechanisms of $\chi$ production in the Universe, provided there is a weak-scale state to allow late time $\chi$ scattering\footnote{In the context of electroweak baryogenesis models, asymmetry generation by Majorana fermion DM scattering has been considered for example in \cite{Cline:2017qpe}, where (unlike here) a chemical potential is first generated in a dark sector and subsequently transferred to the SM sector.}.

Starting with an assumed primordial $\chi$ background, one can use the interaction
\be
\mathcal{L} \supset- F_{\alpha I} \bar{L}_\alpha H \chi_I + \mathrm{h.c.}
\ee
to generate the baryon asymmetry via $\chi\bar L \rightarrow  H$.  That is, rather than introduce $\Phi_i$ at all, one can exploit the neutrino-portal coupling, forbidden in our  $Z_2$-symmetric models.
This takes us to a version of the ARS scenario in which we allow an unspecified source of $\chi$ production, presumably broadening the viable parameter space.  In this scenario, however, X-ray constraints on $\chi \rightarrow \nu \gamma$  rule out the DM being composed of those $\chi$ mass eigenstates that participate directly in the asymmetry generation.  In the models with $\Phi_i$, a sufficiently small neutrino-portal coupling can leave our DM abundance and baryon asymmetry calculations unaltered while still  having potentially observable consequences, as discussed in Sec.~\ref{sec:z2violate}.
\begin{figure*}
        \includegraphics[width=3.4in]{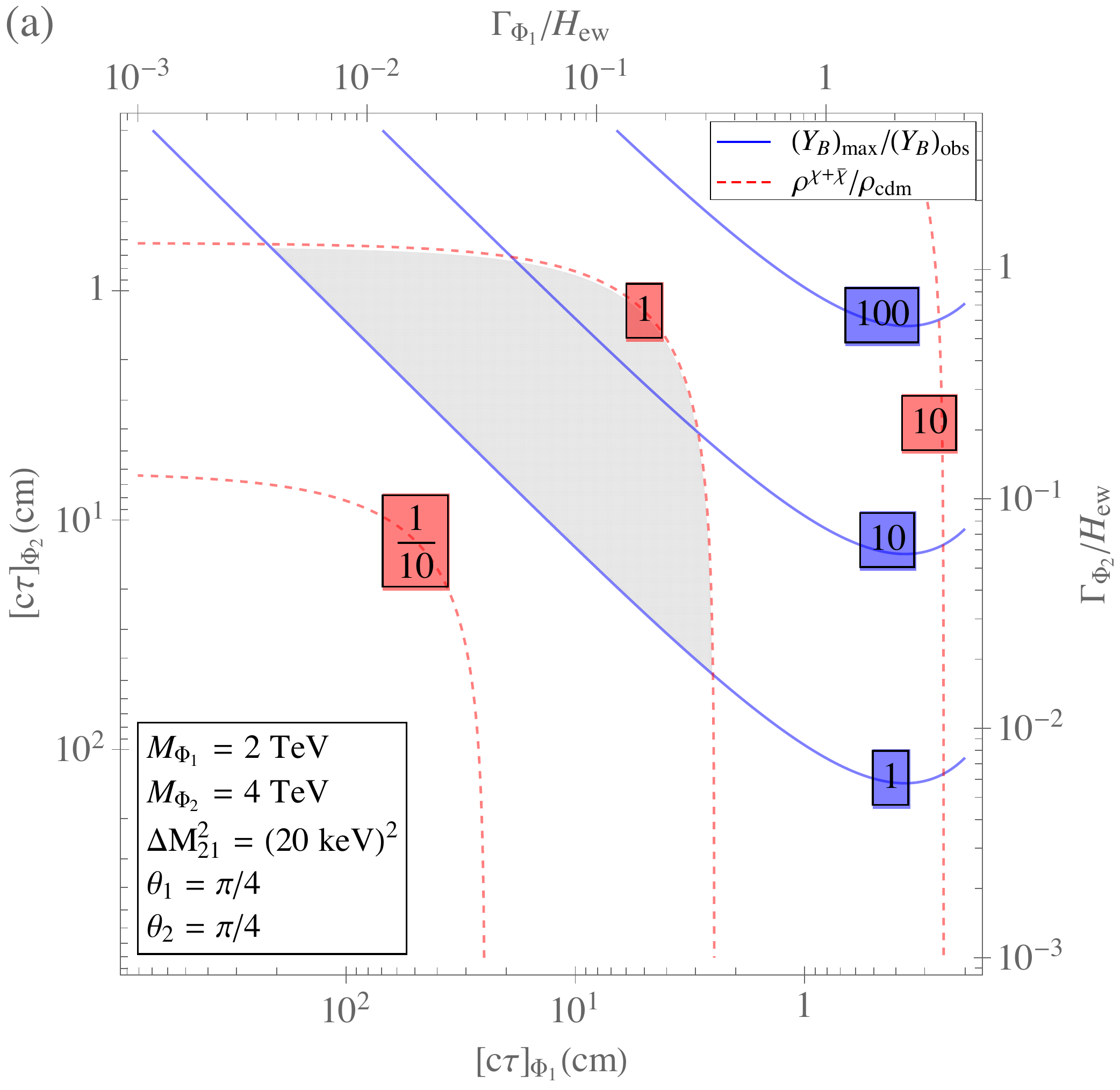}
        \quad\quad\quad
                  \includegraphics[width=3.05in]{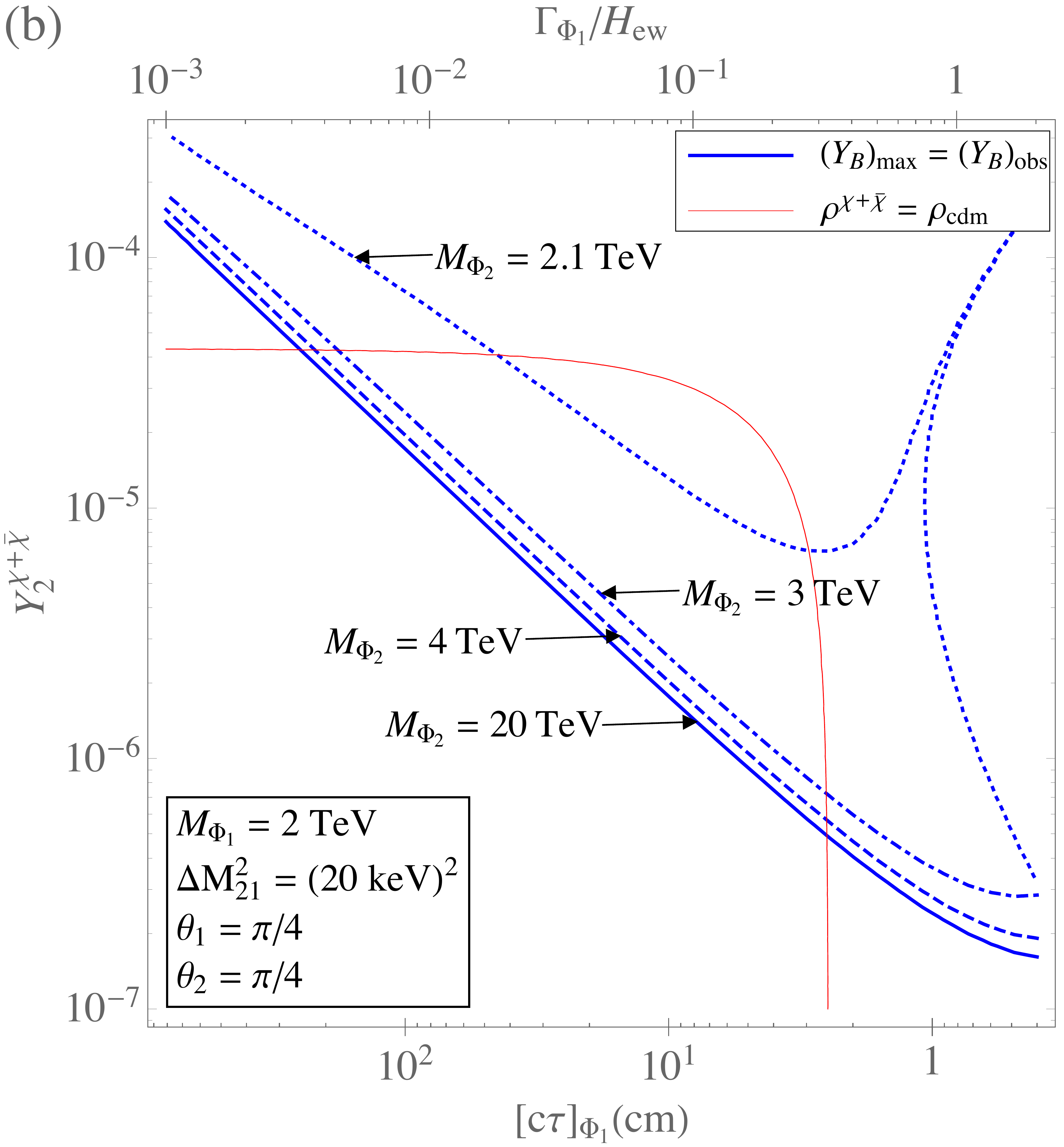}
                     \caption{
For the inputs indicated, contours of $(\YB)_\text{max}$ (blue) and $\rhotot$ (red), expressed in the left plot as ratios relative to the observed values.   In (a), $M_{\Phi_2}$ is held fixed, and the region in which a large enough baryon asymmetry can be achieved without over-producing DM is shaded;  (b) compares contours for various $M_{\Phi_2}$ as a function of the lifetime of $\Phi_1$ and the total $\chi+\overline\chi$ abundance produced in the decays of $\Phi_2$, $Y_2^{\chi+\bar\chi}$. 
For both plots, the baryon asymmetry is maximized for $|{\mathcal J}|= 1$, and we take the $\Phi$ quantum numbers to be those of $u_R$.  
For the DM   contours we assume
$\Mchitwo^2 \simeq \Msplit$ and neglect the energy density stored in the lighter $\chi$ mass eigenstate.  
    }
    \label{fig:tautauplot_1}
\end{figure*}

\section{Baryogenesis and Dark Matter with Two Scalars}
\label{sec:twoscalar}
In this section, we explore the parameters for which the freeze-in baryogenesis model of Sec.~(\ref{sec:calculation}) can simultaneously account for the baryon asymmetry and DM, finding a generic preference for masses $\lesssim5$ TeV and lifetimes $c\tau\gtrsim0.1$ cm for the lightest scalar $\Phi_1$.

Given fixed values of the other parameters, we can use Eqs.~\eqref{eqn:chi_Y_from_Phi_decay} and \eqref{eq:final_asymmetry} to calculate the DM abundance and $\YB$ as functions of $(\Gamma_{\Phi_1} , \Gamma_{\Phi_2})$, or equivalently, 
$(c\tau_{\Phi_1} , c\tau_{\Phi_2})$.
Fig.~\ref{fig:tautauplot_1}(a) shows results for benchmark parameters $M_{\Phi_1}=2$ TeV, $M_{\Phi_2}=4$ TeV, $\Mchione \ll \Mchitwo=20$ keV, and  $\theta_1=\theta_2=\pi/4$. 
The maximum possible baryon asymmetry consistent with these inputs, $(Y_B)_\text{max}$, is realized by choosing the other angles to give ${\mathcal J} = 1$ (as discussed below, the DM abundance depends on $\theta_1$ and $\theta_2$, but not on the other parameters determining ${\mathcal J}$). 

Fig.~\ref{fig:tautauplot_1}(a) shows that, for these inputs, a $\YB$ of around twenty times the observed value is possible, without over-producing DM.  Furthermore, to avoid over-production of DM,  both $\Phi$ particles must be  long-lived on collider scales, and the $\chi, {\overline \chi}$ abundances must  remain well below  equilibrium values, ensuring the validity of our perturbative calculation.

As already emphasized,  $M_{\Phi_2}$ can be much larger than $M_{\Phi_1}$  without suppressing the magnitude of the baryon asymmetry,  provided  the abundance of $\chi$ produced in $\Phi_2$ decays, $Y_2^{\chi+\bar\chi}$, is held fixed by increasing the couplings to $\Phi_2$.  This is evident in  Fig.~\ref{fig:tautauplot_1}(b), which shows contours of baryon asymmetry and $\chi/{\overline \chi}$ energy density in the $(c\tau_{\Phi_1},\Ytwotot)$ plane for various $M_{\Phi_2}$.
  
For the case with approximately  degenerate scalars  ($M_{\Phi_2} = 2.1$ TeV), the results are sensitive to our assumption of no $\Phi_1 \Phi_2^* \leftrightarrow \Phi_1^* \Phi_2$ scattering.  We see two $(Y_B)_\text{max}$ contours for that scenario, corresponding to whether the $\Phi/\Phi^*$ asymmetry at sphaleron decoupling is stored dominantly in $\Phi_1$ (with ${\mathcal J} = 1$) or $\Phi_2$ (with ${\mathcal J} = -1$). The asymmetry is almost entirely in $\Phi_1$ in the viable parameter space consistent with the DM constraint.

We now provide additional details underlying all of the results of this section, including Fig.~\ref{fig:tautauplot_1}.
First, in calculating the baryon asymmetry, we
replace the survival function in Eq.~(\ref{eq:final_asymmetry}) with its $z_1=0$ value,  $S_{\Phi_i}(0)$, which can then be taken outside the integrals in Eq.~(\ref{eqn:final_I_fn}).   That is, we approximate $\Phi$ production to be  at the time of reheating for the purpose of estimating washout via $\Phi$ decay, while still taking into account time dilation.  Given that asymmetry production by $\chi$ scattering dominantly occurs at temperatures well above $T_\text{ew}$, this is a reasonable approach.  We show in Appendix~\ref{sec:app_compare_calcs} that more careful treatments give similar results  (see Fig.~\ref{fig:full_survival_fig}).

Regarding the DM constraint, we adopt a $Z_2$-symmetric model, so that both $\chi$ mass eigenstates are  stable on cosmological time scales. We therefore require\footnote{Following Ref.~\cite{Tanabashi:2018oca}, we take $\Omega_{\rm cdm}h^2 = 0.1186$ and $\Omega_{B}h^2 = 0.02226$.  This gives
\be
\frac{\rho_{\rm cdm}}{s} = \left(  \frac{\Omega_{\rm cdm} h^2}{\Omega_{B}h^2}  \right)m_N \YB = 4.32 \times 10^{-10} \text{ GeV}
\ee
for $\YB = 8.65 \times 10^{-11}$ and a nucleon mass $m_N = 0.938$ GeV.  
 }
 \be
 \frac{\rhotot}{s}
\le
\frac{\rho_{\rm cdm}}{s}
= 
4.32 \times 10^{-10} \text{ GeV}.
\label{eqn:DMconstraint}
\ee
We are particularly interested in the case where this bound is saturated and the $\chi$ particles make up  all of the DM.

The total $\chi+\bar\chi$ abundance from the decay of the scalar $\Phi_i$, $Y^{\chi+\bar\chi}_{i}$, is given by Eq.~\eqref{eqn:chi_Y_from_Phi_decay}. Since the mixing angles $\theta_i$ defined in Eq.~\eqref{eq:mixing_angles} parametrize the relative couplings of $\Phi_i$ to the two $\chi$ mass eigenstates,  we find 
\be
\label{eqn:rho_tot}
 \frac{\rhotot}{s} = \overline{M}_\chi^{(\Phi_1)}  Y^{\chi+\bar\chi}_{1} +\overline{M}_\chi^{(\Phi_2)}  Y^{\chi+\bar\chi}_{2},
 \ee
where 
\be
\overline{M}_\chi^{(\Phi_i)} = \cos^2 \theta_i \Mchione + \sin^2 \theta_i  \Mchitwo
\ee
is the average mass of $\chi$ and $\overline{\chi}$  particles produced in $\Phi_i^{(*)}$ decays, weighted by abundance.

To simplify our analysis we focus first on the case in which the $\chi$ masses are hierarchical, 
\be
\Mchitwo \gg \Mchione,
\ee
to an extent that we can neglect $\Mchione$ entirely.  We consider the implications of having larger $\Mchione$ toward the end of this Section.   
For the remaining parameters, taking $\chi_1$ to be effectively massless maximizes the space that gives the correct baryon asymmetry consistent with the DM constraint of Eq.~\eqref{eqn:DMconstraint}.   In this hierarchical regime we take
\be
\Delta M_{21}^2 \simeq \Mchitwo^2
\ee
and
\be
\overline{M}_\chi^{(\Phi_i)}  \simeq  \sin^2 \theta_i  \Mchitwo, 
\label{eqn:Mchi_ave_hierarchical}
\ee
giving
\be
\Mchitwo( \sin^2 \theta_1 Y^{\chi+\bar\chi}_{1}+\sin^2 \theta_2 Y^{\chi+\bar\chi}_{2} )\leq \frac{\rho_{\rm cdm}}{s}
\quad
\label{eqn:DMconstraint2}
\ee
as our DM constraint. 
For the parameters adopted in Fig.~\ref{fig:tautauplot_1}(a), this translates roughly to $Y^{\chi+\bar\chi} \leq 4\times 10^{-5}$, consistent with our earlier claim that the DM constraint requires the $\chi$ particles to remain well out of equilibrium.

\subsubsection*{The decoupled-$\Phi_2$ regime}

Fig.~\ref{fig:tautauplot_1}(b) shows that, while the baryon asymmetry is reduced as $M_{\Phi_2}$ approaches $M_{\Phi_1}$, the masses need to be  close  to get a strong suppression.  Because we get qualitatively similar results for  $M_{\Phi_2} \gg M_{\Phi_1}$ as for modest hierarchies, $M_{\Phi_2} \gtrsim 2 M_{\Phi_1}$,  we work  in the ``decoupled-$\Phi_2$'' regime for the remainder of this section.  In this regime, the generation of the asymmetry can be factorized into the production of a $\chi$ abundance, which oscillates and then scatters into into $\Phi_1$ at a much later time. 

More precisely, we adopt an approximate expression for $\YB$  that applies when
 \be
M_{\Phi_2}  \gg M_{\Phi_1} \gg T_{\rm ew}
\label{eqn:largephimass}
\ee
and
\be
\frac{\Delta M_{21}^2 M_0}{M_{\Phi_2} ^3} \ll 1
\label{eqn:oscafterphi2}
 \ee
 are both satisfied.   
When Eq.~\eqref{eqn:oscafterphi2} is satisfied,  $\chi$ oscillations develop 
at temperatures $T\ll M_{\Phi_2}$.  We can then ignore the term with $I_{21}$ in Eq.~\eqref{eq:final_asymmetry}, because inverse decays of  $\Phi_2$ are highly Boltzmann-suppressed by the time oscillations begin.  Given that the $\Phi_2^{(*)}$ population annihilates away at high temperatures, we can furthermore take $t=0$ at the moment of $\chi$ production, when calculating the oscillation effect.  In the $I_{12}$ integral  of Eq.~\eqref{eqn:final_I_fn}, this amounts to neglecting the term with $z_2$ in the sine function.  Finally, Eq.~\eqref{eqn:largephimass}  allows us to extend the $z_1$ and $z_2$ integrations in Eq.~\eqref{eqn:final_I_fn} to infinity, because Boltzmann suppressions  of the $\Phi_2$ decay and $\Phi_1$ inverse decay rates effectively cut off the integrals at lower values of $z_1$ and $z_2$ in any case.
Using Eq.~\eqref{eqn:chi_Y_from_Phi_decay}, we can then approximate Eq.~\eqref{eq:final_asymmetry} by
\begin{multline}
\YB   \simeq 
\frac{8 g_\Phi \mathcal{K}_B \mathcal{J}}{3\pi^{1/2}\mathcal{K}_\Phi}
\Ytwotot \\
\times
  \left( \frac{T_{\rm ew}}{M_{\Phi_1}}\right)^{2}
 \left( 
 \frac{\Gamma_{\Phi_1}}{H_{\rm ew}}
 \right)
 S_{\Phi_1}(0)
\tilde{I}_{12} (\tilde{\beta}_\text{osc} ),
   \label{eqn:YB_large_Mphi_alt}
\end{multline}
with 
\begin{multline}
\tilde{I}_{12} (\tilde{\beta}_\text{osc} )= 
 \int_0^\infty \! \!  \! \! 
  dy \;y\, e^{-y} 
  \int_0^{\infty}\! \!  \! \! 
  dx\; x^{1/2} e^{-x}\\
  \times
   \sin\left[ {\tilde \beta}_\text{osc}\; x^{3/2} y^{1/2}\right]   
      \label{eqn:large_Mphi_int_alt}
\end{multline}
and 
\be
{\tilde \beta}_\text{osc}  =\frac{4\Delta M_{21}^2 M_0}{3M_{\Phi_1}^3}.
\label{eqn:beta_tilde}
\ee

\begin{figure}[t] 
          \includegraphics[width=3in]{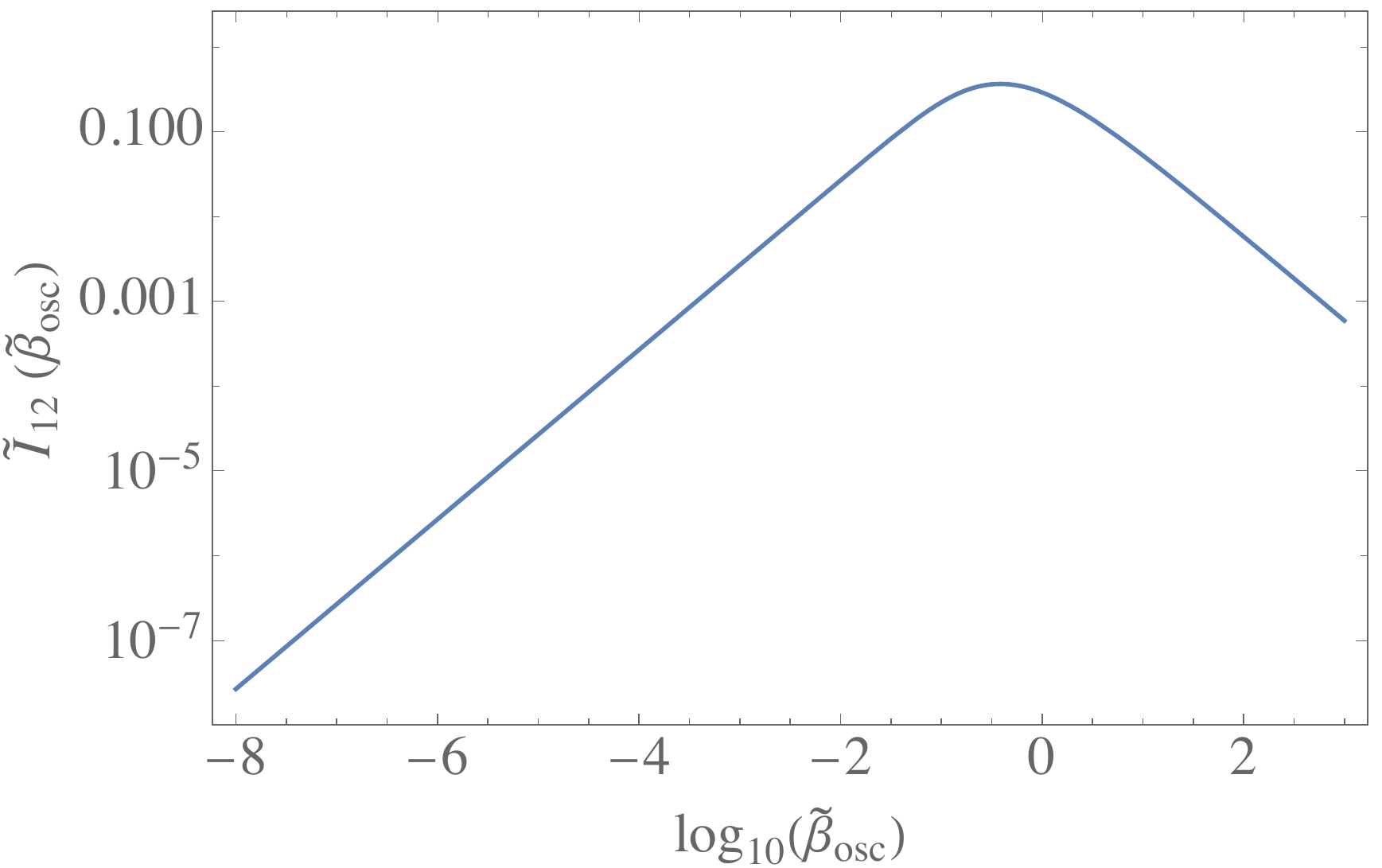}
   \caption{The function $\tilde{I}_{12} (\tilde{\beta}_\text{osc} )$ appearing in Eq.~(\ref{eqn:YB_large_Mphi_alt}), which gives the baryon asymmetry in the 
   decoupled-$\Phi_2$ regime.    }
   \label{fig:betaoscplot}
\end{figure}

Fig.~\ref{fig:betaoscplot} shows a plot of $\tilde{I}_{12} (\tilde{\beta}_\text{osc} )$.
The asymptotic behavior  is
\be
\tilde{I}_{12} (\tilde{\beta}_\text{osc} ) \simeq 
\begin{cases}
(3\sqrt{\pi}/2) \;\tilde{\beta}_\text{osc}  &  \tilde{\beta}_\text{osc} \ll1, \\
\sqrt{\pi}/(3 \; \tilde{\beta}_\text{osc})&  \tilde{\beta}_\text{osc} \gg 1,
\end{cases}
\label{eqn:aymp_beta}
\ee
and the maximum value $( \tilde{I}_{12})_\text{max} \simeq 0.364$ is attained for $\tilde\beta_\text{osc} \simeq 0.385$, corresponding to
\be
\left(
\sqrt{\Delta M_{21}^2}\right)_{\text{max }\tilde{I}_{12}} \simeq  20 \text{ keV}  \times \left( \frac{M_{\Phi_1}}{\text{TeV}} \right)^{3/2}.
\label{eqn:optimal_beta_osc}
\ee
Increasing or decreasing $\sqrt{\Delta M_{21}^2}$ by an order of magnitude from this value shifts ${\tilde \beta}_\text{osc}$ by two orders of magnitude and suppresses $\tilde{I}_{12}$ by a factor of roughly $\sim 3\times 10^{-2}$.  

\subsubsection*{Numerical results in the decoupled-$\Phi_2$ regime}
\begin{figure*} 
        \includegraphics[width=3.3in]{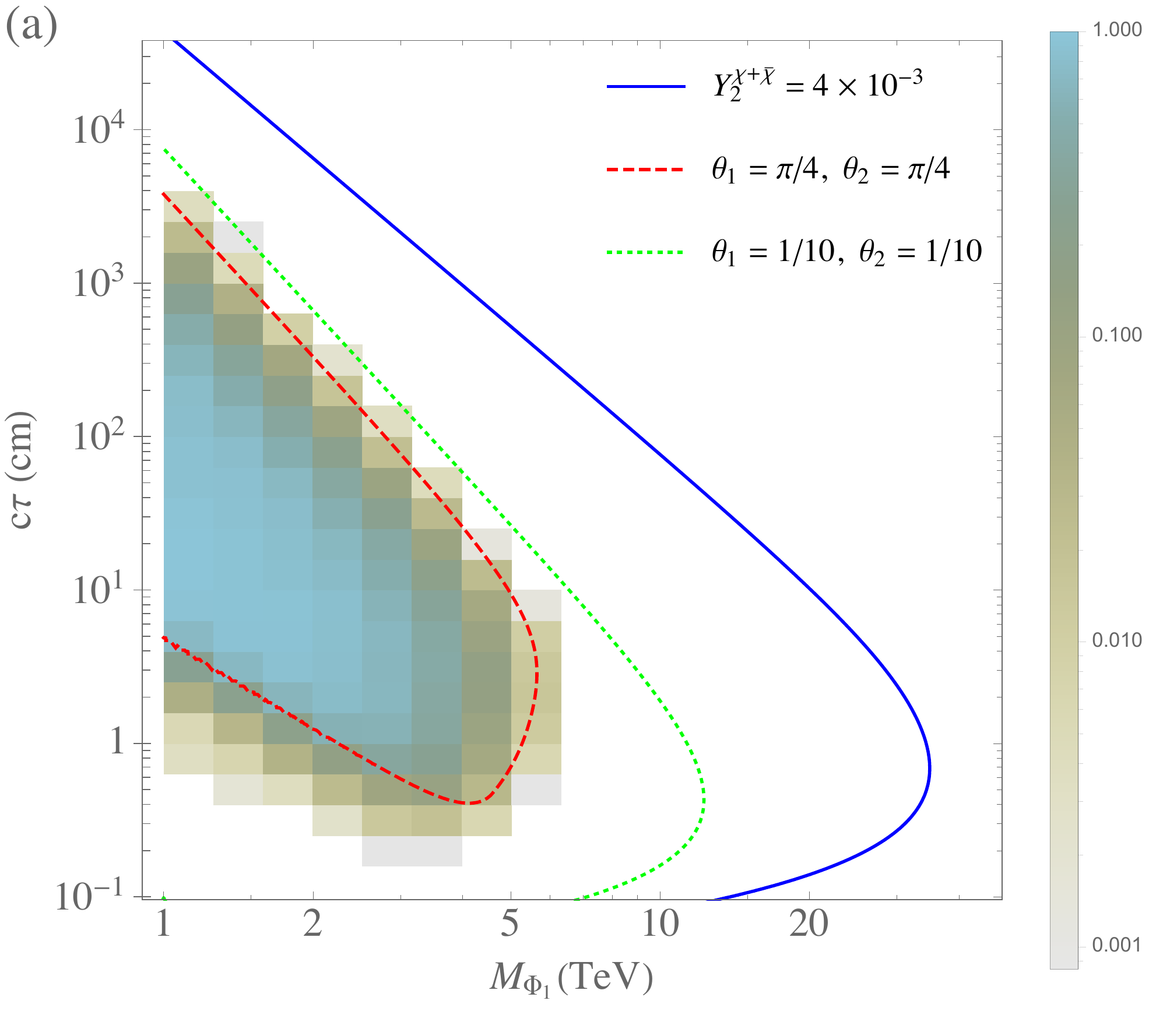}
        \quad\quad
     \includegraphics[width=3.3in]{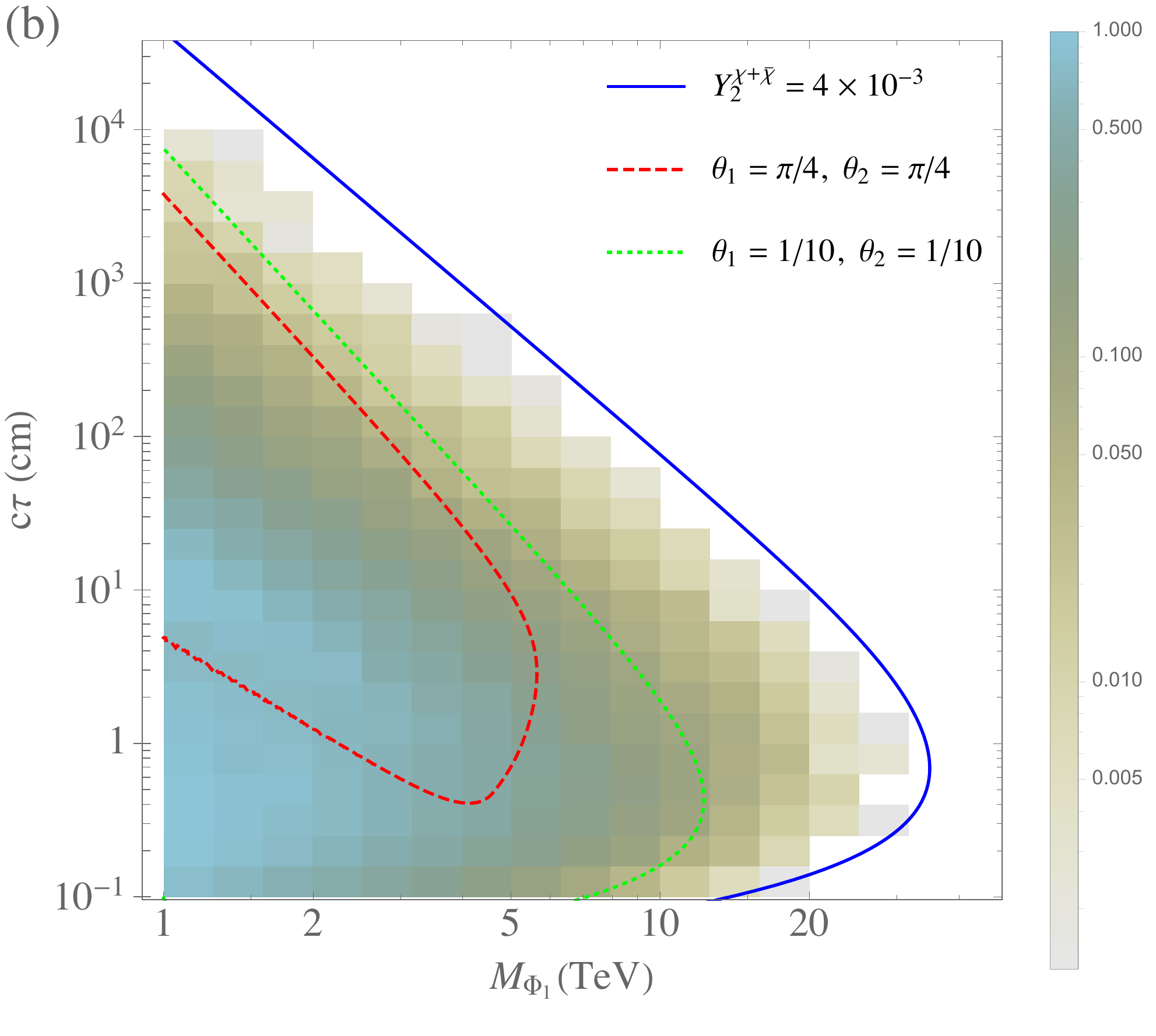}
   \caption{   
   The shading shows relative frequencies
   for points that give the correct baryon asymmetry and DM abundance
    in two different random scans over couplings and masses, as described in the text. In the ``overall-scale'' scan of (a), the magnitude of each $F^1_{\alpha I}$ is generated by multiplying a log-distributed overall scale by a random number in the range $[0,1]$, and similarly for each $F^2_{\alpha I}$. In the ``uncorrelated-couplings'' scan of (b)  the magnitudes of each $F^i_{\alpha I}$ are taken to be log-distributed and fully uncorrelated.
    The contours (identical for the two plots) enclose regions in the $(M_{\Phi_1},c\tau)$ plane that can give a large enough baryon asymmetry while satisfying the DM constraint of Eq.~\eqref{eqn:DMconstraint2}, under three different coupling assumptions.
   We take the $\Phi$ quantum numbers to be those of $u_R$.  
    }
   \label{fig:max_ranges_various_angles}
\end{figure*}
To get a sense of where the  model's most promising parameter space lies,
we perform two random scans  over the couplings and masses that determine $\rhotot$ and $Y_B$ in the decoupled-$\Phi_2$ regime, using Eqs.~(\ref{eqn:chi_Y_from_Phi_decay}), (\ref{eqn:rho_tot}), (\ref{eqn:Mchi_ave_hierarchical}), and (\ref{eqn:YB_large_Mphi_alt}).  The shaded regions in
Fig.~\ref{fig:max_ranges_various_angles} show the preferred
$\Phi_1$ parameter space that emerges.  

We now explain how these scans were performed. First, we impose an upper bound on $\Ytwotot$.
Because the Sakharov conditions require a departure from equilibrium  \cite{Sakharov:1967dj}, at least some linear combination of $\chi$ states must be out of equilibrium at the time of inverse decay into $\Phi_1$. Baryogenesis can still occur if a linear combination of $\chi$ states \emph{does} come into equilibrium. For example, it is possible that   $\Phi_2^{(*)}$ decays thermalize some linear combination of $\chi_1$ and $\chi_2$ (and the associated CP-conjugate state),
in which case the total abundance of $\chi$ and ${\overline \chi}$ particles left over after $\Phi_2$ annihilation approaches the  equilibrium value for a single mass eigenstate,
\be
\Yeqtot =
 \frac{135 \zeta(3)}{4\pi^4 g_*} \simeq 4 \times 10^{-3}.
 \label{eqn:Yeq}
\ee
We impose this value as our upper bound on  $\Ytwotot$,
\be
\Ytwotot < 4 \times 10^{-3},
\label{eqn:Y2bound}
\ee
 even though one can imagine viable scenarios in which a larger-than-equilibrium abundance is produced at high temperatures\footnote{  
For example, if the background of $\chi/{\overline \chi}$ particles is produced non-thermally by a source that couples to a single linear combination of $\chi_1$ and $\chi_2$, the orthogonal linear combination of $\chi_1$ and $\chi_2$ could remain  out of equilibrium given a sufficiently long oscillation timescale.}.  

We also require
\be
\Mchitwo > 10 \text{ keV},
\ee
to approximately satisfy structure-formation constraints (for example, the authors of Ref.~\cite{Heeck:2017xbu} argue that Lyman-$\alpha$ constrains $M_\chi\gtrsim12$ keV provided $\chi+\overline\chi$ production occurs at electroweak temperatures).

 Taking three active flavors, we start with randomly generated $F^1$ and $F^2$ coupling matrices and a randomly generated $M_{\Phi_1} > 1$ TeV (with a flat prior in the logarithm of $M_{\Phi_1}$). 
 In all cases, we take $M_{\Phi_2} \gg M_{\Phi_1}$.  
 We include points in Fig.~\ref{fig:max_ranges_various_angles} if we can find a rescaling of the $F^2$ matrix and a value for $\Mchitwo^2 = \Msplit$ that give the observed DM density and  baryon asymmetry, subject to the constraints $Y_2^{\chi}< 4\times 10^{-3}$ and $\Mchitwo > 10$ keV.   
 Figs.~\ref{fig:max_ranges_various_angles}(a) and \ref{fig:max_ranges_various_angles}(b) differ only in how the  initial $F^i$ couplings are generated.

Fig.~\ref{fig:max_ranges_various_angles}(a) is based on the ``overall-scale'' scan:  overall scales of the coupling matrices $F^1$ and   $F^2$ are randomized with flat priors in the logarithms of those scales,  the magnitude of each individual coupling $F^i_{\alpha I}$ is obtained by multiplying the appropriate overall scale by a random number in the range $[0,1]$, and each coupling is finally assigned a random complex phase.

Fig.~\ref{fig:max_ranges_various_angles}(b) is based on the ``uncorrelated-couplings'' scan:
  absolute values of couplings $F^i_{\alpha I}$ are independently randomized  with  flat priors in the logarithms of those absolute values, and each coupling is then assigned a random complex phase.  
  
 To different degrees,  both scans prefer $M_{\Phi_1}$ to be in the $\sim 1-\text{few}$ TeV range, making this a promising scenario  with respect to collider searches.  
  The uncorrelated-couplings scan  tends to produce  larger hierarchies among coupling matrix elements, making small $\Phi$ branching ratios to $\chi_2$ less rare and avoiding  over-production of DM. This produces a broader distribution in $(M_{\Phi_1}, c\tau_{\Phi_1})$ space.    
  
 To interpret the  numerical scan results, we also analytically identify  viable regions in the $(M_{\Phi_1},c\tau)$ plane under specific coupling assumptions.
We marginalize over the other, unspecified parameters to find the maximum baryon asymmetry subject to the DM constraint; for further details see Appendix~\ref{sec:app_parameters}. 

If we set  $\Ytwotot$ at what we take to be its maximum allowed value, thereby saturating  Eq.~\eqref{eqn:Y2bound}, 
the observed baryon asymmetry can be attained for points within the blue contour of Fig.~\ref{fig:max_ranges_various_angles}.   For this maximum value of $\Ytwotot$, we avoid over-production of DM only for $\theta_2 \lesssim 0.1$, corresponding to a $\Phi_2$ that decays preferentially to the massless $\chi_1$, as opposed to $\chi_2$.  

We can alternatively adopt fixed values for the mixing angles $\theta_1$ and $\theta_2$.  Allowed regions  lie within the red contour of Fig.~\ref{fig:max_ranges_various_angles} for $\theta_1=\theta_2 = \pi/4$ (in which case $\Phi_1$ and $\Phi_2$ both decay to $\chi_1$ and $\chi_2$ with equal probabilities) and within the green contour for $\theta_1=\theta_2 = 1/10$ (in which case $\Phi_1$ and $\Phi_2$ both decay predominantly to $\chi_1$).  The $\Mchitwo >  10 \text{ keV}$ constraint  has a significant impact on the $\theta_1=\theta_2= \pi/4$  parameter space, cutting out a region with smaller $M_{\Phi_1}$ and $c\tau_{\Phi_1}$ where the model would predict an over-abundance of DM.

In Fig.~\ref{fig:max_ranges_various_angles}(a),  the  bulk of the scan points are enclosed within the red, equal-mixing contour, consistent with the fact that the overall-scale scan leads to more anarchic coupling structures and consequently large mixing angles among all states.

\begin{figure}[h]
   \centering
     \includegraphics[width=3in]{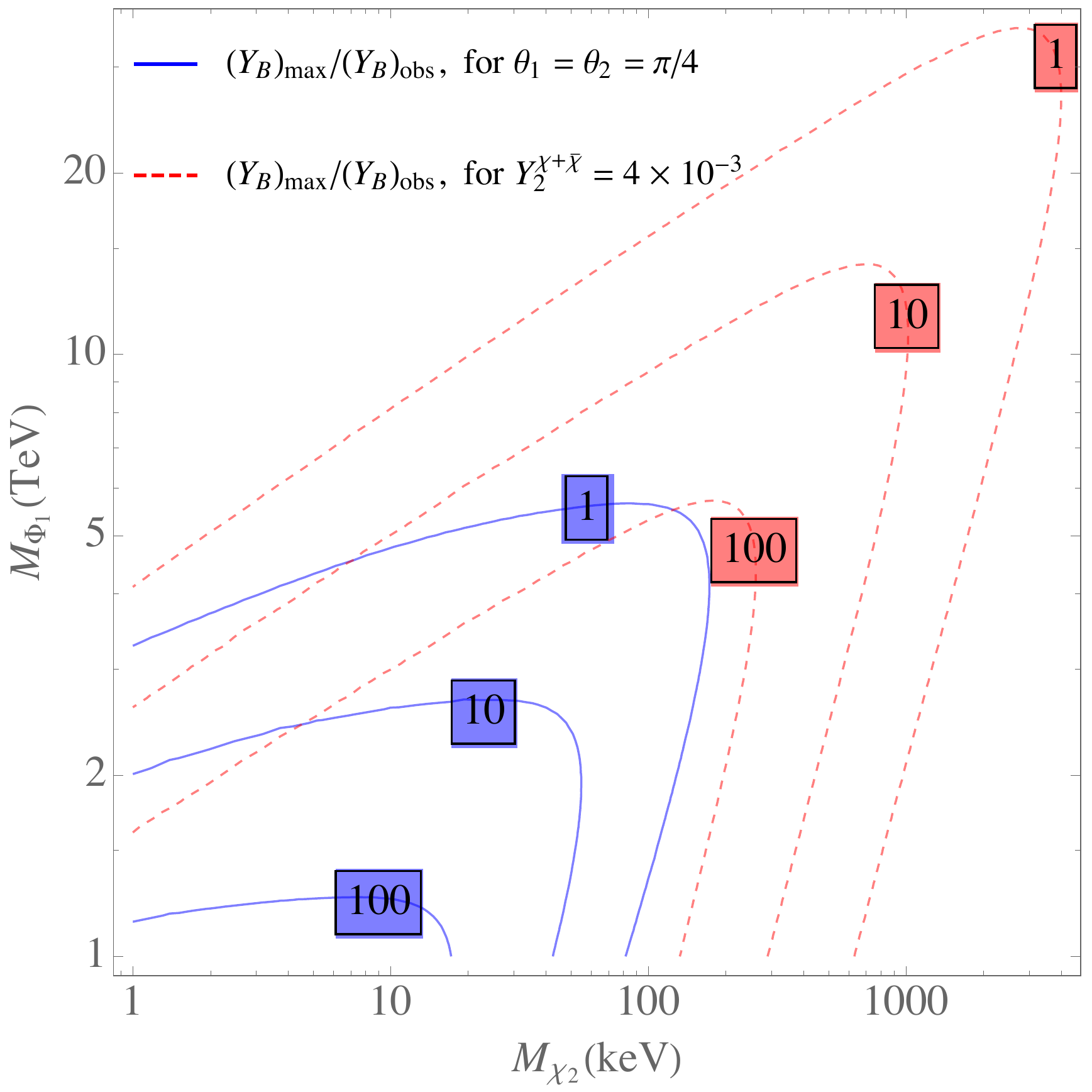}
   \caption{The maximum baryon asymmetry consistent with the DM constraint of Eq.~\eqref{eqn:DMconstraint2}, as a function of $M_{\Phi_1}$ and $\Mchitwo$, in two different coupling scenarios.  
We assume
$\Mchitwo^2 \simeq \Msplit$ and neglect the energy density stored in the lighter $\chi$ mass eigenstate.
For this particular plot, we have not imposed structure formation constraints  relevant for $\Mchitwo \lesssim 10$ keV.
We take the $\Phi$ quantum numbers to be those of $u_R$.  
}
   \label{fig:Mchi_Mphi_plot}
\end{figure}

We also see in Fig.~\ref{fig:max_ranges_various_angles} that the viable parameter space is restricted to $M_{\Phi_1} \lesssim 6$ TeV and $c\tau_{\Phi_1} \gtrsim 4$ mm if $\Phi$ decays produce $\chi_1$ and $\chi_2$ in equal abundance (red contour), but that the allowed values expand to $M_{\Phi_1} \lesssim 35$ TeV and 
into the sub-mm decay regime 
if $\Phi$ decays mainly produce very light $\chi_1$ particles (blue contour).    In Fig.~\ref{fig:Mchi_Mphi_plot} we find that the corresponding ranges  for the $\chi_2$ mass are $\Mchitwo \lesssim 150$ keV and $\Mchitwo \lesssim 4$ MeV, respectively.  

In Figs.~\ref{fig:max_ranges_various_angles} and~\ref{fig:Mchi_Mphi_plot} we take $\chi_1$ to be effectively massless by equating $\Mchitwo^2 = \Msplit$ and neglecting the energy density stored in the lighter $\chi$ mass eigenstate. 
\begin{figure*}
   \centering
          \includegraphics[width=3in]{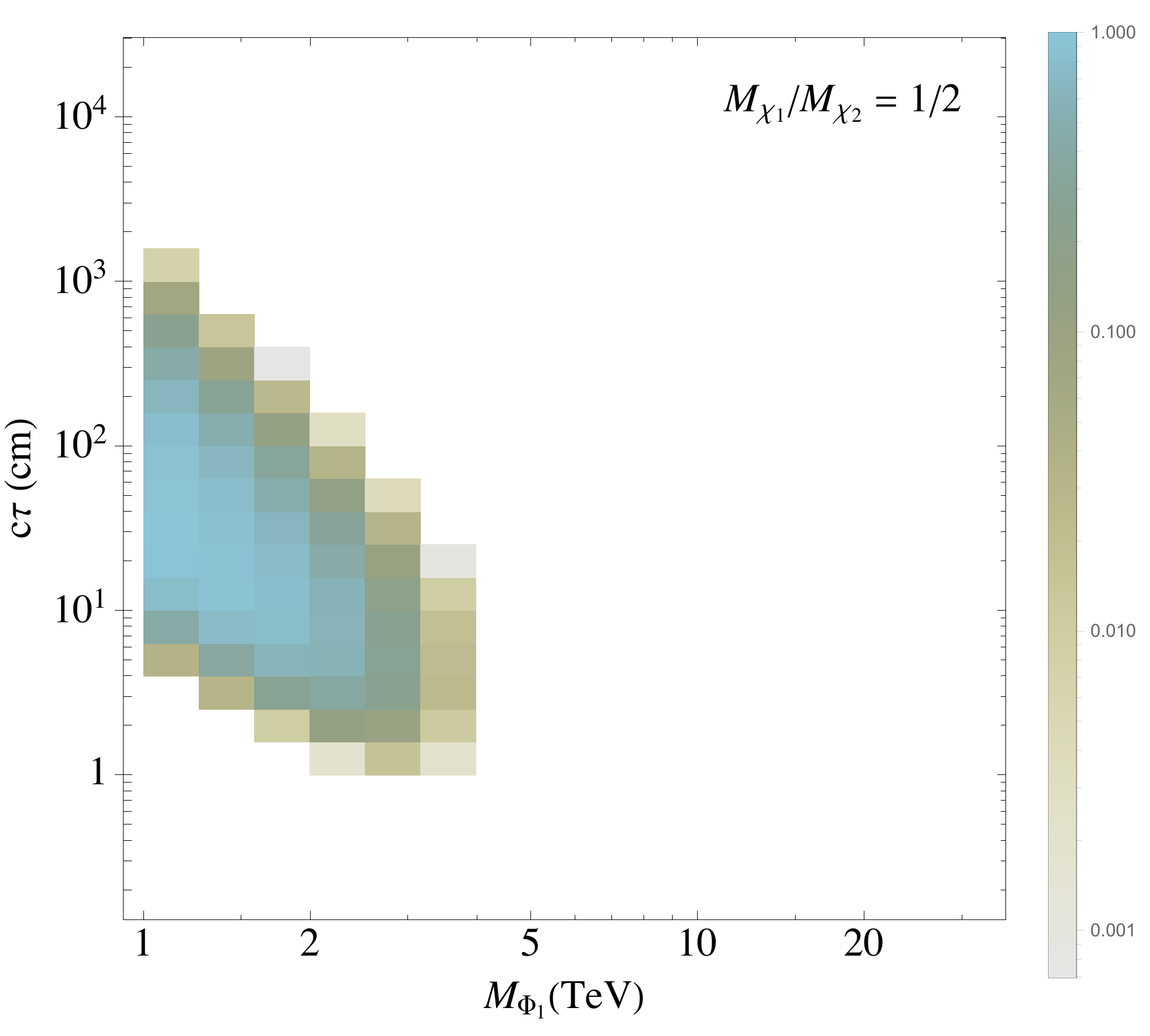}
           \quad \quad \quad
        \includegraphics[width=3in]{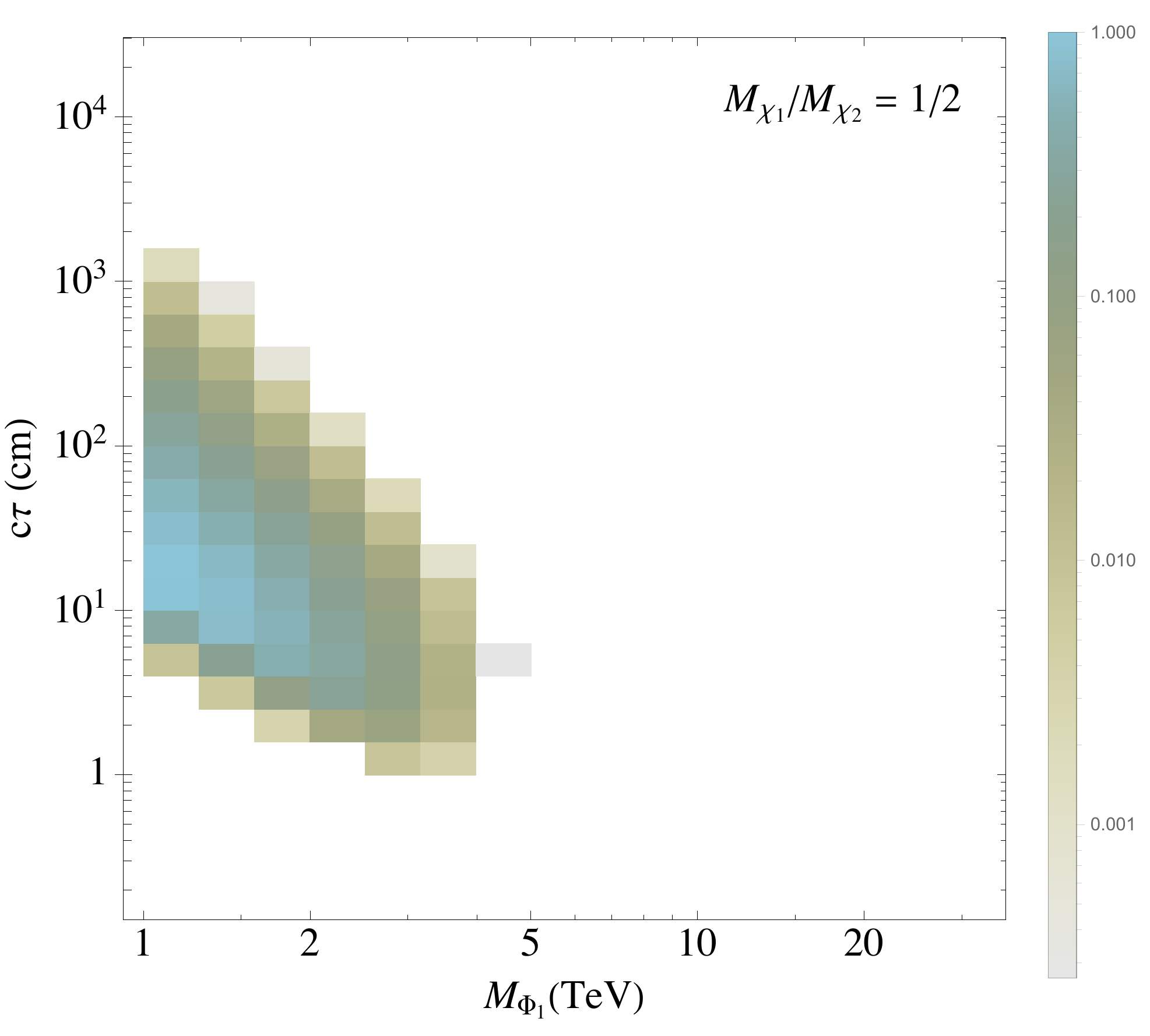}\\
          \includegraphics[width=3in]{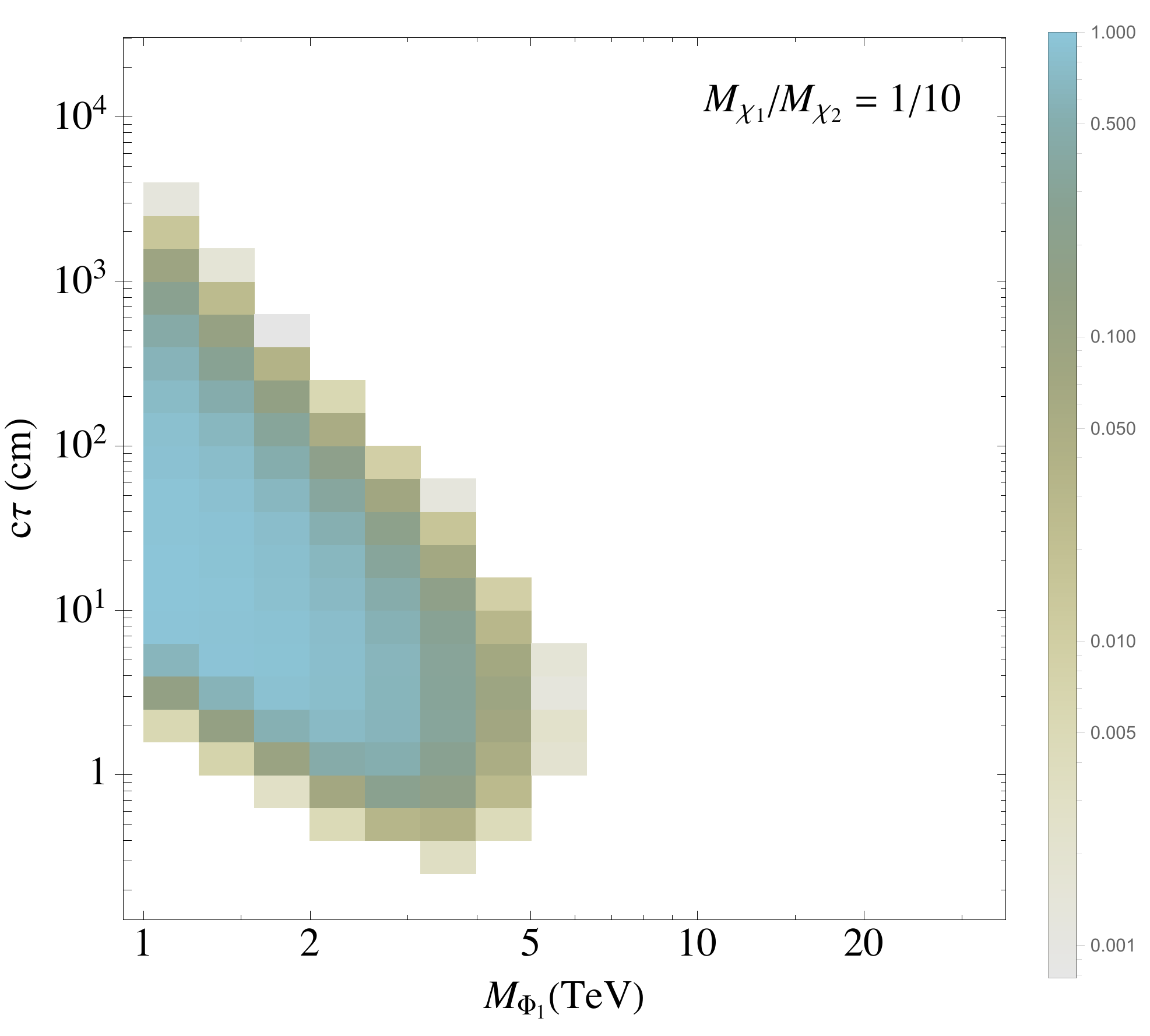}
           \quad \quad \quad
        \includegraphics[width=3in]{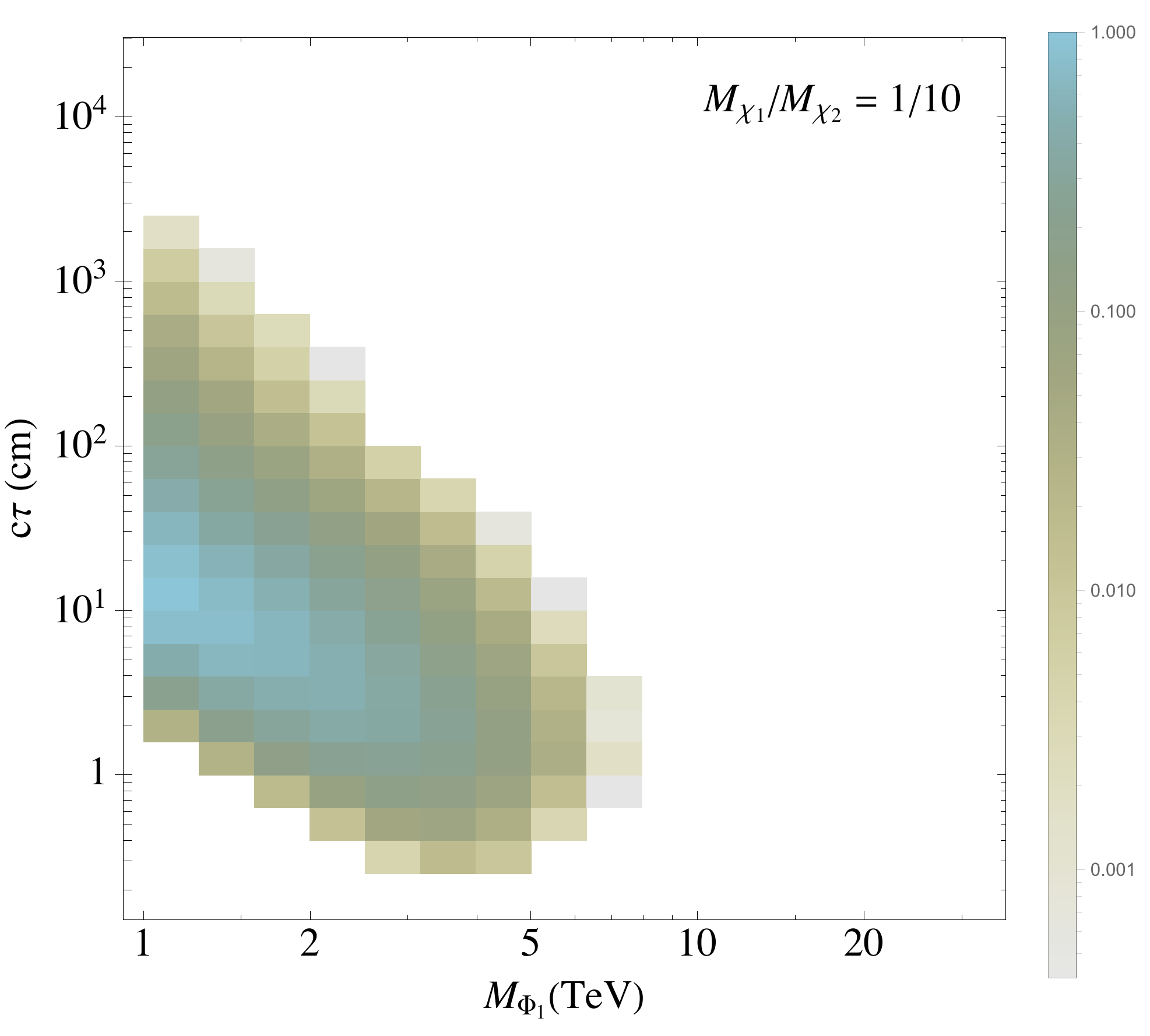}\\
          \includegraphics[width=3in]{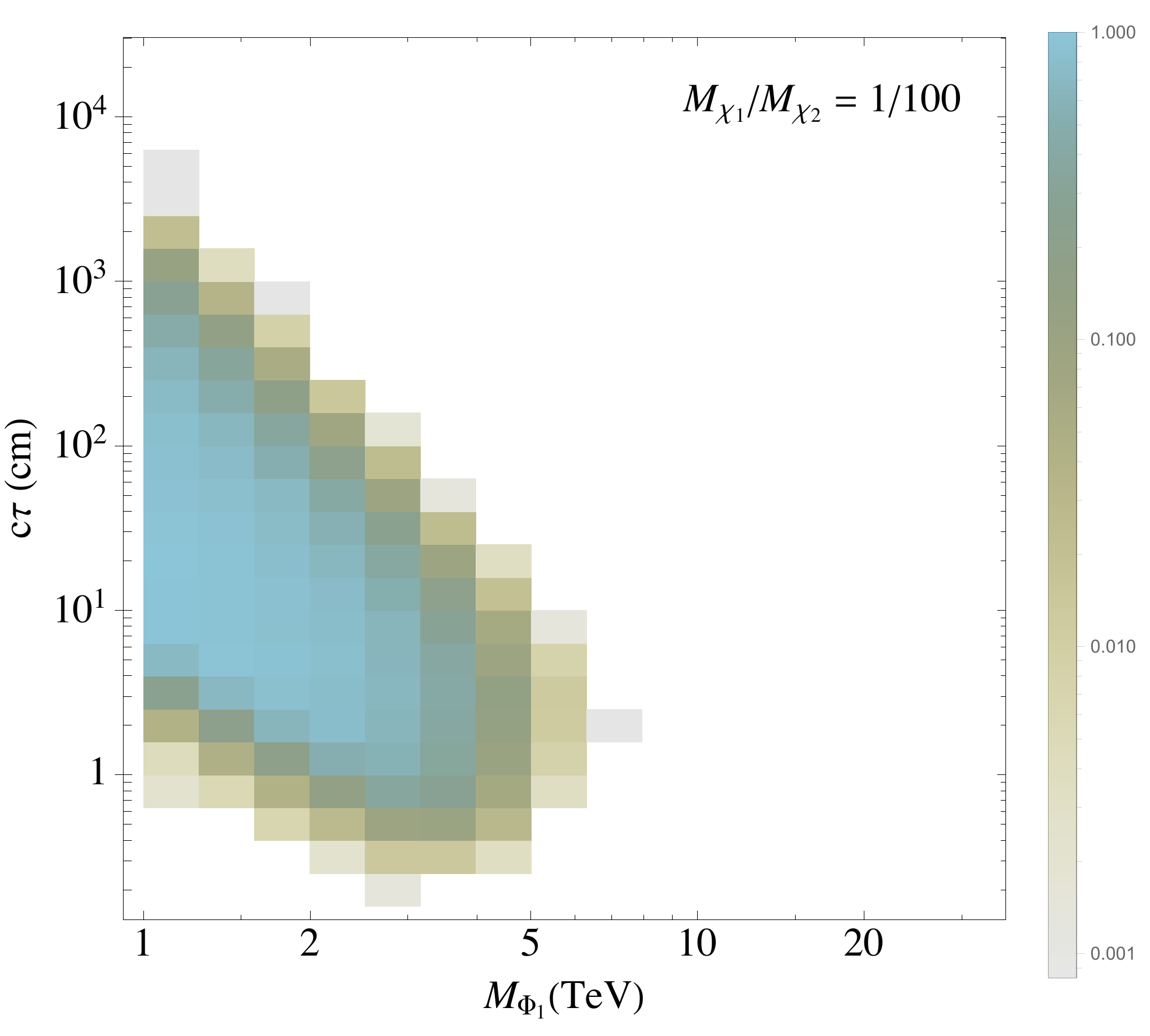}
           \quad \quad \quad
        \includegraphics[width=3in]{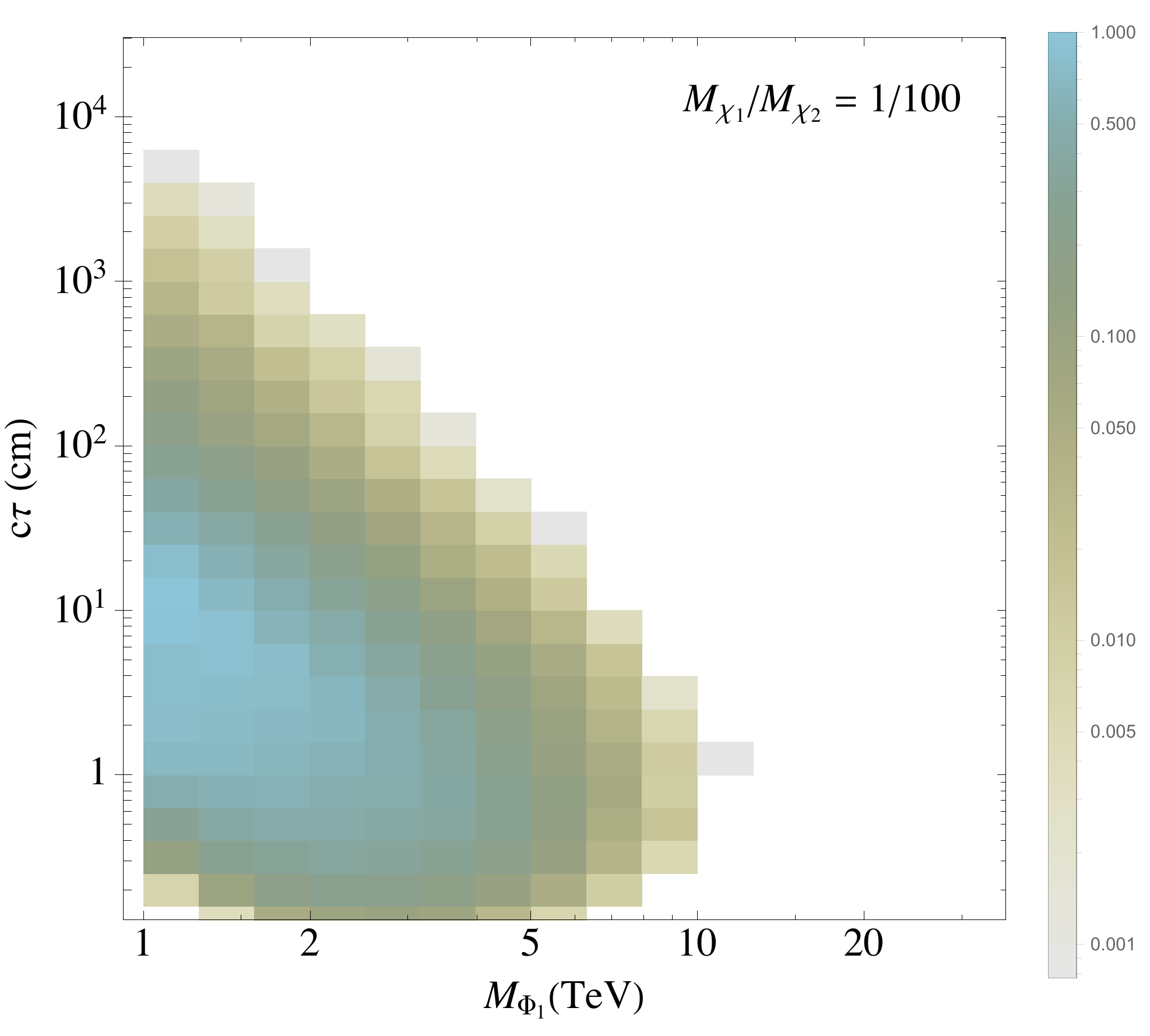}
   \caption{Results of repeating the two scans of Fig.~\ref{fig:max_ranges_various_angles} with $\Mchione/\Mchitwo = 1/2$ (top),
   $\Mchione/\Mchitwo = 1/10$ (middle), and
   $\Mchione/\Mchitwo = 1/100$ (bottom).  For points with $0<\Mchione< 10$ keV, we restrict the $\chi_1$ energy density to be no larger than 1/3  of the total DM energy density to evade structure-formation constraints \cite{Boyarsky:2008xj,Kamada:2016vsc}.  
   }
   \label{fig:scans_nonzero_mchi1_rhothird}
\end{figure*}
 As shown in Fig.~\ref{fig:scans_nonzero_mchi1_rhothird}, the viable $(M_{\Phi_1}, c\tau_{\Phi_1})$ space shrinks further if we adopt different assumptions for $\Mchione$.  As a result, the prospects for testing the model become even more promising.

The bound  $M_{\Phi_1} \lesssim 35 \text{ TeV}$ applies
when $\Phi_{1,2}$ have the same quantum numbers as  $u_R$, but the results are roughly the same in the  $Q_L$ and $d_R$ cases.  While the upper end of this mass range is likely to be inaccessible even at a 100 TeV collider \cite{Cohen:2013xda}, it is at the very least a firm upper bound in the two-scalar model.   As the scan results suggest, saturating this  upper bound requires a special alignment of parameters, while more generic parameters typically prefer values of $M_{\Phi_1}$ that are more accessible at colliders.
 
Finally, if we consider the baryon asymmetry alone and abandon the DM constraint of Eq.~\eqref{eqn:DMconstraint2},
Eq.~\eqref{eqn:YB_large_Mphi_alt} places a very weak  upper bound on the mass of the lighter scalar,
\be
M_{\Phi_1} \lesssim 570\,\,\mathrm{TeV}  \left(\frac{g_\Phi \mathcal{K}_B }{\mathcal{K}_\Phi} \right)^{1/2} \left(\frac{\Ytwotot}{4 \times 10^{-3} } \right)^{1/2}.
\ee
If we impose the bound of Eq.~\eqref{eqn:Y2bound}, the two trailing factors are order-one or smaller.

\section{Baryogenesis and Dark Matter with a Single Scalar}
\label{sec:single_scalar}

Having thoroughly explored the parameter space giving rise to baryogenesis and DM in a model with two scalars, we now return to the more ARS-like scenario with a single scalar. We consider the same model as Sec.~\ref{sec:calculation}, but now including only a single scalar, $\Phi$, with couplings to all quark generations:
\be\label{eq:single_scalar_model}
\mathcal{L} &\supset& - \frac{M_I}{2} \bar\chi_I^{\rm c}\chi_I -(F_{\alpha I} \,\bar Q_\alpha \Phi\chi_I + \mathrm{h.c.}).
\ee
As argued in Sec.~\ref{sec:qualitative}, the baryon asymmetry is expected to be smaller with only a single scalar, and indeed we show in this section that the viable parameter space for baryogenesis and DM is much more constrained than in Sec.~\ref{sec:twoscalar}. This is due to the fact that the baryon asymmetry arises at  higher order in perturbation theory, either $\mathcal{O}(F^4 y_t^2)$ or $\mathcal{O}(F^6)$, where $y_t$ is the SM top quark Yukawa coupling.

We start by discussing why the $\mathcal{O}(F^4)$ asymmetry vanishes if we neglect SM Yukawa couplings.
 Generating an asymmetry relies on processes such as $\Phi\rightarrow Q_\alpha\bar\chi,Q_\beta\bar\chi\rightarrow \Phi$ differing in rate\footnote{We take the ``rate'' for $\Phi\rightarrow Q_\alpha\bar\chi,Q_\beta\bar\chi\rightarrow \Phi$ to mean the contribution to the  $Q_\beta\bar\chi \rightarrow \Phi$ rate at some fixed inverse-decay time, due to $\bar \chi$ particles that were produced in association with $Q_\alpha$.
 }
  from the equivalent processes for $\bar\Phi$. To determine the total asymmetry, we must sum over all quark flavors $\alpha$, $\beta$, 
   which we can organize into a sum over pairs of processes with the two quark flavors switched, such as the
 pair consisting of the ``$(1,2)$'' process   $\Phi\rightarrow Q_1\bar\chi ,\bar\chi Q_2\rightarrow \Phi$ and the ``$(2,1)$'' process $\Phi\rightarrow Q_2\bar\chi ,\bar\chi Q_1\rightarrow\Phi$.
In the absence of SM Yukawa couplings, flavor dependence enters only through the $F_{\alpha I}$ Yukawas themselves.  The $(1,2)$ and $(2,1)$ rates are then related by $F\leftrightarrow F^*$, and their sum is thus symmetric under $F\leftrightarrow F^*$, guaranteeing a vanishing asymmetry once we include the $CP$-conjugate processes.  
 
SM Yukawa couplings spoil this cancelation. Most significantly, the large top Yukawa coupling produces flavor non-universality in the quark thermal masses, leaving, for example,  less available phase space for (inverse) decays involving $t_R$ than for those involving $u_R$.   Thermal mass effects are more important at high temperatures, and so  flavor dependence of the kinematics tends to be more of an issue in the decays than in the inverse decays (thermal-mass effects are unimportant for inverse decays that occur at $T\ll M_\Phi$, for example).
Within the pair  $\Phi\rightarrow t_{\rm R} \bar\chi ,u_{\rm R}\bar\chi \rightarrow\Phi$ and  $\Phi\rightarrow u_{\rm R}\bar\chi ,t_{\rm R}\bar\chi \rightarrow \Phi$, the $(t_{\rm R},u_{\rm R})$ process  is therefore kinematically suppressed relative to $(u_{\rm R},t_{\rm R})$, and the two rates are no longer related by $F\leftrightarrow F^*$.  This source of asymmetry is consistent with $CPT$ which, due to the expansion of the universe, only relates equal-time rates 
for processes that can be approximated as instantaneous.
Since the resulting asymmetry vanishes in the flavor-universal limit $y_t\rightarrow0$, we find the resulting asymmetry is $\mathcal{O}(F^4y_t^2)$. 

 Note that an asymmetry at this order requires more than just flavor-non-universal  quark or lepton masses.  It requires flavor-non-universal {\it temperature dependence} of the $\Phi$-decay reaction densities, which arises in our scenario due to the large tree-level $\Phi$ mass.  
 
 Indeed, the asymmetry vanishes if we can express the $\Phi \rightarrow Q_\alpha \bar\chi$ reaction densities as $\gamma_\alpha(T) = \tilde\gamma_\alpha\,f(T)$, where 
 $\tilde\gamma_\alpha$ are flavor-dependent constants and
 $f(T)$ is the same for every flavor. 
In that case the rates for  $\Phi\rightarrow Q_\alpha\bar\chi_I,Q_\beta\bar\chi_I\rightarrow \Phi$ and  $\Phi\rightarrow Q_\beta\bar\chi_I,Q_\alpha\bar\chi_I\rightarrow \Phi$ share the common factor $f(T_1) f(T_2)$, where $T_2$ and $T_1$ are the decay and  inverse-decay temperatures, respectively.  As for the case with flavor-universal quark masses, the two rates differ only by $F \leftrightarrow F^*$, and the asymmetry vanishes at this order.  

This is of particular relevance to 
ARS leptogenesis, where $\Phi$ is the SM Higgs. In this case, no fields have significant tree-level masses in the unbroken phase, all reaction densities must scale like $\gamma_\alpha(T) = \tilde\gamma_\alpha\,T^4$ from dimensional analysis, and consequently the $\mathcal{O}(F^4)$ asymmetry vanishes even in the presence of a $\tau$ thermal mass. This is not quite true at $T\sim T_{\rm ew}$, since at this point the Higgs tree-level mass is relevant, and so a small effect may be observed there. The $\mathcal{O}(F^4y_t^2)$ asymmetry we find in our  model therefore crucially depends on the tree-level $\Phi$ mass, and the asymmetry generation at this order is highly suppressed for $T\gg M_\Phi$.

In the rest of this section, we  consider two separate, equally motivated cases:~in the first, the top quark couples appreciably to $\chi$ and $\Phi$, and an asymmetry is generated according to the flavor-non-universal mechanism described above. In the second, the top quark does not couple appreciably to $\chi$ and $\Phi$, in which case the smallness of the light-quark Yukawa couplings leads to the dominant asymmetry production occurring instead at $\mathcal{O}(F^6)$.

\subsection{Top-Mass-Induced Asymmetry}
\label{sec:top_thermal_mass}
\begin{figure*}
\includegraphics[width=3.5in]{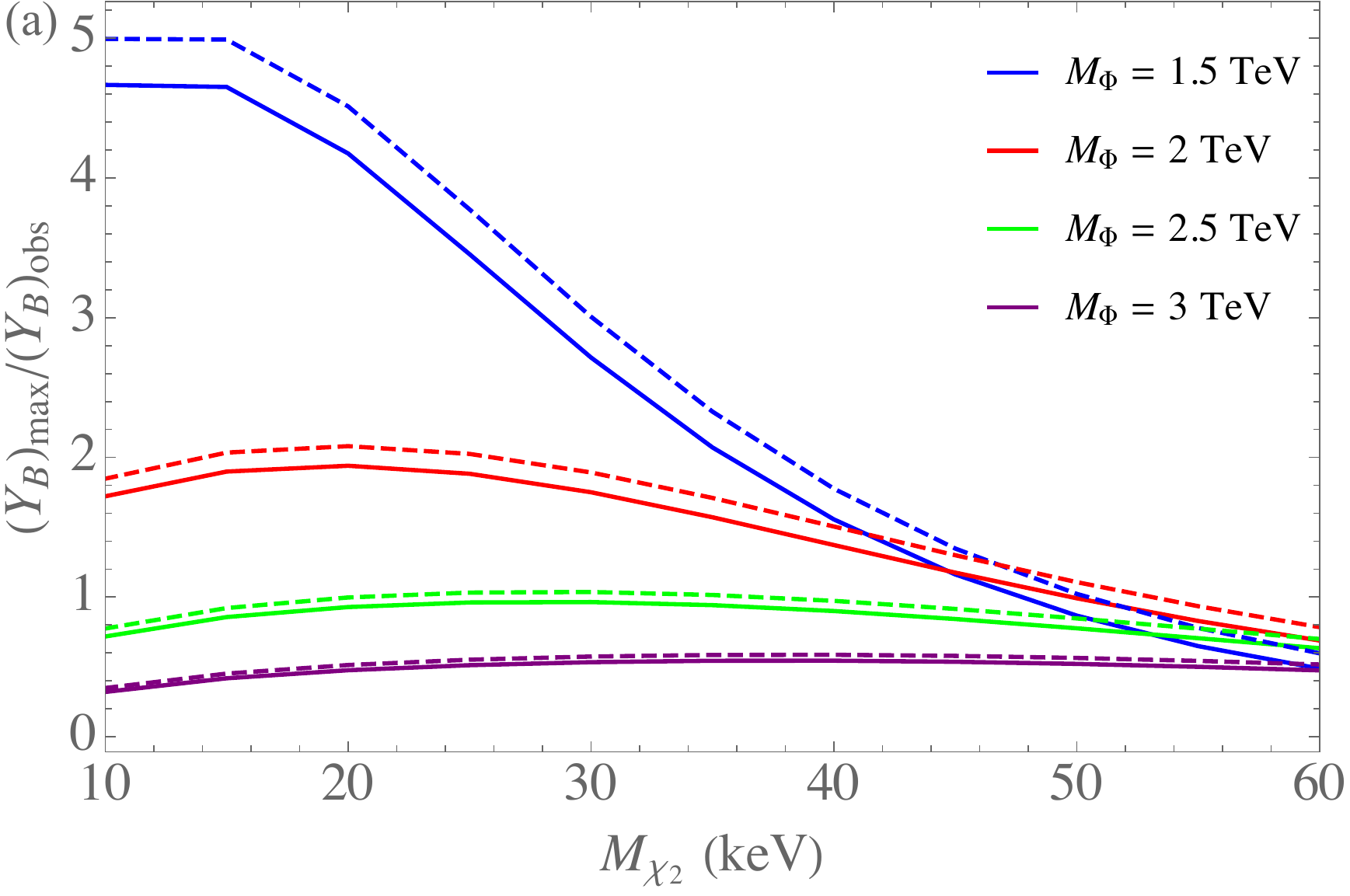}
\quad\quad
\includegraphics[width=3in]{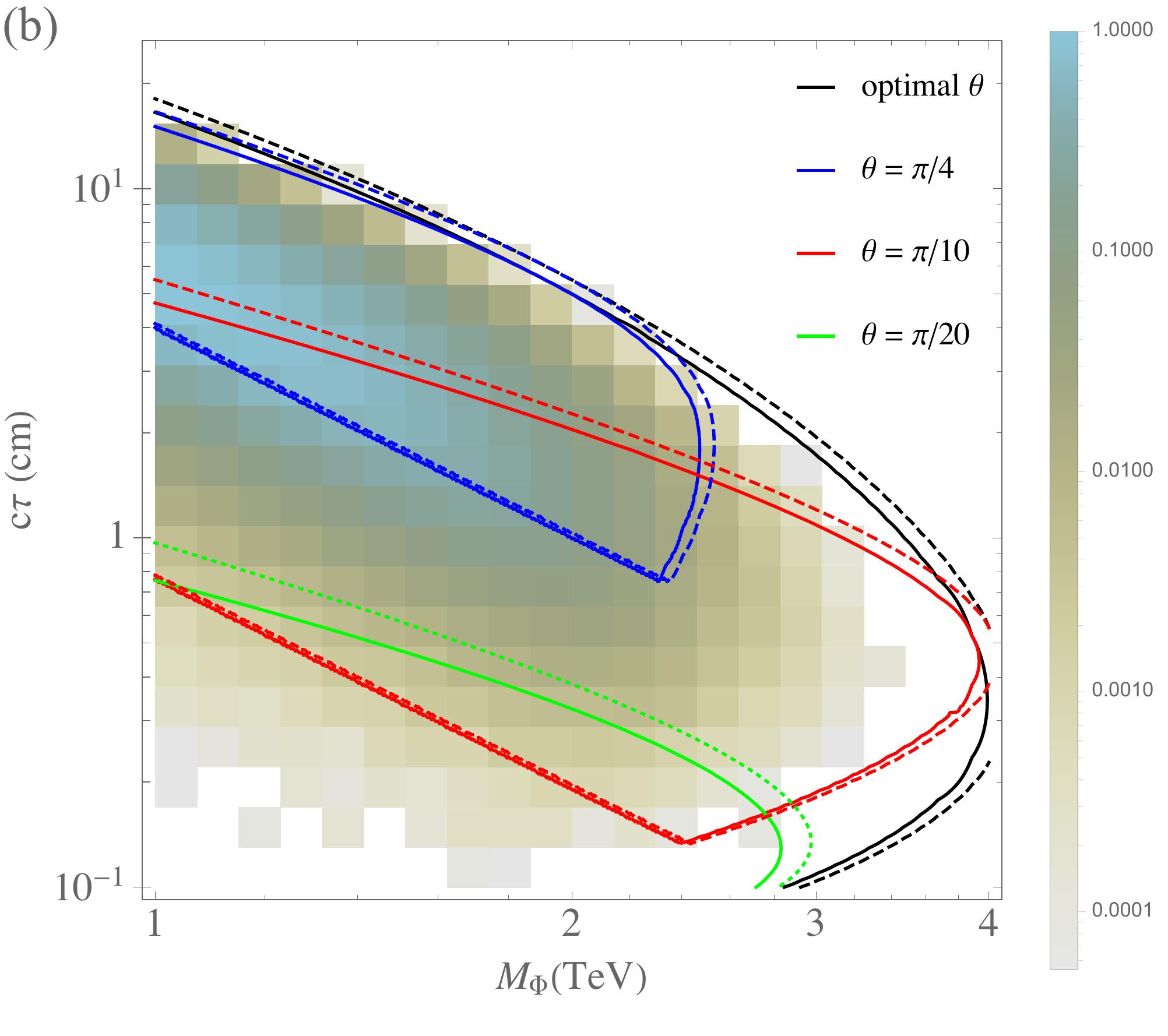}
\caption{Results for the single-scalar scenario with top-mass-induced asymmetry. For both plots, we take $\Mchione = 0$ and impose $\rhotot = \rho_\text{cdm}$.  In (a), we take $\theta = \pi/4$ and show the maximum possible baryon asymmetry for various values of $M_\Phi$ and $\Mchitwo$.  
In (b), $(M_\Phi,c\tau)$ points for which the maximum possible baryon asymmetry is equal to $(Y_B)_\text{obs}$  lie on the contours shown.  
We include $\theta$ among the adjustable parameters for the black contours,  whereas  we fix $\theta$ at the indicated values for the others.  As described in the text, the shading shows relative frequencies
   for points that give $Y_B \ge (Y_B)_\text{obs}$ in a random scan over couplings and masses.
For the solid (dashed) contours of both plots we take $A_\text{self} = 0$ (1/3) in calculating the scalar thermal mass; see Eq.~(\ref{eqn:thermalphimass}).
}
\label{fig:top_assisted_single_scalar}
\end{figure*}
Here we consider in detail the single-scalar scenario in which the asymmetry arises at $\mathcal{O}(F^4y_t^2)$, adopting couplings to $u_R$-type quarks for concreteness.
We treat thermal masses as described for the two-scalar case in Appendix~\ref{sec:thermal_mass_appendix}, with further details given below.
That is, we use  Eqs.~(\ref{eqn:thermalQmass0}) and (\ref{eqn:thermalQmass2}), except that for the top quark we include a Yukawa contribution \cite{Weldon:1982bn}:
\begin{equation}
\overline{M}_{u_{R,3}}^2 = \left(\frac{1}{3} g_3^2 + \frac{1}{9} g_1^2+ \frac{1}{4} y_t^2\right)T^2.
\label{eqn:thermal_yuk}
\end{equation}
This expression is based on the finite-temperature quark dispersion relation in the high-momentum regime.   
Here and below, bars over masses indicate that thermal contributions are included.  

Our results for this scenario, with the $\rhotot = \rho_\text{cdm}$ constraint in place,  are summarized in Fig.~\ref{fig:top_assisted_single_scalar}.  
As described below, the angle $\theta$ parametrizes the  relative overall strength of $\Phi$'s couplings to $\chi_1$ (taken to be massless)  versus $\chi_2$ (taken to have $\Mchitwo >10$ keV).
In Fig.~\ref{fig:top_assisted_single_scalar}(a) we take $\theta=\pi/4$, corresponding to  equal-strength couplings to $\chi_1$  and $\chi_2$.  We see that  there exists parameter space in which a realistic baryon asymmetry and DM abundance can simultaneously be realized,  
with $M_{\Phi} \lesssim 2.5$ TeV and $\Mchitwo \lesssim 60$ keV. 

 Fig.~\ref{fig:top_assisted_single_scalar}(b) shows the viable $(M_\Phi,c\tau)$ parameter space.  For the black contours we maximize $Y_B$ with respect to all other parameters including $\theta$, while for the other contours we consider fixed values of $\theta$.

As we did for the two-scalar scenario, we also perform a random scan for a rough, qualitative determination of the preferred parameter space, also shown in Fig.~\ref{fig:top_assisted_single_scalar}(b).   We start with randomly generated $M_{\Phi_1} > 1$ TeV and $\Mchitwo> 10$ keV (with flat priors in the logarithms of these masses).
Taking three active flavors,  we generate a random coupling texture by 
assigning each $F^i_{\alpha I}$ a random number in the range $[0,1]$ multiplied by a random complex phase.  This texture for  $F$ determines $\sin\theta$ (along with all other relevant mixing angles and phases), allowing the overall scale of $F$, and therefore $\Gamma_\Phi$, to be determined by the $\rhotot = \rho_\text{cdm}$ requirement.  We keep points with $Y_B \ge (Y_B)_\text{obs}$.  

Taken together, the results of Fig.~\ref{fig:top_assisted_single_scalar} show that the viable parameter space is significantly more limited than for the two-scalar case, making the prospects for conclusively testing this scenario at colliders particularly favorable.  While the contours of Fig.~\ref{fig:top_assisted_single_scalar}(a)  show that it is in principle possible for the $\Phi$ mass to be as large as $\sim 4$ TeV, the scan suggest that generic patterns of couplings prefer smaller masses, $M_\Phi \lesssim 2$ TeV. 
For this scenario, $\Phi$ must couple both to the top quark and at least one flavor of light quark.  Production of $\Phi \Phi^*$  at colliders would lead to  various final states involving isolated jets from $\Phi \rightarrow q \bar\chi$ and/or top quarks from $\Phi \rightarrow t \bar\chi$.
 
A caveat regarding Fig.~\ref{fig:top_assisted_single_scalar} is that, unlike for the two-scalar scenario,  the baryon asymmetry arises here as an intrinsically thermal-mass-related effect.  We have adopted a rather crude quasiparticle approximation to obtain our results; for example, our use of the high-momentum limit of the quark dispersion relation might be called into question given the relatively large coefficients appearing in Eq.~(\ref{eqn:thermal_yuk}).  However, even when we adopt the low-momentum dispersion relation (thereby reducing quark thermal masses-squared by a factor of two), the model can still satisfy the DM and baryon asymmetry constraints.    
 A more refined finite-temperature field theory calculation 
might give a more robust determination of the viable parameter space. 
 
We now provide additional details on our calculation of the $\mathcal{O}(F^4y_t^2)$ asymmetry.  Following our treatment of 
the two-scalar case in Appendix~\ref{sec:thermal_mass_appendix},
we define the flavor-dependent functions
\be
\rho_{\alpha}(z) = 1-\frac{\overline{M}^2_{Q_\alpha}(z)}{\overline{M}^2_{\Phi}(z)}
\ee 
along with 
\be
\tau(z) \equiv \frac{\overline{M}_\Phi(z)}{M_\Phi(z)}.
\ee
By retracing the steps of the perturbative calculation of Sec.~\ref{sec:calculation}, but  with a single scalar, and with temperature-dependent masses included, we find that the baryon asymmetry can be expressed as
\begin{multline}
\YB = \frac{45g_\Phi^2}{4\pi^4g_*}\,\frac{\mathcal{K}_B}{\mathcal{K}_\Phi}\frac{M_{\Phi}^4}{T_{\rm ew}^2 H_{\rm ew}^2} \sum_{\gamma,\delta}
\frac{4\; \text{Im}\left[
F_{\gamma 1}{F_{\gamma 2}}^{\!\!*} 
F_{\delta 2} {F_{\delta 1}}^{\!\!*}
\right]}{\left(16 \pi\right)^2}
   \\
 \times
  \int_0^\infty \!\!\!\! dy \;\frac{e^{-y}}{y^2} 
  \int_0^1 \!\!\! dz\;  S_{\Phi}(z)\;z^2 \;
  \tau^2(z)
  \;\rho_\gamma (z)
    \;  e^{ - \frac{1-\rho_\gamma(z)}{\rho_\gamma(z)}y  }\\
    \times      e^{-\alpha \frac{z^2}{y}   \tau^2(z)\rho_\gamma(z)}
   \int_0^{z} \!\!\!dz'\;{z'}^2  
     \tau^2(z')\rho_\delta(z')
      \;  e^{ - \frac{1-\rho_\delta(z')}{\rho_\delta(z')}y  }\\
      \times 
        e^{-\alpha \frac{z'^2}{y}   \tau^2(z')\rho_\delta(z') }\;
    \sin\left[ \beta_\text{osc} \left( \frac{z^3-z'^3}{y} \right)\right],
 \label{eqn:YB_ss_1}
\end{multline}
where the survival function $S_\Phi(z)$ can be obtained from 
Eq.~(\ref{eqn:thermal_survival}), with the substitutions $M_{\Phi_i} \rightarrow M_\Phi$, $\tau_i \rightarrow \tau$, and 
\begin{equation}
\Gamma_{\Phi_i} \rho_i^2 \rightarrow \sum_\gamma \frac{\left(F {F}^\dagger \right)_{\gamma \gamma} M_{\Phi} \rho_{\gamma}^2}{16 \pi}.
\end{equation}
Applying the same  substitutions in Eq.~(\ref{eqn:chi_Y_from_Phi_decay_WTM}) gives the DM abundance.

As a consistency check on our calculations, we can compare Eq.~(\ref{eqn:YB_ss_1}) with the contribution to $Y_B$ coming from $\Phi_2^{(*)}$ decay followed by $\Phi_1^{(*)}$ production,  in the two-scalar case with flavor-universal thermal masses.  The two-scalar result is given by Eqs.~(\ref{eqn:final_YB}) and (\ref{eqn:YB_WTM}), with only the $I_{12}$ term included.  We reproduce those expressions by starting with the single-scalar $Y_B$ of Eq.~(\ref{eqn:YB_ss_1}) and making the appropriate substitutions:
$M_\Phi^4 \rightarrow M_{\Phi_1}^2 M_{\Phi_2}^2$, $S_\Phi(z) \rightarrow S_{\Phi_1}(z)$, $\tau(z) \rightarrow \tau_1(z)$, $\tau(z') \rightarrow \tau_2(z')$, $\rho_\gamma(z) \rightarrow \rho_1(z)$, and  $\rho_\delta(z') \rightarrow \rho_2(z')$.

If we neglect the $z$-dependence in the survival function by taking $S_\Phi(z) \rightarrow S_\Phi(0)$, Eq.~(\ref{eqn:YB_ss_1}) can be simplified somewhat by exploiting the $z \leftrightarrow z'$ symmetry of the integrand.   We find
\begin{multline}
\YB \simeq \frac{45g_\Phi^2}{4\pi^4g_*}\,\frac{\mathcal{K}_B}{\mathcal{K}_\Phi}\frac{M_{\Phi}^4}{T_{\rm ew}^2 H_{\rm ew}^2}
S_{\Phi}(0)
 \sum_{\gamma<\delta}
\frac{4\; 
\text{Im}\left[
F_{\gamma 1}{F_{\gamma 2}}^{\!\!*} 
F_{\delta 2} {F_{\delta 1}}^{\!\!*}
\right]}{\left(16 \pi\right)^2}
   \\
 \times
  \int_0^\infty \!\!\!\! dy \;\frac{e^{-y}}{y^2} 
  {\rm Im} \left[ \mathcal{H}_\gamma(y) \mathcal{H}_\delta^*(y)\right],
   \label{eqn:YB_ss_2}
  \end{multline}
  where
  \begin{multline}
  \mathcal{H}_\gamma(y) = 
  \int_0^1 \!\!\! dz \;z^2 
  \tau^2(z)
  \rho_\gamma (z)
     e^{ - \frac{1-\rho_\gamma(z)}{\rho_\gamma(z)}y  }
     \\ \times    
       e^{-\alpha \frac{z^2}{y}   \tau^2(z)\rho_\gamma(z)}
   e^{i \beta_\text{osc}  \frac{z^3}{y} }.
  \end{multline}

Because we take into account only third-generation Yukawa couplings, the symmetry between the first two quark generations allows us to rewrite the baryon asymmetry as
\begin{multline}
\YB \simeq \frac{45g_\Phi^2}{16\pi^4g_*}\,\frac{\mathcal{K}_B}{\mathcal{K}_\Phi}
\tilde{ \mathcal{J}}
\left(\frac{M_{\Phi}}{T_{\rm ew}}\right)^2
\left(\frac{\Gamma_{\Phi}}{H_{\rm ew}}\right)^2
S_{\Phi}(0)\\
\times
  \int_0^\infty \!\!\!\! dy \;\frac{e^{-y}}{y^2} 
  {\rm Im} \left[ \mathcal{H}_{12}(y) \mathcal{H}_3^*(y)\right],
   \label{eqn:YB_ss_3}
  \end{multline}
  where we neglect Yukawa coupling contributions to thermal masses in 
  $\mathcal{H}_{12}$, and where
\be\label{eq:jarlskog_invariant_topcase}
\tilde{ \mathcal{J}} &=& \sin^2 2\theta 
\sin2\rho_1
\sin2\rho_2
\cos^2\gamma
\sin(\phi_3-\phi_{12}),
\ee
with
\be\label{eq:mixing_angles_topcase}
\cos\theta &=& \sqrt{\frac{({F}^\dagger {F})_{11}}{\mathrm{Tr}({F}^\dagger F)}},\\
\cos\rho_I &=& \frac{|F_{3I}|}{\sqrt{({F}^\dagger F)_{II}}},\\
\cos\gamma &  = & 
\frac{\left| \sum_{\alpha = 1,2} F_{\alpha 1}^* F_{\alpha 2} \right|}{\sqrt{ \sum_{\alpha = 1,2} |F_{\alpha 1}|^2  \sum_{\beta = 1,2} |F_{\beta 2}|^2 }}
\\
\phi_{12} &=& \arg\left(\sum_{\alpha = 1,2}{F_{\alpha 1}^* F_{\alpha 2}} \right) \\
\phi_3 &=& \arg(F_{3 1}^* F_{3 2} ). 
\ee
The angles $\theta$, $\rho_I$, and $\gamma$ all lie in the first quadrant. 
Analogously to the two-scalar case, $\theta$ parametrizes the relative overall strength  of the  $\chi_1$ couplings compared to those of  $\chi_2$.  Each $\rho_I$ angle reflects the coupling strength of $\chi_I$ to the third-generation quark, relative to overall $\chi_I$ coupling strength.   The angle $\gamma$  parametrizes the degree of alignment between the couplings of $\chi_1$ and $\chi_2$ within the first two generations.   Finally, $\phi_{12}$ and $\phi_3$ are  relative phases between the couplings of $\chi_1$ and $\chi_2$ to the first two generations and to the third generation, respectively.  

Fig.~\ref{fig:top_assisted_single_scalar} is based on the $Y_B$ expression of Eq.~(\ref{eqn:YB_ss_3}) and the DM abundance of Eq.~(\ref{eqn:chi_Y_from_Phi_decay_WTM}), modified for the single-scalar scenario as described above.  For  $\theta \ll1$ and $\Mchione \ll \Mchitwo$, both $\rhotot$ and $Y_B$ are approximately proportional to $\theta^2$.  This differs from the two-scalar model,  where $Y_B \sim \theta_1$ for small $\theta_1$.  In Fig.~\ref{fig:top_assisted_single_scalar}(b),  
 the viable $(M_\Phi,c\tau)$ parameter space is consequently
 not enhanced much by suppressing the $\chi_1$ couplings relative to  those of $\chi_2$, in contrast with the two-scalar scenario. 

\subsection{Asymmetry with Flavor-Universal Masses}\label{sec:ars_compare}

We now turn to the scenario where the top quark has a vanishing coupling to $\chi_I$, $F_{3I}=0$. In this case, the asymmetry can arise only at $\mathcal{O}(F^6)$, which is the same order as ARS leptogenesis. 
There are crucial differences between the 
 model in Eq.~\eqref{eq:single_scalar_model} and the conventional ARS model.  In the absence of neutrino masses, $B/3-L_\alpha$ is conserved for all three lepton flavors in the SM, while different quark flavors come into chemical equilibrium at temperatures $T\gg T_{\rm ew}$. The dominant source for the  asymmetry in ARS leptogenesis relies on the accumulation of asymmetries in individual lepton flavors, even though the total asymmetry sums to zero; these flavor asymmetries are then converted to a total lepton asymmetry by washout processes. For quarks, however, all flavors have equal chemical potentials, and the flavor asymmetries are therefore driven to zero by SM scattering processes. The standard ARS results therefore do not hold in the case where the $\chi$ fields couple predominantly to quarks.

As recently pointed out in Ref.~\cite{Abada:2018oly}, however, there exists an additional source term for the baryon asymmetry at $\mathcal{O}(F^6)$, and it is non-zero even for vanishing initial quark flavor chemical potentials. When  SM-Yukawa effects in the reaction densities are negligible, this is the dominant source for the baryon asymmetry in the case of QCD-charged $\Phi$.  Here we perform a systematic study of its effects.

The source in question requires {\it three} or more $\chi$ fields.  
We find that the asymmetry it produces is sufficient to account for the observed baryon asymmetry over a relatively restricted part of parameter space.
In particular, we find that $M_\Phi\lesssim2.5$ TeV to obtain the observed baryon asymmetry. The model therefore faces strong constraints from collider probes of $\Phi$, and the bulk of the parameter space can be tested with current experiments.

To study the  $\mathcal{O}(F^6)$ asymmetry, we turn to kinetic equations that give the evolution of density matrices for the various particle abundances; see Refs.~\cite{Hambye:2017elz,Abada:2018oly} and references therein. 
 We continue to focus on rates for processes that conserve 
$U(1)_{\chi - \Phi}$,
since violations of this symmetry include Majorana mass insertions that are subdominant at high temperature given the small masses for $\chi$ in our model \cite{Asaka:2005pn}. 
We are thus led to the single-scalar versions of Eqs.~(\ref{eqn:chiY}), (\ref{eqn:chibarY}), and (\ref{eqn:YUniversal}), the kinetic equations presented in Appendix~\ref{sec:kinetic_two_scalar} for the two-scalar case.
As we explain there, these are momentum-integrated equations that assume a thermal ansatz for the $\chi$ momentum distribution.
This treatment of the momentum dependence simplifies the analytic calculation of the $\mathcal{O}(F^6)$ asymmetry, and the two-scalar results of Appendix \ref{sec:thermal_ansatz_appendix} suggest it should give correct $Y_B$ values to within a factor of two.  
We refer the reader to Appendix~\ref{sec:kinetic_two_scalar} for notational background and other details regarding the kinetic equations.

We  assume there are no pre-existing asymmetries and solve the equations iteratively assuming small coupling $F$.   
Given our assumption of flavor-universal quark chemical potentials, there is no asymmetry at $\mathcal{O}(F^4)$, and consequently the   asymmetry at $\mathcal{O}(F^6)$ does not depend on washout terms.
Eq.~(\ref{eqn:chiY}) then becomes
\be\label{eq:density_matrix_evolution}
\frac{dY_{IJ}^\chi}{d\ln z} &=& -\frac{1}{2}\left\{  \tilde\gamma_0, Y^\chi - Y^\chi_{\rm eq}\right\},
\ee
where, as described in Appendix~\ref{sec:kinetic_two_scalar} and specifically Eq.~\eqref{eq:gamma_tilde}, $ \tilde\gamma_0$ is a dimensionless reaction-density matrix for $\chi$ production, defined in the interaction picture. $Y_{IJ}^\chi$ is proportional to the density matrix for $\chi$, containing information on both the abundance and phases for the coherent $\chi$ states.  It is convenient to define the dimensionless function $\bar\gamma(z)$,   obtained from the $\tilde\gamma_0$ reaction density of Eq.~\eqref{eq:gamma_tilde} by stripping off the $(F^\dagger F)_{IJ}$ and oscillation factors:
\be
\left[\tilde\gamma_0(z)\right]_{IJ} &=& (F^\dagger F)_{IJ}\,e^{i\Delta M_{IJ}^2z^3/3\mu_{\rm osc}^2}\,\bar\gamma(z),
\label{eqn:strip_flavor}
\ee
where 
\be
\mu_{\rm osc}^2 = \frac{2 T_\text{ew}^3}{M_0} \left\langle \frac{T}{E_\chi}\right\rangle^{-1} =\frac{36 \zeta(3) T_\text{ew}^3}{\pi^2 M_0} \simeq (3.75 \text{ keV})^2. \quad\quad
\ee

The total $\chi -  \bar\chi$ asymmetry, $\delta Y^\chi$, is found  by
 taking the trace of the difference in $\chi$ and $\bar\chi$ density matrices:  $\delta Y^\chi = \text{Tr} [Y^\chi - Y^{\bar\chi}]$.
The $U(1)_{\chi-\Phi}$ symmetry guarantees that  $\delta Y^\chi$ is the same as $\delta Y^\Phi \equiv Y^\Phi - Y^{\Phi^{*}}$.  
Following the discussion for the two-scalar case leading to Eqs.~(\ref{eqn:spec1}) and~(\ref{eqn:spec2}), 
the final baryon asymmetry can therefore be calculated as
\be
Y_B = \frac{{\mathcal K}_B}{{\mathcal K}_\Phi} \delta Y^\chi(z=1).
\ee 
Starting with the initial condition $Y^\chi=Y^{\bar\chi}=0$, we can obtain at $\mathcal{O}(F^2)$ the $\chi$ abundances
\be\label{eq:chi_abundance2}
Y^\chi_{IJ}(z) &=& (F^\dagger F)_{IJ}\,Y_{\rm eq}^\chi\,\int_{0}^{z}\,\frac{dz_1}{z_1} \bar\gamma(z_1)\,e^{i\Delta M_{IJ}^2z_1^3/3\mu_{\rm osc}^2}.
\quad\quad
\ee
Because  $Y^{\bar\chi}$ is obtained by switching $F \leftrightarrow F^*$, it is clear that no asymmetry arises at $\mathcal{O}(F^2)$, and straightforward to show  that none arises at $\mathcal{O}(F^4)$ either.

We determine the $\mathcal{O}(F^6)$ contribution to the baryon asymmetry by computing the $\chi$ asymmetry iteratively using  Eq.~\eqref{eq:density_matrix_evolution} three times. In this way we calculate the asymmetry to be
\be
\Ychi(z) &=& \frac{Y^\chi_{\rm eq}}{4}\int_0^z   \frac{dz_3}{z_3}      \int_0^{z_3}     \frac{dz_2}{z_2}     \int_0^{z_2}\frac{dz_1}{z_1}\nonumber\\
&&\mathrm{Tr}\left[\left\{\tilde\gamma(z_3),\left\{\tilde\gamma(z_2),\tilde\gamma(z_1)\right\}\right\}-(F\rightarrow F^*)\right].\quad\quad
\ee
Because of the cyclic property of trace, the integrand is fully symmetric under any permutation of the variables of integration. We can thus use the following identity for symmetric integrands, $\mathcal S$:
\be
\int_0^z \!dz_3\int_0^{z_3} \!dz_2\int_0^{z_2} \!dz_1\,\mathcal{S}(z_1,z_2,z_3)&&\nonumber \\
= \frac{1}{3!}\int_0^z\int_0^z\int_0^z\,dz_3\,dz_2\,dz_1\,&&\mathcal{S}(z_1,z_2,z_3),
\ee
which permits us to factorize our integral into a product of three integrals. With appropriate use of the symmetry of the integrand and re-labelling of variables of integration, we get 
\be
\Ychi(z) &=& \frac{iY^\chi_{\rm eq}}{3}\int_0^z\int_0^{z}\int_0^{z}\frac{dz_3}{z_3}\frac{dz_2}{z_2}\frac{dz_1}{z_1}\bar\gamma(z_1)\bar\gamma(z_2)\bar\gamma(z_3)\nonumber\\
&&{}\sum_{I,J,K}e^{i(\Delta M_{IJ}^2z_1^3+\Delta M_{JK}^2z_2^3+\Delta M_{KI}^2z_3^3)/3\mu_{\rm osc}^2}\nonumber\\
&&{}\quad\quad\mathrm{Im}\left[(F^\dagger F)_{IJ}(F^\dagger F)_{JK}(F^\dagger F)_{KI}\right].
\ee
The summed quantity is non-zero for $I\neq J\neq K$; thus, we need three $\chi$ particles to get a non-zero asymmetry (in agreement with the finding of Ref.~\cite{Abada:2018oly}). We assume for simplicity that there are precisely three $\chi$ fields. We must then sum separately over even and odd cyclic permutations of $\{1,2,3\}$. The permutations within each equivalence class are identical due to the symmetry of the integrand under permutations of $\{z_1,z_2,z_3\}$, while interchanging even and odd permutations is equivalent to complex conjugating both the oscillation factor and $F$. We thus obtain our final expression for the $\chi$ asymmetry: 
\be
\Ychi(z) &=& -2Y^\chi_{\rm eq}\,\mathrm{Im}\left[(F^\dagger F)_{12}(F^\dagger F)_{23}(F^\dagger F)_{31}\right]\nonumber\\
&&\quad\mathrm{Im}\left[\tilde f_{12}(z)\tilde f_{23}(z)\tilde f_{31}(z)\right],
\label{eq:final_singlescalar}
\ee
where
\be
\tilde f_{IJ}(z) &=& \int_0^z\,\frac{dz'}{z'}\,\bar\gamma(z')\,e^{i\Delta M_{IJ}^2 {z'}^3/3\mu_{\rm osc}^2}.
\label{eq:final_singlescalar2}
\ee

Because $M_\Phi \gg T_{\rm ew}$ in our QCD-triplet $\Phi$ model, the production of the baryon asymmetry is dominated by interactions that occur when the $\Phi$ mass is dominated by its tree-level value, rather than by the thermal corrections typical in the conventional ARS scenario.  Thus, for our subsequent numerical work we neglect all thermal masses, which leads to the following expression for 
the dimensionless reaction density $\bar\gamma(z)$: 
\be
\bar\gamma(z) & = & \frac{g_\Phi M_\Phi^2 T_{\rm ew}^2}{32\pi^3Y_{\rm eq}^\chi s_{\rm ew} H_{\rm ew}} \,z^3\int_0^\infty\frac{du}{e^u+1} \nonumber\\
&&\int_{M_\Phi^2z^2/(4uT_{\rm ew}^2)}^\infty dw\left(\frac{1}{e^w+1}+\frac{1}{e^{u+w}-1}\right).
\quad\quad
\label{eq:reaction_actual}
\ee

 In our final expression for the baryon asymmetry, we  include the effects of $\Phi$ decays prior to the electroweak phase transition on the asymmetry through the inclusion of a survival factor analogous to Eq.~\eqref{eqn:survival}. Note that the survival factor actually modifies the integrand so that it is no longer symmetric with respect to interchange of $z_3$ with $z_1$ or $z_2$. This makes the calculation considerably more complicated. However, because the asymmetry is predominantly produced at $T\gtrsim1$ TeV and the greatest sensitivity of the survival factor is to decays at $T_{\rm ew}\ll 1$ TeV, we can to a good approximation assume the asymmetry is produced very early on and 
take $S_\Phi(0)$ as our survival factor, calculated using
 Eq.~(\ref{eqn:survival}).
  This re-establishes the symmetry of the integrand, and the baryon asymmetry is consequently 
\be
\YB(z) &\simeq& -\frac{2 \mathcal{K}_B}{\mathcal{K}_\Phi}Y^\chi_{\rm eq}
\,\mathrm{Im}\left[(F^\dagger F)_{12}(F^\dagger F)_{23}(F^\dagger F)_{31}\right]\nonumber\\
&&
\times S_\Phi(0)\,\mathrm{Im}\left[\tilde f_{12}(z)\tilde f_{23}(z)\tilde f_{31}(z)\right].
\label{eqn:single_scalar_YB}
\ee
Below, we compare this perturbative result to a fully numerical solution to the kinetic equations and find good agreement in the weak-washout regime; see Fig.~(\ref{fig:singlescalar_comparepert}).

\subsubsection*{Baryon Asymmetry \& Dark Matter}

We now proceed to study the parameter space over which the baryon asymmetry can be obtained. We also investigate whether this parameter space is consistent with obtaining the correct abundance of DM, finding that it is 
unlikely that $\chi$ can have the correct DM abundance if we impose the observed baryon asymmetry.

As we will show in Sec.~\ref{sec:signals}, the most constraining aspects of the model with a QCD-charged scalar are the direct limits from colliders, $M_\Phi \gtrsim1$ TeV. As a result, the dominant epoch of $\chi$ production and inverse decay is $T\sim1$ TeV, with a corresponding optimal mass splitting of $\Delta M^2\sim(10\,\,\mathrm{keV})^2$ corresponding to oscillations at $T\sim M_\Phi$. This leads to an upper bound on the asymmetry for fixed $F$.

Since the baryon asymmetry arises at $\mathcal{O}(F^6)$ in the case of a single scalar, the baryon asymmetry is much smaller than in Sec.~\ref{sec:calculation} for the same Yukawa couplings. Alternatively, the Yukawa couplings must be larger to accommodate the observed baryon asymmetry, leading to the over-production of $\chi$ in this scenario relative to the DM abundance. This over-production can satisfy cosmological constraints on $N_{\rm eff}$ if $\chi$ decays to lighter species, but this is incompatible with  $\chi$ being the DM.

To set up our numerical studies, we parametrize the coupling factor of Eq.~(\ref{eqn:single_scalar_YB})
in a manner analogous to what we did for the two-scalar model in Eq.~\eqref{eq:jarlskog_invariant},
\be\label{eq:singlescalar_Jarlskog}
4\,\mathrm{Im}\left[(F^\dagger F)_{12}(F^\dagger F)_{23}(F^\dagger F)_{31}\right] &=& \mathcal{J} \left[\mathrm{Tr}(F^\dagger F)\right]^3,
\quad\quad
\ee
where $\mathcal{J}$ is a Jarlskog-like invariant. In Appendix \ref{app:singlescalar_param}, we show that $\mathcal{J}\le1/27$. However, the optimal choice for the baryon asymmetry may lead to an over-production of DM. To demonstrate this, we introduce the angle
\be
\cos\theta_1 &=& \sqrt{\frac{(F^\dagger F)_{11}}{\mathrm{Tr}F^\dagger F}},
\ee
 which is analogous to the $\theta_i$ angles of the two-scalar model and describes how strongly $\Phi$ is coupled to $\chi_1$ relative to the two heavier species. Similarly, a related quantity
 \be
 \cos\theta_2 &=& \frac{1}{\sin\theta_1}\sqrt{\frac{(F^\dagger F)_{22}}{\mathrm{Tr}F^\dagger F}},
 \ee
 specifies the  coupling of $\Phi$ to $\chi_2$ relative to its coupling to $\chi_3$.
  As shown in Appendix \ref{app:singlescalar_param},
 \be\label{eq:approx_Jarlskog_f6}
 \mathcal{J} &\propto& \cos^2\theta_1 \sin^4\theta_1 \sin^2(2\theta_2),
 \ee
  which is maximized for $\cos\theta_1 = 1/\sqrt{3}$ and $\theta_2=\pi/4$. However, the DM abundance depends on
  \be
 \rhotot
 &\propto& M_1 \cos^2\theta_1+ M_2\sin^2\theta_1\cos^2\theta_2 \nonumber\\
  &&\quad{}+ M_3\sin^2\theta_1\sin^2\theta_2.
  \ee 
  If $M_1\rightarrow0$, then it may be preferable to have $\theta_1\ll1$, which suppresses the baryon asymmetry but also prevents an over-abundance of DM. For our numerical studies, we fix $\theta_2=\pi/4$ but scan over all possible values of $\theta_1$ to uncover the largest possible parameter space consistent with both the observed baryon asymmetry and DM abundance. For a complete description of the other parameters in $\mathcal{J}$ and their optimal values, see Appendix \ref{app:singlescalar_param}.

Since  $\Gamma_\Phi = \mathrm{Tr}(F^\dagger F) M_\Phi / 16 \pi$, it is possible to express the asymmetry in terms of $M_\Phi$, $\Gamma_\Phi$ (or, equivalently, $c\tau_\Phi$), and the mass splittings $\Delta M_{IJ}^2$. To simplify the numerical study, we parametrize the $\chi$ masses and splittings through a single parameter, $M_\chi$:~$M_1=0$, $M_2=M_\chi$, and $M_3=2M_\chi$. This minimizes the DM abundance while keeping an equal mass splitting between states.

We now investigate the possibility that the $\chi$ fields constitute the DM. In this case, we fix the values of $M_\Phi$, $M_\chi$, and $\mathcal{J}$ as described above; requiring the $\chi$ abundances as calculated in Eq.~\eqref{eq:chi_abundance2} to match the observed DM abundance dictates the value of $\mathrm{Tr}(F^\dagger F)$, which in turn can be used to calculate the baryon asymmetry. This is the maximum baryon asymmetry subject to the requirement of obtaining the DM abundance, since the baryon asymmetry can always be made smaller with smaller values of the $CP$-violating phases in $\mathcal{J}$.

\begin{figure}
   \centering
          \includegraphics[width=3.4in]{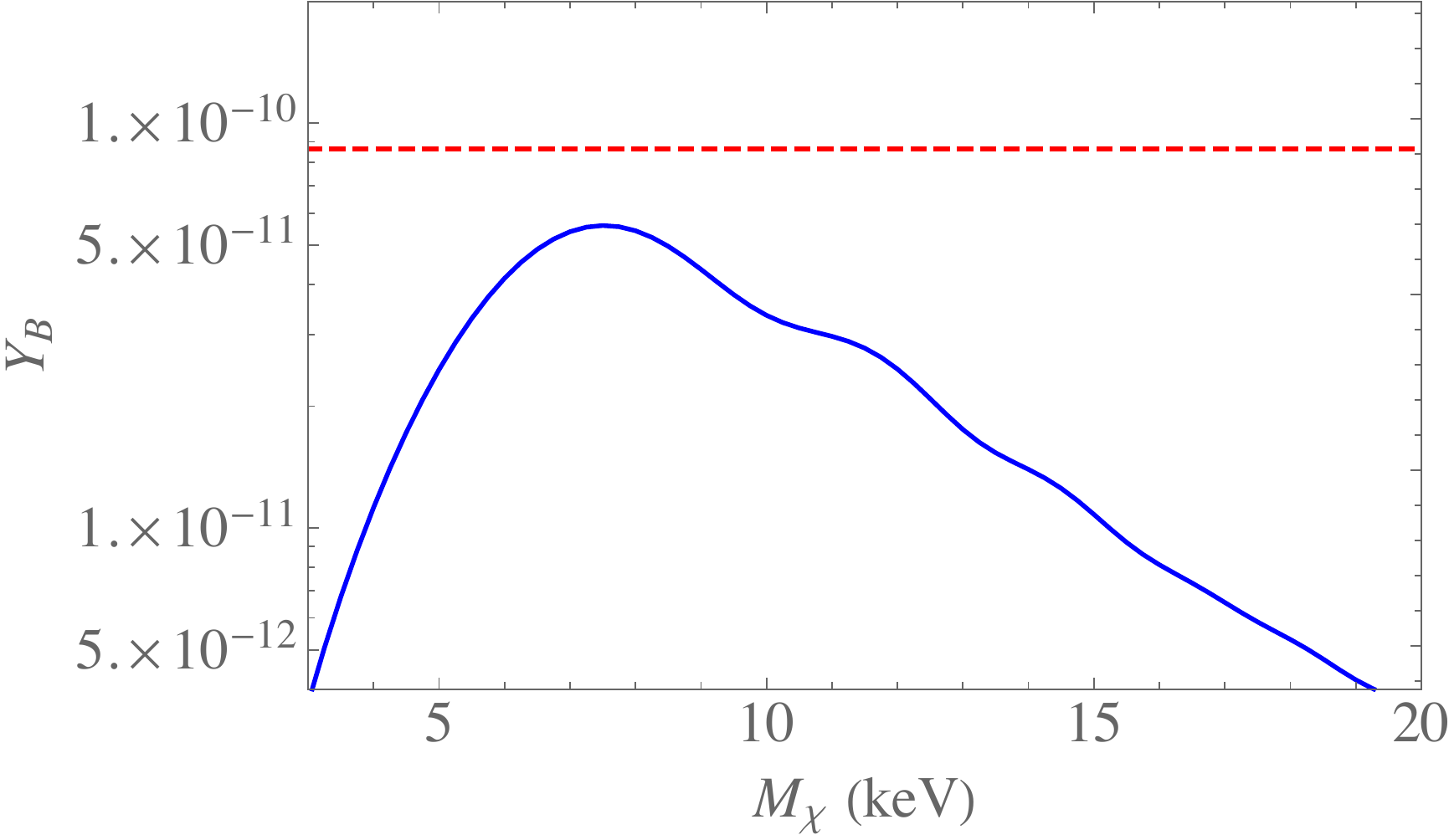}
   \caption{
   For $\theta_2 = \pi/4$, $M_\Phi=1$ TeV, and equally-spaced $\chi$ masses $M_1=0$, $M_2=M_\chi$, and $M_3=2M_\chi$,  the maximum baryon asymmetry that can be obtained in the single-scalar scenario of Sec.~\ref{sec:ars_compare} when we impose the requirement that the total $\chi$ abundance match the observed DM abundance. The observed baryon asymmetry is indicated with a dashed red line.
   }
   \label{fig:singlescalar_DM_baryo}
\end{figure}

We show in Fig.~\ref{fig:singlescalar_DM_baryo} our results for a scalar mass of $M_\Phi=1$ TeV. 
We see that, even for the favorable parameters we have chosen, the maximum possible baryon asymmetry is $\YB\approx 6\times10^{-11}$, below the observed value. 
Since $M_\Phi\gtrsim1$ TeV from current collider constraints (see Sec.~\ref{sec:signals}), we conclude that 
simultaneously accounting for DM and the baryon asymmetry is impossible, or at least very difficult,  in this particular scenario. 
The lack of viable parameter space is directly linked to the fact that the asymmetry arises at $\mathcal{O}(F^6)$ with a single scalar while the DM abundance is still established at $\mathcal{O}(F^2)$, a serious obstacle to simultaneously satisfying both observed abundances.

In examining our results, we have found that the largest baryon asymmetry consistent with the DM abundance is associated with large mixing angles $\theta_1\sim1$. This is in contrast with our findings in the two-scalar model, where there was a larger parameter space associated with small mixings.  Since we now have two massive $\chi$ states, and $\Phi$ must couple to all three of them to generate a baryon asymmetry, it is difficult to get an appreciable asymmetry without significantly populating the heavier states. Additionally, according to Eq.~\eqref{eq:approx_Jarlskog_f6} the baryon asymmetry for small mixing angles is proportional to $\theta_1^4$, which is a significant suppression. Indeed, as we show in the next section, it is difficult to obtain the observed baryon asymmetry even for maximal mixing and removing all constraints from DM.

\subsubsection*{Baryon Asymmetry Without Dark Matter}
\label{sec:no_DM}

Above, we found that imposing the requirement that $\chi$ constitute the DM yields a baryon asymmetry that is too small. Alternatively, obtaining the correct baryon asymmetry leads to an over-abundance of $\chi$. This is not necessarily a problem:~since the $\chi$ hidden sector is relatively poorly constrained, it is possible that there exist additional states to which the heavier $\chi$ fields could decay. For example, we could imagine a model with a new massless singlet scalar $\varphi$ such that $\chi_{2,3}\rightarrow \chi_1\varphi$ prior to recombination. Because all of the $\chi$ fields have sub-thermal number densities and they are produced at $T\sim$TeV, such a scenario is safely within cosmological limits from the effective number of neutrinos ($N_{\rm eff}$) provided $\chi_1$ is sufficiently light. 

While it may be possible for $\chi$ or some other hidden sectors fields to be DM in this scenario, the details depend sensitively on the content and structure of the hidden sectors. Because of the loss of predictive power with respect to the DM abundance in this scenario, we instead take a different approach:~we simply assume that the $\chi$ fields decay to (nearly) massless particles that are safe from cosmological limits, and abandon the requirement of obtaining the DM density. We then explore which parameters can still give rise to the baryon asymmetry based on the production and oscillations of $\chi$.

As in Sec.~\ref{sec:calculation}, the baryon asymmetry depends on several physical parameters:~the mass of the scalar, $M_\Phi$; the Yukawa couplings, or alternatively the $\Phi$ lifetime $c\tau_\Phi = 1/\Gamma_\Phi$; the mass splittings, $\Delta M_{IJ}^2$, and the $CP$ phases as encoded in $\mathcal{J}$. Because we wish to explore the most expansive parameter space that gives rise to the observed baryon asymmetry, we set $\mathcal{J}$ to its maximal single-scalar value of $1/27$ (see Appendix \ref{app:singlescalar_param}). Once again, we parametrize the $\chi$ masses and splittings through a single parameter, $M_\chi$:~$M_1=0$, $M_2=M_\chi$, and $M_3=2M_\chi$.

Up until now, we have employed a perturbative analysis as outlined in Sec.~\ref{sec:mechanism}, which is valid in the out-of-equilibrium, weak-washout regime. The requirement that $\chi$ constitute the DM situates us safely within the perturbative regime. Once we relax this assumption, however, it is possible that $\chi$ attains a near-equilibrium abundance and baryogenesis can still occur. For example, if $\chi$ decouples at $T\gtrsim100$ GeV, then each Weyl fermion only contributes 0.05 to $N_{\rm eff}$ \cite{Brust:2013xpv}. As a result, we must consider the possibility that the $\chi$ particles come close to equilibrium.

It is perhaps surprising that the strong-washout regime would be relevant at all for baryogenesis, since the asymmetry appears to be exponentially damped. However, if the $CP$-violating source and washout terms are both large, then the asymmetry can reach a quasi-steady-state solution where $d\YB/dz = 0$ due to a cancellation between source and washout terms:~if the kinetic equations have the form $d\YB/dz = S(z) - W(z) \YB$, we see that a quasi-steady-state solution is obtained with $\YB = S(z) / W(z)$. In this case, the asymmetry is \emph{not} exponentially suppressed. 

To generate a sizable asymmetry in the strong washout regime, we need a large source of $CP$-violation down to $T\sim T_{\rm ew}$; since the production rate of $\Phi$ is suppressed by $e^{-M_\Phi/T_{\rm ew}}$, this suggests that $M_\Phi$ cannot be too much larger than the electroweak scale for the strong washout regime to be relevant. Furthermore, we must have $Y_\chi \neq Y_\chi^{\rm eq}$. For the optimal benchmark outlined in Appendix \ref{app:singlescalar_param} with $\mathcal{J}=1/27$, $F^\dagger F$ has a zero eigenvalue, meaning that there is a linear combination of $\chi$ states that does not interact with $\Phi$. This is valid \emph{until} oscillations become important, in which case the final $\chi$ state is brought into equilibrium. For sufficiently small $M_\chi$, this can lead to an appreciable $CP$-violating rate even at $T\sim T_{\rm ew}$.

There is one final effect we must consider:~sphaleron decoupling. In the strong-washout regime, the baryon asymmetry is being continually generated and destroyed at $T\sim T_{\rm ew}$ and so the final baryon asymmetry depends sensitively on the effects of sphaleron decoupling. In other words, it is perhaps too simplistic to assume that sphaleron decoupling is instantaneous at $T_{\rm ew}\approx130$ GeV. 
To go beyond this instantaneous approximation, we follow Ref.~\cite{Eijima:2017cxr}; since there are no new chiral states that couple to sphalerons in our model, the rates relating baryon and lepton number are the same as in the SM. The effect is a gradual decoupling of sphalerons as we approach $T_{\rm ew}$.

\begin{figure*}
   \centering
          \includegraphics[width=3in]{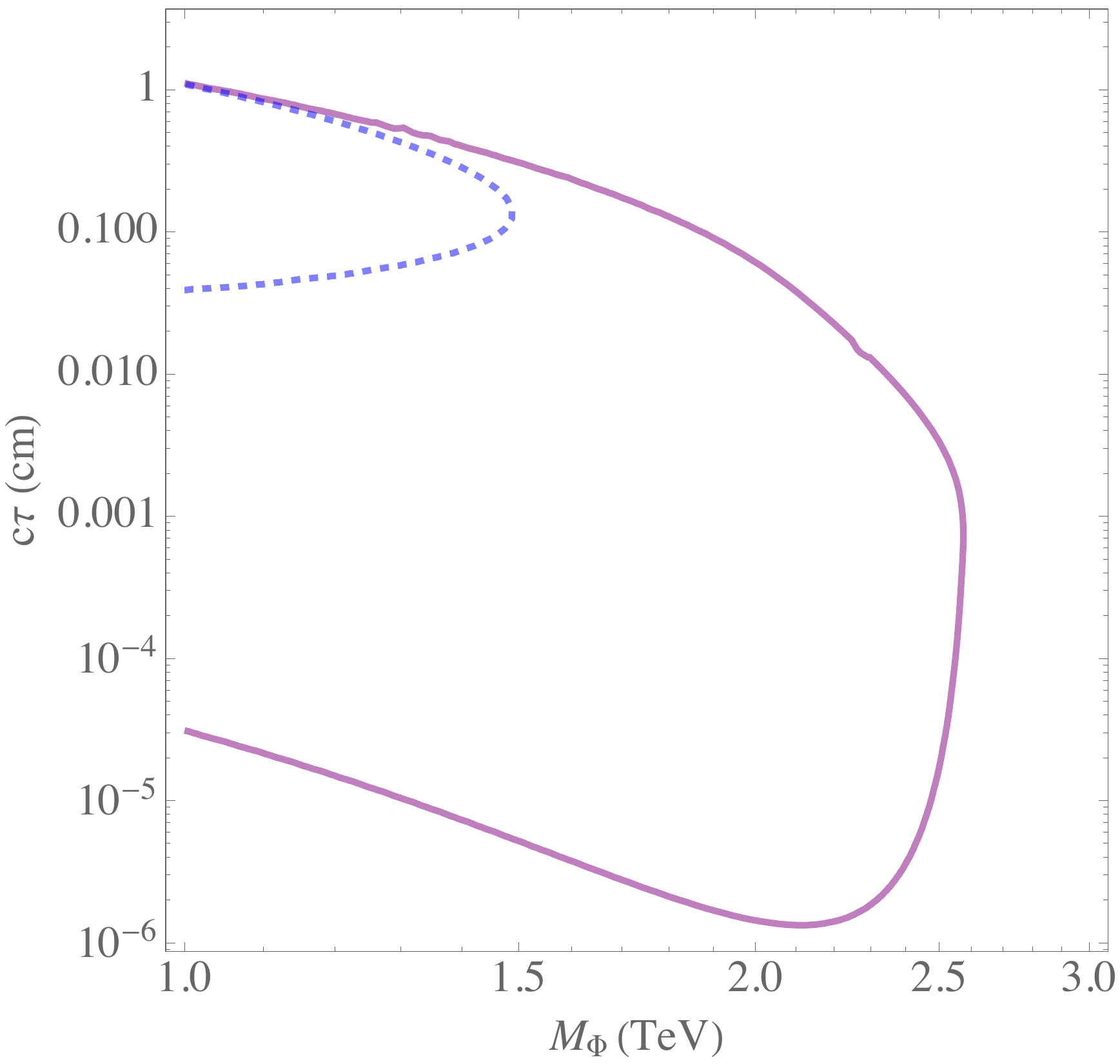}
          \quad \quad \quad
                    \includegraphics[width=3in]{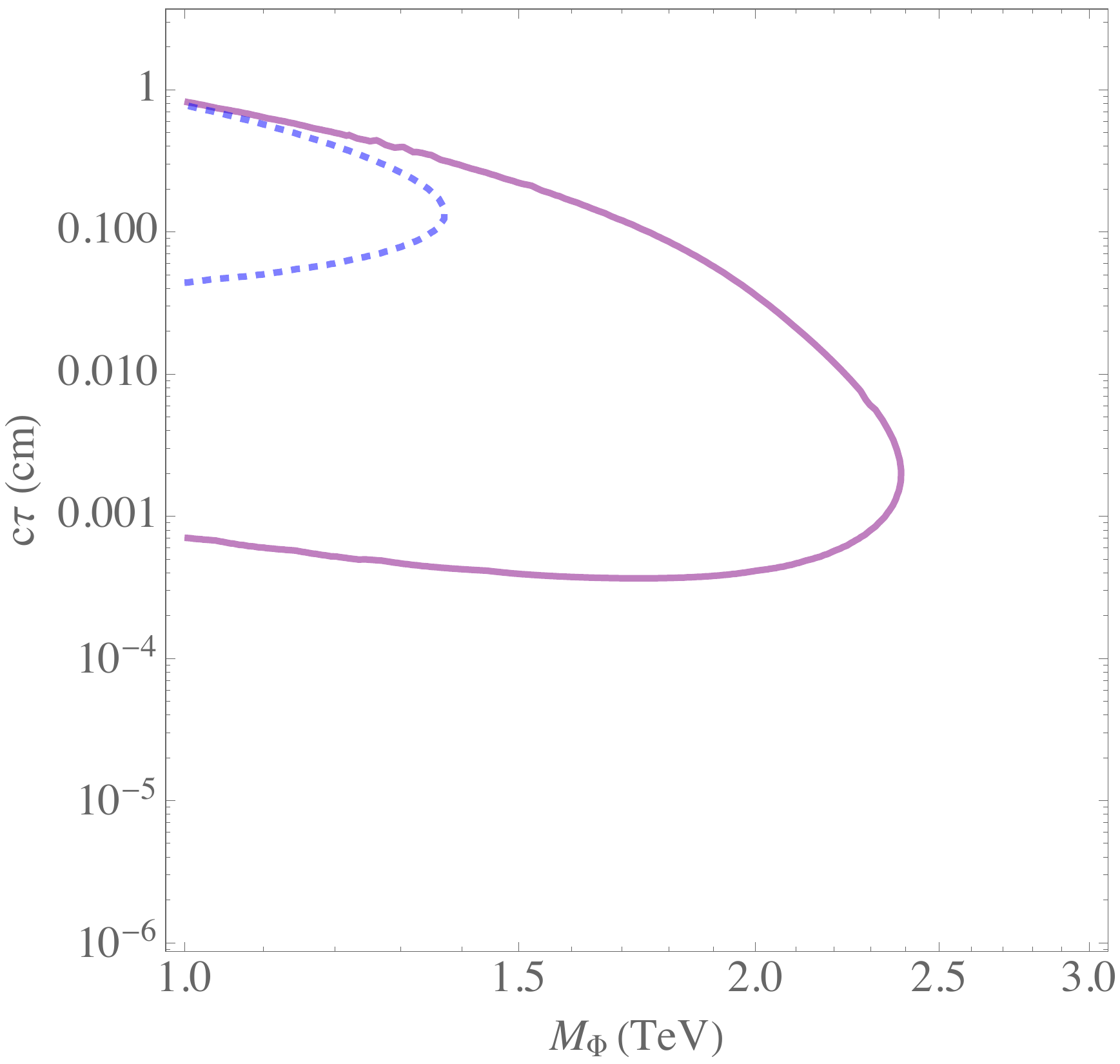}
   \caption{
For the single-scalar model with no DM constraint, comparison of parameters giving rise to the observed baryon asymmetry based on (blue, dashed) perturbative analysis, and (purple, solid) solution of kinetic equations. The $\chi$ masses are $M_1=0$, $M_2=5$ keV, $M_3=10$ keV. (Left) Couplings correspond to optimal $CP$-violating parameters from Appendix \ref{app:singlescalar_param}.  (Right) Couplings correspond to the modified benchmark from  Appendix \ref{app:singlescalar_param}.
   }
   \label{fig:singlescalar_comparepert}
\end{figure*}

Putting all of these effects together, we solve the kinetic equations with a thermal ansatz for the $\chi$ energies to compute the baryon asymmetry, including washout and back-reaction terms. To illustrate the parameter space for which strong washout is relevant, we compare the full solution of the kinetic equations with that of our perturbative analysis in Fig.~\ref{fig:singlescalar_comparepert}. We show results  for both the optimal benchmark couplings from  Appendix \ref{app:singlescalar_param}, as well as for a modified benchmark (also outlined in Appendix \ref{app:singlescalar_param})  which leads to an earlier equilibration time of all $\chi$ interaction eigenstates. Any area to the left of the indicated contours can give rise to the observed baryon asymmetry. The asymmetry in the strong-washout limit is still relevant, but is much reduced relative to the optimal benchmark.

\begin{figure}
   \centering
          \includegraphics[width=3.2in]{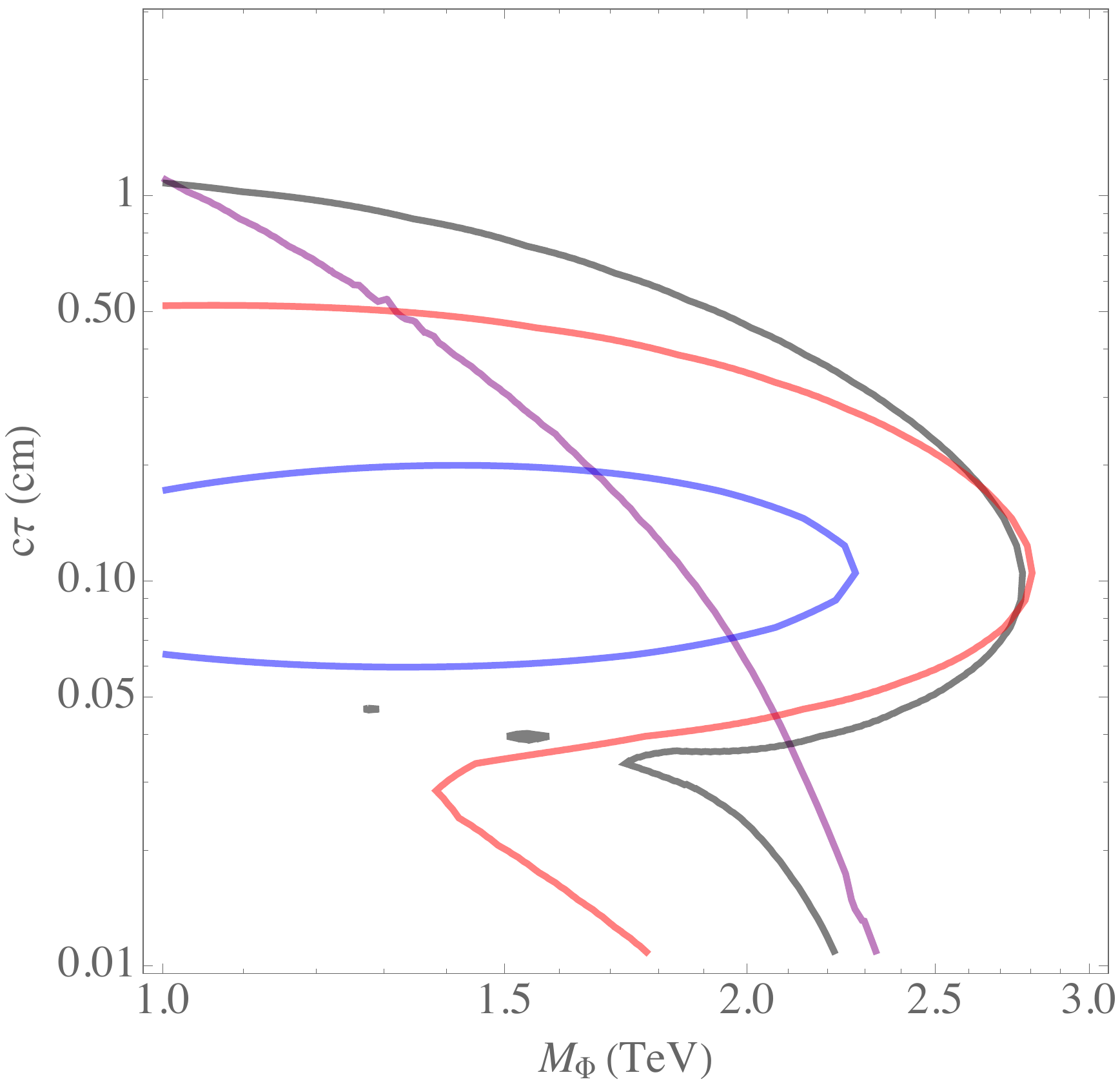}\\
           \includegraphics[width=3.2in]{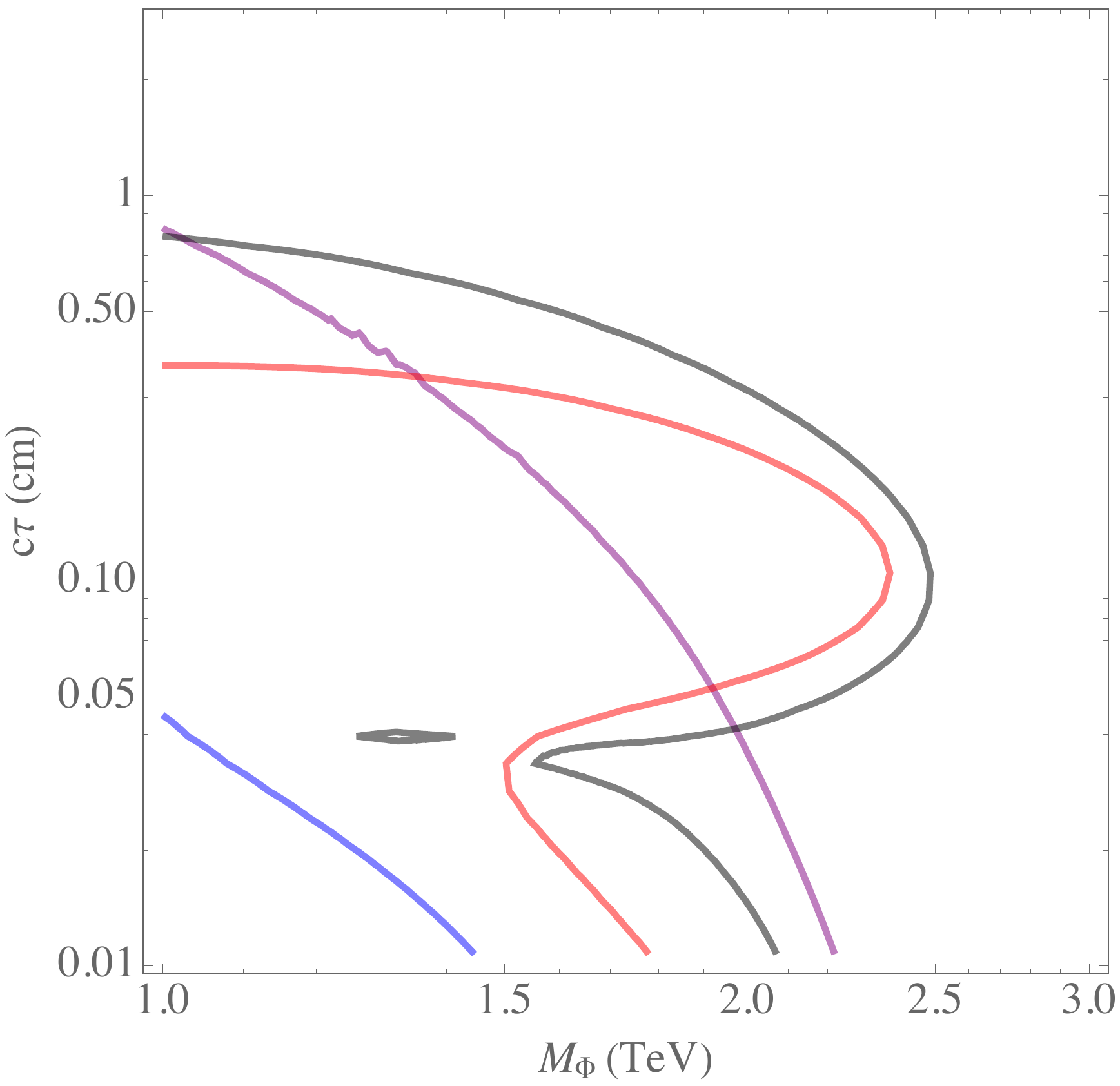}
   \caption{
  Comparison of parameters giving rise to the observed baryon asymmetry for $c\tau>0.01$ cm, based on a full solution of kinetic equations for (top) the optimal $CP$-violating parameters; (bottom) the modified benchmark, both from Appendix  \ref{app:singlescalar_param}. The $\chi$ masses are $M_1=0$, $M_2=M_\chi$, $M_3=2M_\chi$ with:~(purple) 5 keV, (black) 30 keV, (red) 50 keV, (blue) 80 keV. 
   }
   \label{fig:singlescalar_comparechimass_benchmark}
\end{figure}

We show in Fig.~\ref{fig:singlescalar_comparechimass_benchmark} the contours giving rise to the observed baryon asymmetry for different $\chi$ masses using the solution to the full kinetic equations.  We show results for both the optimal $CP$-violating parameters and the modified benchmark (see Appendix \ref{app:singlescalar_param}). We see that it is possible to obtain the observed baryon asymmetry for $M_\phi\lesssim2.5$ TeV depending on the $\Phi$ lifetime. It is evident that the strong-washout limit is relevant for a wide range of $\chi$ masses. We also observe interesting features in the shapes of the contours, which are due to the presence of a multitude of important time scales in the asymmetry generation process, including three oscillation times corresponding to the $\Delta M_{IJ}^2$, as well as the time scale of the decays and inverse decays of $\Phi$. While we have attempted to characterize the precise shapes of the oscillations in the contours in Fig.~\ref{fig:singlescalar_comparechimass_benchmark}, we have been unable to find a simple explanation due to the irreducible complexity of the four different time scales. However, we have checked that our solutions are robust against variations of the methods of performing the numerical integration, as well as under small variations of the initial conditions and parameters, suggesting that our solutions are physically correct. 

\begin{figure}
   \centering
          \includegraphics[width=3.2in]{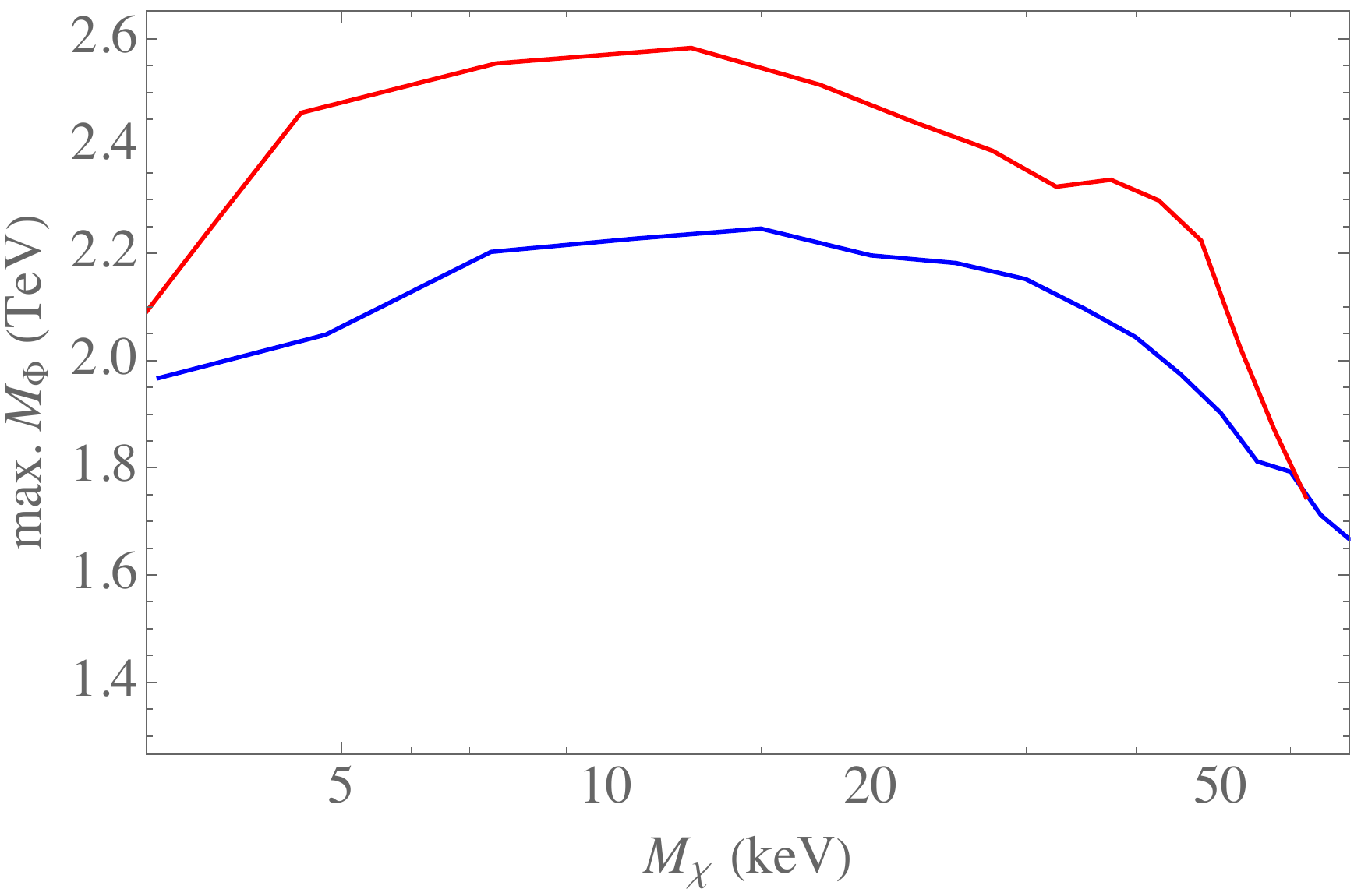}
   \caption{
  Maximum value of $M_\Phi$ that allows successful generation of the baryon asymmetry in the strong washout limit ($c\tau_\Phi<0.01$ cm) as a function of $M_\chi$. The $\chi$ masses are $M_1=0$, $M_2=M_\chi$, $M_3=2M_\chi$. The $CP$-violating parameters are (red) the optimal benchmark, and (blue) the modified benchmark described in Appendix \ref{app:singlescalar_param}. It is evident that  $\Phi$ must be lighter than approximately 2.5 TeV.
   }
   \label{fig:singlescalar_maxMPhi}
\end{figure}

For $c\tau_\Phi<0.01$ cm, the only phenomenologically distinguishable feature of the model is the value of $M_\Phi$:~the $\Phi-\chi-Q$ couplings are sufficiently small as to be difficult to probe directly, and $\Phi$ now decays promptly in a collider experiment, which removes the main experimental consequence of the non-zero lifetime. Therefore, we truncate Fig.~\ref{fig:singlescalar_comparechimass_benchmark} at $c\tau_\Phi=0.01$ cm, and for shorter lifetimes present instead the \emph{maximum} value of $M_\Phi$ that can give rise to the baryon asymmetry with $c\tau_\Phi<0.01$ cm. The maximum $M_\Phi$ values are found via a scan over the $c\tau_\Phi-M_\Phi$ parameter space, and to remove jaggedness associated with the granularity of the scan we perform a running average. We show the maximum value of $M_\Phi$ consistent with the baryon asymmetry for different values of $M_\chi$ in Fig.~\ref{fig:singlescalar_maxMPhi}. We see that the maximum value of $M_\Phi$ is obtained for $M_\chi\sim10$--20 keV, such that oscillations regularly occur at the sphaleron decoupling temperature but are not too fast. We also see that the modified benchmark permits a smaller range for $M_\Phi$ at $c\tau_\Phi<0.01$ cm, which is consistent with Fig.~\ref{fig:singlescalar_comparepert}.

To summarize the results of the analysis with a single QCD-triplet scalar, neglecting flavor-dependence in the quark thermal masses, we find that obtaining the observed baryon asymmetry is possible but is apparently incompatible with $\chi$ being DM candidates. The parameter space for a baryon asymmetry is quite limited, with the scalar having a mass $M_\Phi \lesssim2.5$ TeV, with lifetimes ranging from prompt to the centimeter scale. This gives the model excellent prospects for discovery or exclusion in the high-luminosity phase of the LHC.

\subsubsection*{Comparison of Asymmetry with ARS Leptogenesis}

We conclude this section by comparing two potential $\mathcal{O}(F^6)$ contributions to the baryon asymmetry: (1) the standard ARS contribution involving flavor-dependent washout, and (2)  the contribution studied above and identified in Ref.~\cite{Abada:2018oly},  which survives when the active fermions have flavor-universal chemical potentials, {\em i.e.} the relevant one in the
QCD-triplet scalar case.  

To make this comparison we need to go beyond the kinetic equations developed in Appendix~\ref{sec:kinetic_two_scalar}, which assume flavor-universal chemical potentials for the active fermions.  We instead consider
\be\label{eq:density_matrix_evolution_flavor}
\frac{dY_{IJ}^\chi}{d\ln z} &=& -\frac{1}{2}\left\{ \sum_\alpha \tilde\gamma_\alpha, Y^\chi - Y^\chi_{\rm eq}\right\},\\
\frac{d \;\Ya}{d\ln z} &=& \mathrm{Tr}\left[\tilde\gamma_\alpha Y^\chi - \tilde\gamma_\alpha^* Y^{\bar\chi}\right] - \Ya\left(\frac{Y_{\rm eq}^\chi}{Y_{\rm eq}^\alpha}\right)\mathrm{Tr}\,\tilde\gamma_\alpha^{\rm w},\quad
\ee
where $\Ya$ is the asymmetry in the SM fermion flavor $\alpha$; because of rapid scattering, we do not have to keep track of oscillations or preserve off-diagonal components of the $\Ya$ density matrix.  Note that we now include a flavor-specific washout reaction density 
$\tilde\gamma_\alpha^{\rm w}$ along with $\tilde\gamma^0_{\alpha}$, the flavor-specific version of the reaction density considered earlier.  In place of Eq.~(\ref{eqn:strip_flavor}), we can write these  $\chi$-space matrices as
\be
\tilde\gamma^{0,\text{w}}_\alpha(z)_{IJ} &\equiv& F_{\alpha I}^* F_{\alpha J}\,e^{i\Delta M_{IJ}^2z^3/3\mu_{\rm osc}^2}\,\bar\gamma^{0,\text{w}}(z),
\ee
where $\bar\gamma^0(z)$ and $\bar\gamma^\text{w}(z)$ are dimensionless and flavor-universal functions of temperature.  

The standard analytic results for ARS assume that the scattering processes for $\chi$ production are dominated by $2\leftrightarrow2$ processes and decays of the SM Higgs where the dominant contribution to the Higgs mass comes from thermal processes. In this case, all reaction densities are by dimensional analysis proportional to $T^4$, and because the barred reaction densities are obtained from these by stripping off the coupling factors and dividing by $Y_{\rm eq}^\chi \,s(z) H(z)$, we have
\be\label{eq:ARS_gammabar}
\bar\gamma^{0,\text{w}} (z) &=& \frac{\alpha_\chi^{0,\text{w}} T_{\rm ew}^4}{Y_{\rm eq}^\chi s_{\rm ew} H_{\rm ew}} \,z,
\ee
where $\alpha_\chi^{0}$ and $\alpha_\chi^{\text{w}}$ are dimensionless constants.  To make a meaningful comparison we use this form for the reaction densities to calculate both  $\mathcal{O}(F^6)$ contributions.

\paragraph{Conventional ARS Contribution.}~At $\mathcal{O}(F^4)$, the $\chi$ abundance leads to an asymmetry in SM fermions due to a difference in $\chi$ and $\bar\chi$ rates for flavor $\alpha$:
\be
\Ya(z) &=& \int_0^z\,\frac{dz_2}{z_2} \mathrm{Tr}\left[\tilde\gamma^0_\alpha(z_2) Y^\chi(z_2) - {\tilde\gamma}_\alpha^{0\,*}(z_2) Y^{\bar\chi}(z_2)\right]\nonumber\\
&=& 4\sum_{I<J}\mathrm{Im}\left[F_{\alpha J}F_{\alpha I}^*(F^\dagger F)_{JI}\right]\int_0^z\,\,\frac{dz_2}{z_2}\int_0^{z_2}\,\frac{dz_1}{z_1}\nonumber\\
&&{}\quad\quad\bar\gamma^0(z_1)\bar\gamma^0(z_2)\sin\left[\frac{\Delta M_{JI}^2(z_2^3-z_1^3)}{3\mu_{\rm osc}^2}\right] Y_\chi^{\rm eq}.\quad\quad
\ee
However, as argued in Sec.~\ref{sec:qualitative}, the total 
asymmetry is zero because
\be
\delta Y^\chi
=-
\sum_\alpha\,\Ya \propto \sum_\alpha \left[F_{\alpha I}F_{\alpha J}^*(F^\dagger F)_{IJ}\right] = 0.
\ee
A total asymmetry arises at $\mathcal{O}(F^6)$ because
\be
\frac{d}{d\ln z} 
\sum_\alpha\,\Ya
&=& -\left(\frac{Y_{\rm eq}^\chi}{Y_{\rm eq}^\alpha}\right)\sum_\alpha \Ya \mathrm{Tr}\,\tilde\gamma_\alpha^{\rm w}.
\ee
Integrating to $z_{\rm ew}=1$, we obtain
\be
\delta Y^\chi &=& \sum_\alpha\int_0^1\left(\frac{Y_{\rm eq}^\chi}{Y_{\rm eq}^\alpha}\right)\frac{dz}{z} \Ya(z) \mathrm{Tr}\,\tilde\gamma_\alpha^{\rm w}(z)\\
&=& \sum_\alpha(FF^\dagger)_{\alpha\alpha}\left(\frac{Y_{\rm eq}^\chi}{Y_{\rm eq}^\alpha}\right)\int_0^1\,\frac{dz}{z} \Ya(z) \bar\gamma^{\rm w}(z).
\ee

Thanks to Eq.~(\ref{eq:ARS_gammabar}), the analytic results simplify significantly because the factors of $z$ cancel in the integrals and we have, restricting ourselves to only two $\chi$ particles,
\be
\int_0^z\!\!dz_2\,\int_0^{z_2}\!\!dz_1\sin\left[\frac{\Delta M_{21}^2(z_2^3-z_1^3)}{3\mu_{\rm osc}^2}\right]\approx 1.4\left(\frac{\mu_{\rm osc}^2}{\Delta M_{21}^2}\right)^{2/3}\quad
\ee
for $\Delta M_{21}^2/3\mu_{\rm osc}^2 z^3\gg1$ (\emph{i.e.,} many oscillations prior to time $z$) \cite{Hambye:2017elz}. We then have
\be
\delta Y^\chi &=& \frac{5.6\alpha_\chi^{\rm w}({\alpha_\chi^0})^2 T_{\rm ew}^{12}}{(Y_{\rm eq}^\chi s_{\rm ew} H_{\rm ew})^3}\left(\frac{\mu_{\rm osc}^2}{\Delta M_{21}^2}\right)^{2/3}Y_{\rm eq}^\chi\nonumber\\
&&\quad\sum_\alpha\,\frac{Y_{\rm eq}^\chi}{Y_{\rm eq}^\alpha}(FF^\dagger)_{\alpha\alpha}\,\mathrm{Im}\left[F_{\alpha1}F_{\alpha2}^*(F^\dagger F)_{12}\right],\quad\quad\label{eq:ARS_final}
\ee
which agrees with Eq.~(A8) of Ref.~\cite{Hambye:2017elz} for the case where $\chi$ is coupled to SM leptons. 

\paragraph{Contribution without flavor-dependent washout.
}
We now compare the ARS result with what we get when we evaluate Eqs.~(\ref{eq:final_singlescalar})
and (\ref{eq:final_singlescalar2}), plugging in Eq.~(\ref{eq:ARS_gammabar}) as the barred reaction density.  
Defining $\beta_{IJ}\equiv \Delta M_{IJ}^2 / 3\mu_{\rm osc}^2$, and taking the limit of many oscillations ($z^3\beta_{IJ}\gg1$), we find
\be
\tilde f_{IJ}(z) \rightarrow \frac{\alpha_\chi^0 T_{\rm ew}^4}{Y_{\rm eq}^\chi s_{\rm ew}H_{\rm ew}}e^{i\pi/6}\frac{\Gamma(4/3)}{\beta_{IJ}^{1/3}},
\ee
 where we have selected mass orderings such that $\beta_{IJ}>0$ to simplify the phases. This gives us
\be
\Ychi &\approx& 2.1\frac{(\alpha_\chi^0)^3 T_{\rm ew}^{12}Y^\chi_{\rm eq}\mu_{\rm osc}^2\,\mathrm{Im}\left[(F^\dagger F)_{12}(F^\dagger F)_{23}(F^\dagger F)_{31}\right]}{(Y^\chi_{\rm eq} s_{\rm ew} H_{\rm ew})^3(\Delta M_{21}^2\Delta M_{32}^2\Delta M_{31}^2)^{1/3}}.
\nonumber
\ee
Beyond $\mathcal{O}(1)$ factors, the relative factor of $\alpha_\chi^0/\alpha_\chi^{\text w}$, and the fact that this contribution relies on a distinct combination of Yukawa couplings from the ARS asymmetry in Eq.~\eqref{eq:ARS_final}, we find that the new contribution suffers from a $(\mu_{\rm osc}^2/\Delta M^2)^{1/3}$ suppression\footnote{This is a suppression because $\mu_{\rm osc}^2 < \Delta M^2$; otherwise, the assumption of many oscillations prior to the electroweak phase transition is not satisfied and these results do not hold.} relative to the ARS contribution; otherwise, the asymmetries from the two terms are comparable.

\section{Signals of Freeze-In Baryogenesis}
\label{sec:signals}

\subsection{Collider Signatures}

The models proposed in this paper have very specific phenomenological signatures that follow naturally from the observed baryon asymmetry. They arise from the existence of one or more QCD-triplet scalars, $\Phi$, with lifetimes often governed by
\be
\Gamma_\Phi \lesssim H_{\rm ew} \sim \mathrm{cm}^{-1}.
\ee
Connections between the baryon asymmetry, particle lifetime and the Hubble expansion rate at the electroweak scale arise in other models as well \cite{Cui:2012jh,Barry:2013nva,Cui:2014twa,Ipek:2016bpf}, but in our case the connection between the decay rate of the scalar and $H_{\rm ew}$ is particularly direct.

Particles which travel macroscopic distances before decaying give rise to spectacular signatures at colliders. Because the only truly long-lived particles (LLPs) in the SM have masses $\lesssim5$ GeV, the decay of a TeV-scale LLP has no irreducible backgrounds. However, such decays may not be reconstructed using standard algorithms, and the backgrounds are challenging to characterize. Searching for  LLPs has therefore been identified as a primary opportunity for the discovery of new particles, and a large community of theorists and LHC experimentalists are working on new ways of looking for LLPs \cite{Alimena:2019zri} \footnote{For other reviews of theoretical motivations for LLPs and existing experimental searches, see Refs.~\cite{Curtin:2018mvb,Lee:2018pag}.}. 

Several earlier studies have noted that freeze-in DM models can give rise to LLP signatures \cite{Hall:2009bx,Co:2015pka,Hessler:2016kwm,Ghosh:2017vhe,Brooijmans:2018xbu,Calibbi:2018fqf,Belanger:2018sti,Chakraborti:2019ohe,Barman:2019lvm,No:2019gvl,Aboubrahim:2019kpb,Bae:2020dwf}. In particular, Ref.~\cite{Belanger:2018sti} has done a careful study of models that are  accessible at the LHC. However, general freeze-in DM models  do not necessarily predict states that are accessible  at colliders:~the BSM particles may be very heavy while still giving rise to the observed DM abundance. By contrast, when we require both DM and baryogenesis in a freeze-in model, the parameter space shrinks considerably in both mass and lifetime:~very long lifetimes yield an insufficient abundance of baryons, while very short lifetimes lead to excessive washout and over-production of DM. Furthermore, $\Phi$ masses well above the TeV scale suppress the baryon asymmetry, and so our model largely predicts new scalars that are accessible at current or future colliders and with lifetimes in the 1--100 cm range.

The primary prediction of our model is the existence of one or more scalars $\Phi$, which carry QCD charge and have proper lifetimes ranging from promptly decaying to 10 meters. These particles subsequently decay to a SM quark and an invisible $\chi$ state. This leads to several distinct signatures depending on the decay location, including:
\bi

\item One or more heavy quasi-stable charged particles resulting from $\Phi$ being bound inside a hadronic final state prior to its decay. These states leave tracks with unusual ionization or timing properties that can be distinguished from SM particle tracks;
\item One or two displaced hadronic vertices or jets,  accompanied by missing transverse momentum; 

\item A pair of prompt jets plus missing transverse momentum, in the case where the $\Phi$ decay occurs sufficiently rapidly that its decay point cannot be reliably distinguished from the interaction point.
\ei
Top quarks may be produced in $\Phi$ decays, in which case the signatures only become more striking, with a sizable fraction of events having final-state leptons. In the single-scalar scenario of Sec.~\ref{sec:top_thermal_mass}, for example, $\Phi$ decays would yield a mixture of light quarks and tops.  Below we focus on light-quark signatures, under the assumption that top couplings lead to even stronger constraints.  

\subsubsection*{Heavy Stable Charged Particles}

The most relevant search for heavy stable charged particles (HSCPs) is from ATLAS \cite{Aaboud:2018hdl}. This search uses $36\,\,\mathrm{fb}^{-1}$ of data at 13 TeV. The analysis makes use of the distinctive ionization signature in the inner tracker of slow-moving, massive LLPs; as a result, this search is sensitive to shorter-lifetime LLPs than other HSCP searches. The analysis also provides limits on the scalar mass as a function of $c\tau$. In the case of a scalar with the same charges as $u_{\rm R}$, the search excludes LLPs with $c\tau\gtrsim10$ cm, with the most stringent constraint of 1375 GeV for $\Phi$  that traverses the entire detector.

\subsubsection*{Displaced and Delayed Jets}

There are many different searches for high-mass particles decaying to displaced and delayed jets targeting decays in different parts of the detector and in different kinematic regimes \cite{Aaboud:2017iio,Aaboud:2018aqj,Sirunyan:2018vlw,Aaboud:2019opc,Sirunyan:2019gut}. Most relevant for us are searches most sensitive to LLP lifetimes $c\tau\lesssim1$ m, since these are the parameters that are largely uncovered by the HSCP searches. 

There are two powerful searches that are readily reinterpreted for our model. The first is a search by CMS for delayed jets \cite{Sirunyan:2019gut} with $137\,\,\mathrm{fb}^{-1}$ of data at 13 TeV. This search is most sensitive to heavy LLPs with $c\tau\sim$ 0.1--1 m, since the propagation of the slow LLP over an appreciable time leads to a significant delay for the resulting jets \cite{Liu:2018wte}. The CMS search includes limits on a benchmark model with LLPs that decay to a gluon plus an invisible particle; this is very similar to our model where the LLP decays to a quark plus an invisible particle. We assume there is no appreciable change in the signal efficiency for the quark scenario, and interpret their cross-section limits as a function of LLP mass and lifetime in terms of our signal. The best constraints are for $M_\Phi\lesssim1.6$ TeV for lifetimes of 20 cm, and we truncate the sensitivity at $M_\Phi=1$ TeV since that is as low as Ref.~\cite{Sirunyan:2019gut}  goes in their search.

The second search is a CMS search for displaced jets \cite{Aaboud:2018aqj} based on $35.9\,\,\mathrm{fb}^{-1}$ of data at 13 TeV. This search relies on a  trigger requiring at least two displaced jets, meaning jets that contain less than three prompt tracks and at least one displaced track. Displaced tracks associated with each jet pair in the event are used to construct secondary vertices, which must have a track mass larger than 4 GeV. While the parton produced in $\Phi$ decays is massless, QCD gives a mass associated with the resulting jets, allowing for this selection to be passed.
Importantly, the vertex reconstruction does not require tracks from both jets in a pair to be assigned to the vertex;
as a result, displaced jets arising from separate decays (in our case, from the two $\Phi\rightarrow j\chi$ decays) can still pass the selections, which gives sensitivity to our model. Ref.~\cite{Aaboud:2018aqj} gives cross-section limits for a model where the LLP decays to a gluon and a massless, invisible particle. This is not exactly the same as our model, which gives a quark in the LLP decay, and so we suspect there may be a slightly lower sensitivity to the quark model because the jet mass is smaller. Nevertheless, we expect comparable limits for the two cases, and in the absence of more information for reinterpretation we assume the limits are the same for gluon and quark decays for the purpose of our analysis. CMS presents limits on two LLP masses:~950 GeV and 2400 GeV. The 950 GeV is most appropriate for our model, and so we re-interpret the cross section limits in terms of limits on $M_\Phi$ as a function of $c\tau$. The strongest limit is for $c\tau\sim2$ cm, with constraints on $M_\Phi\gtrsim1.6$ TeV, although there exist constraints on $M_\Phi\gtrsim0.8$ TeV for lifetimes ranging from 0.1--1000 cm.

Other searches, including an ATLAS displaced vertex search \cite{Aaboud:2017iio}, are expected to yield comparable results. ATLAS does not provide an interpretation in terms of a jet plus missing momentum LLP decay, and so it is more involved to re-interpret that search; furthermore, given that the cross section limits in Ref.~\cite{Aaboud:2017iio} are comparable to those from the searches we have used, we expect the results to be qualitatively similar.

\subsubsection*{Prompt Jets and $\slashed{E}_{\rm T}$}

Finally, in the short lifetime limit there exist stringent constraints on $\Phi\rightarrow j\chi$ from searches for jets and missing transverse momentum. The most stringent constraint comes from searches for squarks:~if $\Phi$ decays predominantly to light-flavor quarks, then $M_\Phi\gtrsim1.13$ TeV, while the constraints are slightly stronger if it decays to tops ($M_\Phi\gtrsim1.175$ TeV) or bottoms ($M_\Phi\gtrsim1.25$ TeV) \cite{Sirunyan:2019ctn}. Strictly speaking, these limits only apply in the limit of prompt decays ($c\tau\lesssim10\,\,\mu\mathrm{m}$). If $\Phi$ has a longer lifetime, the sensitivity is expected to degrade, but provided it decays well before the calorimeter the jets should still be reconstructed. 

Recently, there have been more efforts to re-interpret prompt searches in terms of LLP models in order to determine precisely at what lifetimes prompt searches fail, and to identify any possible gaps between prompt and long-lived searches \cite{Alimena:2019zri,ATLAS-CONF-2014-037,ATLAS-CONF-2018-003,Sirunyan:2018ryt,Sirunyan:2018vjp,Aaboud:2018iil}. None of these studies are directly applicable to the $\Phi\rightarrow j+\slashed{E}_{\rm T}$ signature in our model; however, several re-interpret prompt searches for gluinos decaying to $2j+\slashed{E}_{\rm T}$. Since we expect the lifetime dependence of the jet reconstruction efficiency to be roughly independent of the number of jets in the final state, we use the results of Ref.~\cite{Sirunyan:2018vjp} to derive a ratio between the excluded prompt cross section and the excluded cross section at a finite lifetime $c\tau$. We then assume this ratio is the same for our signature, and use this to re-interpret the prompt squark limits of Ref.~\cite{Sirunyan:2019ctn} for finite lifetimes. While this is only an approximate procedure, we expect that it gives the correct  qualitative behavior of the limits for $\Phi\rightarrow q\chi$ decays.

\subsubsection*{Summary of Collider Constraints and Prospects}

\begin{figure}
   \centering
          \includegraphics[width=3.4in]{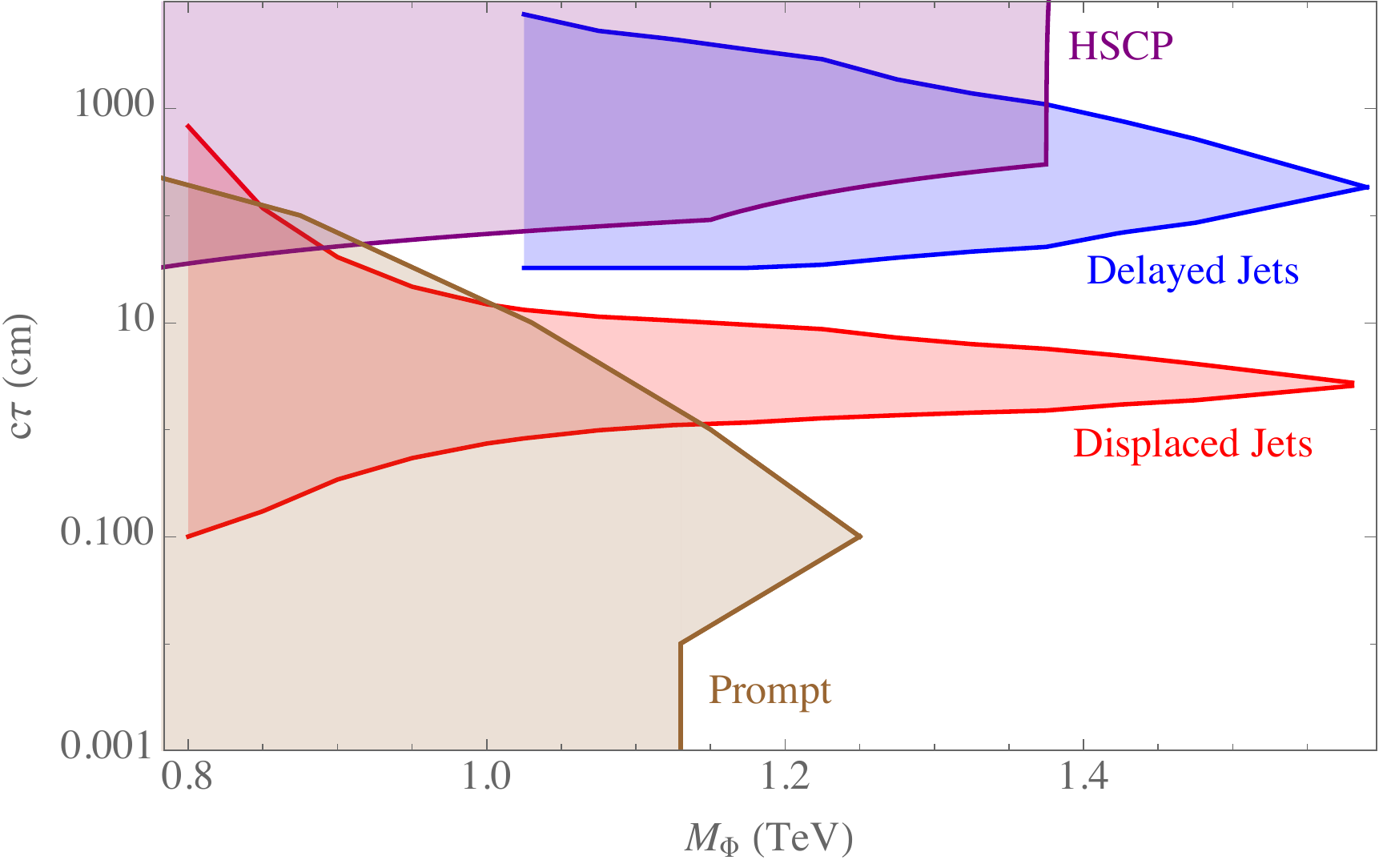}
   \caption{
   Summary of existing collider constraints from (purple) heavy stable charged particle searches; (blue) delayed jet searches; (red) displaced jet searches; (brown) searches for prompt decays to jets and $\slashed{E}_{\rm T}$. Details are provided in the text.
   }
   \label{fig:collider_constraints}
\end{figure}

We summarize the existing collider constraints in Fig.~\ref{fig:collider_constraints}. It is evident that nearly all the parameter space with $M_\Phi\le1$ TeV is ruled out, with the possible exception of a small sliver around $c\tau=10$ cm. However, it is likely that the delayed jet search has some sensitivity below 1 TeV, which would close most of the sliver. It is evident that a combination of prompt and long-lived searches currently gives excellent sensitivity to the freeze-in baryogenesis model with a new QCD-charged scalar. The search of Ref.~\cite{Sirunyan:2019ctn}  is new and, as understanding of the detectors improves, we expect sensitivity could get even better, allowing excellent prospects for discovery. At $\sqrt{s}=14$ TeV, the high-luminosity phase of the LHC should have more than 10 signal events for $M_\Phi\lesssim2.5$ TeV, and this is the upper mass limit of possible sensitivity at the LHC for high-efficiency, low-background searches.

While our reinterpretations of existing searches show good sensitivity to $\Phi\rightarrow q\chi$, most of the existing searches do not directly give results in terms of our simplified model. It may be true, for example, that the smaller mass of the quark-initiated jet could reduce sensitivity relative to a model with gluons. We therefore suggest that the experimental collaborations explicitly include a quark $+\slashed{E}_{\rm T}$ model in their LLP studies, since it is theoretically well motivated and it may be that variants of the existing search strategies could be used to improve  signal efficiency for the quark model. It would also facilitate reinterpretation and give a more accurate understanding of how much of the model space is covered by current and planned collider searches.

\subsection{$Z_2$-violating signals}
\label{sec:z2violate}

So far, we have assumed that the only coupling of $\Phi$ to the SM is via the operator(s) in 
Eq.~\eqref{eq:ARS_model}. 
This is true if there exists a $Z_2$ symmetry under which $\Phi$ and $\chi$ are charged and the SM fields are uncharged. However, one can also imagine a scenario 
without such a symmetry:~in this case, baryon-number-violating terms such as 
\be
\lambda_{ij}\,\Phi^* d^c_i d^c_j+\mathrm{h.c.}
\ee
are allowed, depending on the $\Phi$ quantum numbers (here we have switched to Weyl-spinor notation). Unless the $\lambda_{ij}$ couplings are tiny (that is, not much larger than the $F_{I\alpha}$ couplings responsible for the baryon asymmetry and DM production), this term would lead to rapid $\Phi\rightarrow j j$ decay, leaving prompt signatures in colliders. Because the term violates baryon number, it would also potentially lead to a larger asymmetry that does not depend on spectator effects in the manner of the $Z_2$-symmetric models.

However, such operators   lead to proton decay via off-shell $\Phi$-mediated processes, such as $p\rightarrow \pi^+\chi$. This leads to extremely strong constraints on $\lambda$:~indeed, we have checked that even if $\Phi$ couples exclusively to heavy-flavor quarks at tree-level, there are couplings to light-flavor quarks induced at loop level that violate proton decay constraints unless $|\lambda|\ll |F|$. 
Therefore, our preliminary investigation finds that $Z_2$-violating couplings of $\Phi$ to quarks are constrained to be so small that, if present, they are unlikely to dramatically alter the phenomenology.  

Another potentially important $Z_2$-violating term is the neutrino-portal coupling
\be
\mathcal{L} \supset y \bar{L} H \chi + \mathrm{h.c.}
\ee
This coupling leads to decays $\chi\rightarrow 3\nu$ and, more importantly, $\chi\rightarrow \gamma\nu$, giving rise to X-ray line signatures with  $E_\gamma = M_\chi/2$ if $\chi$ is the DM. This could, for example, explain a possible feature in X-ray data at $E_\gamma \simeq 3.5$ keV (first noted in Refs.~\cite{Bulbul:2014sua,Boyarsky:2014jta}), although there is conflicting evidence (or lack thereof) for the existence of this line in different galaxies and clusters. The coupling $y$ could easily be large enough to account for any X-ray lines that are observed, while being small enough to not otherwise disrupt how our mechanism works. In particular, since $\chi$ is produced at temperatures well above the electroweak scale, it is produced colder than conventional sterile neutrinos via the Dodelson-Widrow mechanism, although a 3.5 keV X-ray line would still be in tension with structure formation constraints that require $M_\chi\gtrsim10$ keV \cite{Heeck:2017xbu}.

\section{Conclusions}
\label{sec:conclusions}

Early-universe oscillations of DM particles, $\chi$, may have played a central role in generating the baryon asymmetry.  In this paper we studied models
in which these oscillations lead to asymmetric rates for $\bar\chi q \rightarrow \Phi$ and $ \chi \bar q \rightarrow \Phi^*$, where $\Phi$ is a 
a QCD-triplet scalar. 
Exploration of the phenomenology for different  BSM-particle spins and SM charges is work in progress.  Together, these various scenarios constitute a rich array of testable low-scale baryogenesis models, which simultaneously explain the DM and baryon abundances and generically predict new long-lived states at colliders.

We considered separately the minimal case, with a single $\chi$ interaction term, and scenarios in which there are \emph{multiple, distinct} ways of producing and annihilating $\chi$ particles.  
The presence of multiple channels tends to greatly enhance the baryon asymmetry.   For concreteness, we demonstrated this enhancement in a model with two QCD-charged scalars, both with couplings to $\chi$.  Alternatively, a primordial out-of-equilibrium abundance of $\chi$ from inflation or dynamics in the very early Universe is sufficient to realize the enhancement.

Along with sub-MeV $\chi$ masses, viable parameter points for DM and baryogenesis typically have  $M_\Phi \sim1-$few TeV, and $c \tau_\Phi \gsim 1$ cm, leading to striking signatures at colliders.  
The DM constraint pushes us into the weak-washout regime, where the asymmetry calculation is analytically tractable and physically transparent.
Independent of DM considerations, the baryon asymmetry in the weak-washout regime is strongly suppressed   for $\Phi$ lifetimes much less than the Hubble time at sphaleron decoupling, because it depends on the $\Phi/\Phi^*$ asymmetry at that time.  This provides a concrete link between cosmological time scales and  long-lived particle searches at colliders.  
\acknowledgments
We are grateful to Marco Drewes, Jason Gallicchio, and Bill Wootters for helpful conversations about decoherence. We thank the organizers and participants of the ``Testing $CP$-Violation for Baryogenesis'' workshop at the Amherst Center for Fundamental Interactions, University of Massachusetts for the stimulating environment in which this work was initiated. BS completed part of this work at the the Kavli Institute for Theoretical Physics (supported by the U.S.~National Science Foundation under Grant PHY-1748958).  DTS completed part of this work at the Aspen Center for Physics, which is supported by National Science Foundation grant PHY-1607611. 
The work of BS is supported by the U.S.~National Science Foundation under Grant PHY-1820770. 
\appendix
\section{Comparison of calculational schemes for the two-scalar model }
\label{sec:app_compare_calcs}
In this appendix we compare various methods of calculating $\YB$ and $\rhotot$ with the  
simplified perturbative calculation from Sec.~\ref{sec:calculation}. 
We first consider two modified perturbative calculations, one  
that includes thermal masses (Sec.~\ref{sec:thermal_mass_appendix}), and a second  that adopts a thermal ansatz for the $\chi$ momentum distribution (Sec.~\ref{sec:thermal_ansatz_appendix}).   We then use that thermal ansatz to go beyond the perturbative framework (Sec.~\ref{sec:kinetic_two_scalar}).  We write down and numerically solve an appropriate system of kinetic equations that incorporates back-reaction and washout effects, thermal masses,  and quantum statistics. We  find that the discrepancies when compared  with the ``minimal'' $\YB$ and $\rhotot$ calculations of  Sec.~\ref{sec:calculation} are typically smaller than $\sim 50 \%$, corresponding to modest differences in the viable parameter regions for baryogenesis and DM.

\subsection{Thermal mass effects}
\label{sec:thermal_mass_appendix}
We approximate quark thermal mass contributions based on the finite-temperature quark dispersion relation in the high-momentum regime.  
Using bars where thermal effects are included,   we  therefore take
\begin{eqnarray}
{\overline M}_{\Phi_i}^2&  = & M_{\Phi_i}^2 + (A_{g}+A^i_\text{self} ) T^2, \label{eqn:thermalphimass}\\
{\overline M}_{Q}^2&  = & A_{g}  T^2, \label{eqn:thermalQmass0}
\end{eqnarray}
where the gauge contributions, identical for $\Phi_i$ and $Q$, are given by~\cite{Weldon:1982bn}
\begin{eqnarray}
Q = Q_L: \quad\quad A_g & = & \frac{1}{3} g_3^2 + \frac{3}{16} g_2^2 + \frac{1}{144} g_1^2,
 \label{eqn:thermalQmass1}
 \\
Q = u_R: \quad\quad A_g & = & \frac{1}{3} g_3^2 + \frac{1}{9} g_1^2,
 \label{eqn:thermalQmass2}
 \\
Q= d_R: \quad\quad A_g & = & \frac{1}{3} g_3^2+ \frac{1}{36} g_1^2.
\label{eqn:thermalQmass3}
\end{eqnarray}
In our analysis of two-scalar models, we neglect contributions to  ${\overline M}_{Q}$ from SM Yukawa couplings,
 leaving a more careful treatment of the top quark for future work.  The coefficient $A^i_\text{self}$  allows us to consider the effects of extra contributions to ${\overline M}_{\Phi_i}$ coming from scalar self-interactions. 

\begin{figure*}
   \centering
            \includegraphics[width=3.3in]{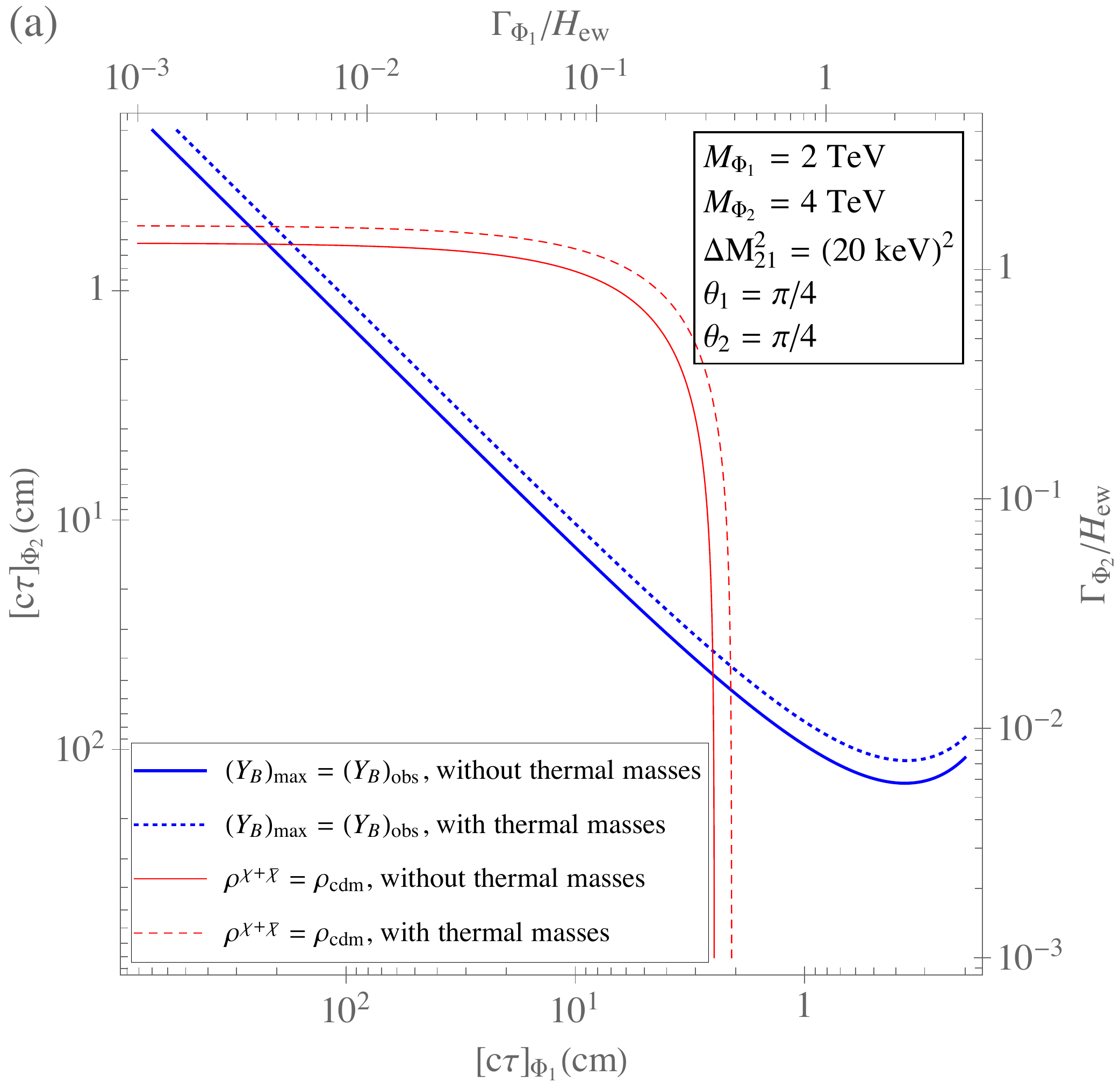}
\quad\quad\quad
          \includegraphics[width=3.3in]{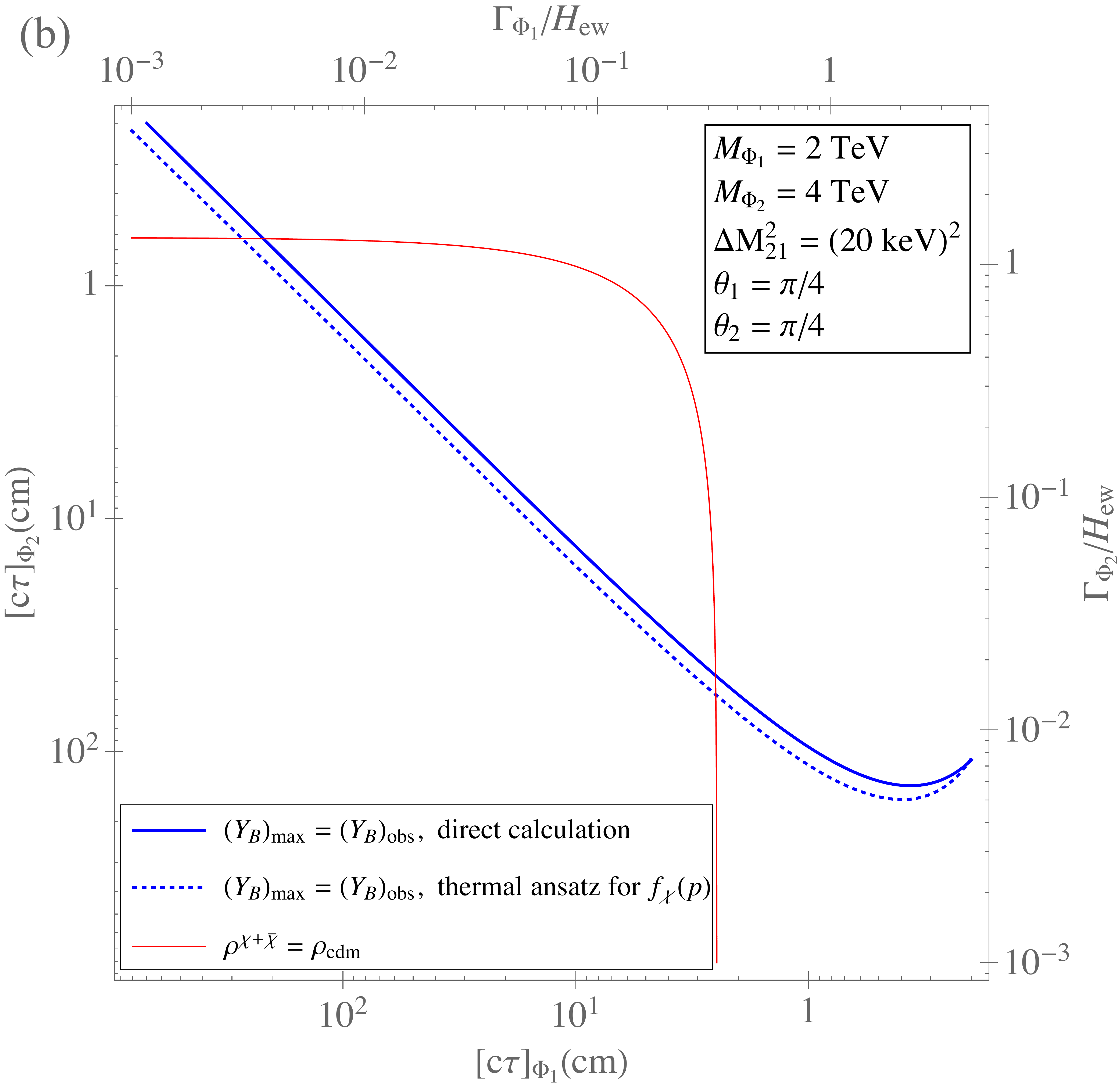}
   \caption{Variations on our perturbative calculations of $\YB$ and  $\rhotot$.  
 In both plots, we adopt Maxwell-Boltzmann statistics and replace the survival function by its $z=0$ value:  $S_{\Phi_i}(z) \rightarrow S_{\Phi_i}(0)$.  The solid contours reproduce our earlier results, see Fig.~\ref{fig:tautauplot_1}(a).  In (a), The dashed contours show the effect of including thermal mass contributions, with $A_\text{self} = 0$.  (Turning on a moderate scalar self-coupling of  $A_\text{self} = 1/3$ barely changes the plot.)  In (b) we adopt a thermal ansatz for the $\chi$ momentum distribution, leading to Eqs.~(\ref{eqn:YBtwoPhi_ansatz_1}) and (\ref{eqn:YBtwoPhi_ansatz_2}).   We use the Maxwell-Boltzmann expressions for $\langle T/E_\chi \rangle$ and $n^\chi_\text{eq}$ in Eqs.~(\ref{eqn:y_rep}) and (\ref{eqn:n_rep}); switching to the Fermi-Dirac ones again produces an almost unnoticeable shift.}
   \label{fig:tautau_appendix}
   \end{figure*}

We incorporate thermal mass effects in our $\YB$ and $\rhotot$ calculations with the help of the dimensionless functions
\be
\tau_i(z) = \frac{\overline{M}_{\Phi_i} (z)}{M_{\Phi_i}},
\ee
and
\be
\rho_i(z) = 1-\frac{\overline{M}_{Q}^2 (z)}{\overline{M}_{\Phi_i} ^2(z)}.
\ee
Neglecting thermal masses amounts to taking $\tau_i \rightarrow 1$ and, given that we neglect  Yukawa contributions to $Q$ masses, $\rho_i \rightarrow 1$.  For the $\Phi_i$ decay widths, these definitions imply
\be
{\overline \Gamma}_{\Phi_i} (z)= \tau_i(z) \rho_i^2(z) \Gamma_{\Phi_i}.
\ee
Using these functions, the final baryon asymmetry can still be represented by  
Eq.~(\ref{eqn:final_YB}), 
except with a modified expression for $I_{ij}$:
\begin{multline}
I_{ij}= 
 \int_0^\infty \!\!\!\! dy \;\frac{e^{-y}}{y^2} 
  \int_0^1 \!\!\! dz\;  S_{\Phi_i}(z)\;z^2 \;
  \tau_i^2(z)
  \;\rho_i(z)
    \;  e^{ - \frac{1-\rho_i(z)}{\rho_i(z)}y  }\\
    \times      e^{-\alpha_i \frac{z^2}{y}   \tau^2_i(z)\rho_i(z)}
   \int_0^{z} \!\!\!dz'\;{z'}^2  
     \tau_j^2(z')\rho_j(z')
      \;  e^{ - \frac{1-\rho_j(z')}{\rho_j(z')}y  }\\
      \times 
        e^{-\alpha_j \frac{z'^2}{y}   \tau^2_j(z')\rho_j(z') }\;
    \sin\left[ \beta_\text{osc} \left( \frac{z^3-z'^3}{y} \right)\right],
     \label{eqn:YB_WTM}
    \end{multline}
   and  with the survival function now given as
\begin{multline}
S_{\Phi_i} (z) =\exp
\Bigg\{
-\frac{\Gamma_{\Phi_i}}{H_\text{ew}}  
\int_{z}^{1} \!\! dz' z'
\tau_i(z') 
\rho_i^2(z')  \\
\times  \frac{\mathcal{K}_1\left( \frac{M_{\Phi_i }}{T_\text{ew}}  \tau_i(z')  z'\right)}{\mathcal{K}_2 \left( \frac{M_{\Phi_i }}{T_\text{ew}} \tau_i(z')  z'\right)}
 \Bigg\}.
 \label{eqn:thermal_survival}
    \end{multline}
We choose to preserve the definition $\alpha_i = (M_{\Phi_i}/2T_{\rm ew})^2$,  which involves the zero-temperature scalar masses.  
The summed abundance of $\chi$ and ${\overline \chi}$ particles from $\Phi^{(*)}_i$ decays, given by Eq.~(\ref{eqn:chi_Y_from_Phi_decay}) in the absence of thermal masses, becomes
\begin{multline}
Y^{\chi +\overline{\chi}}_{i}
 = 
  \frac{45 g_\Phi}{2 \pi^4 g_*} 
\left( 
 \frac{\Gamma_{\Phi_i}}{H_\text{ew}}
 \right)
    \left( \frac{M_{\Phi_i}}{T_\text{ew}}\right)^{2}  \\
    \times
 \int_{0}^\infty \!\! dz \;z^3 \tau^3_i(z) \rho^2_i(z)\,\mathcal{K}_1\left( \frac{M_{\Phi_i }}{T_\text{ew}}  \tau_i(z)  z \right).
\label{eqn:chi_Y_from_Phi_decay_WTM}
\end{multline}
For a particular choice of inputs, Fig.~\ref{fig:tautau_appendix}(a) shows
that including thermal masses slightly shifts the parameter space  allowed by the baryon asymmetry and DM constraints.   The $\rhotot$ and  $\YB$ contours move together somewhat, and in fact the range of viable $\Phi_1$ masses is not significantly affected when we include thermal masses.

\begin{figure*}
   \centering
             \includegraphics[width=3.3in]{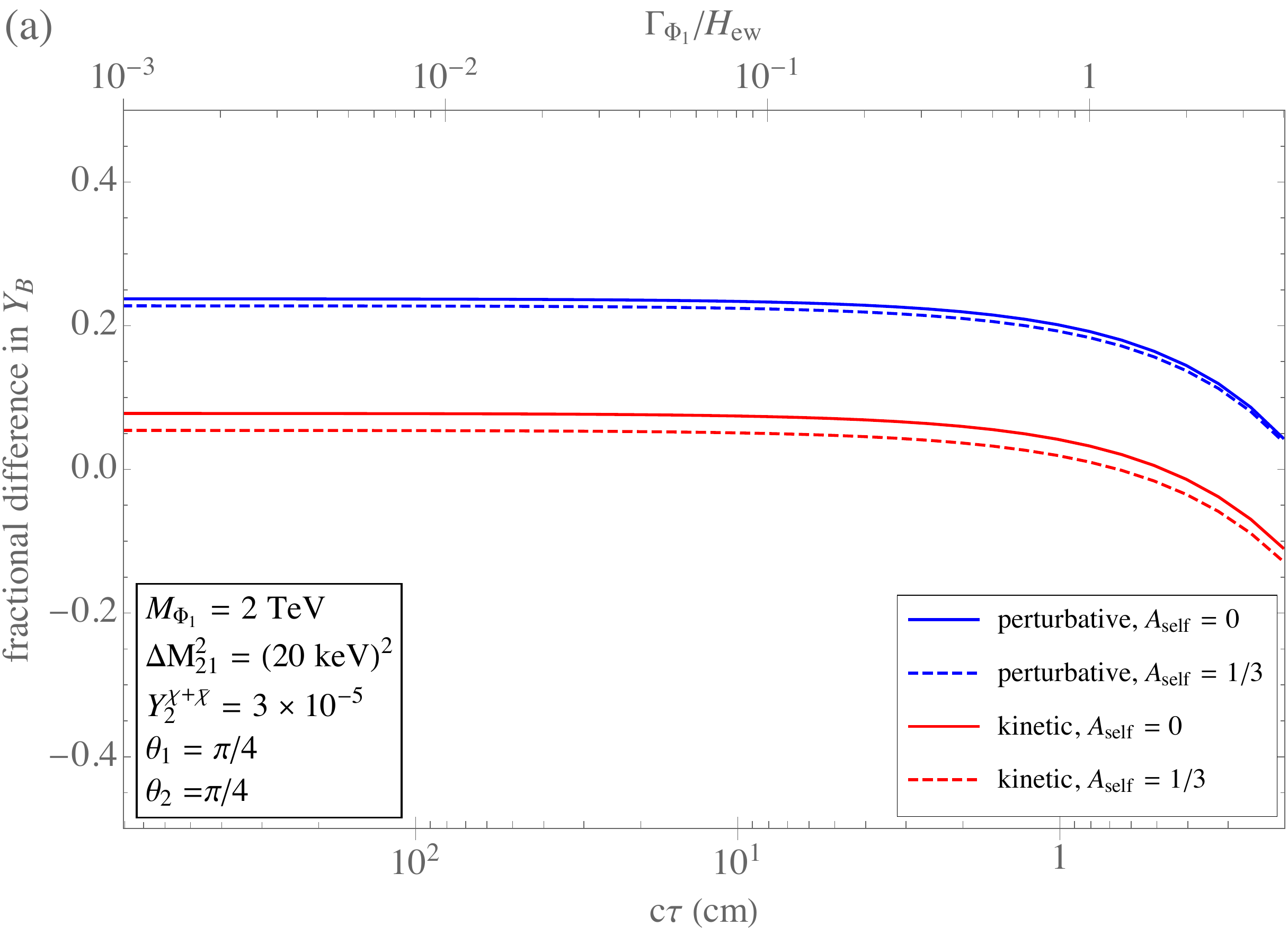}
\quad\quad
          \includegraphics[width=3.3in]{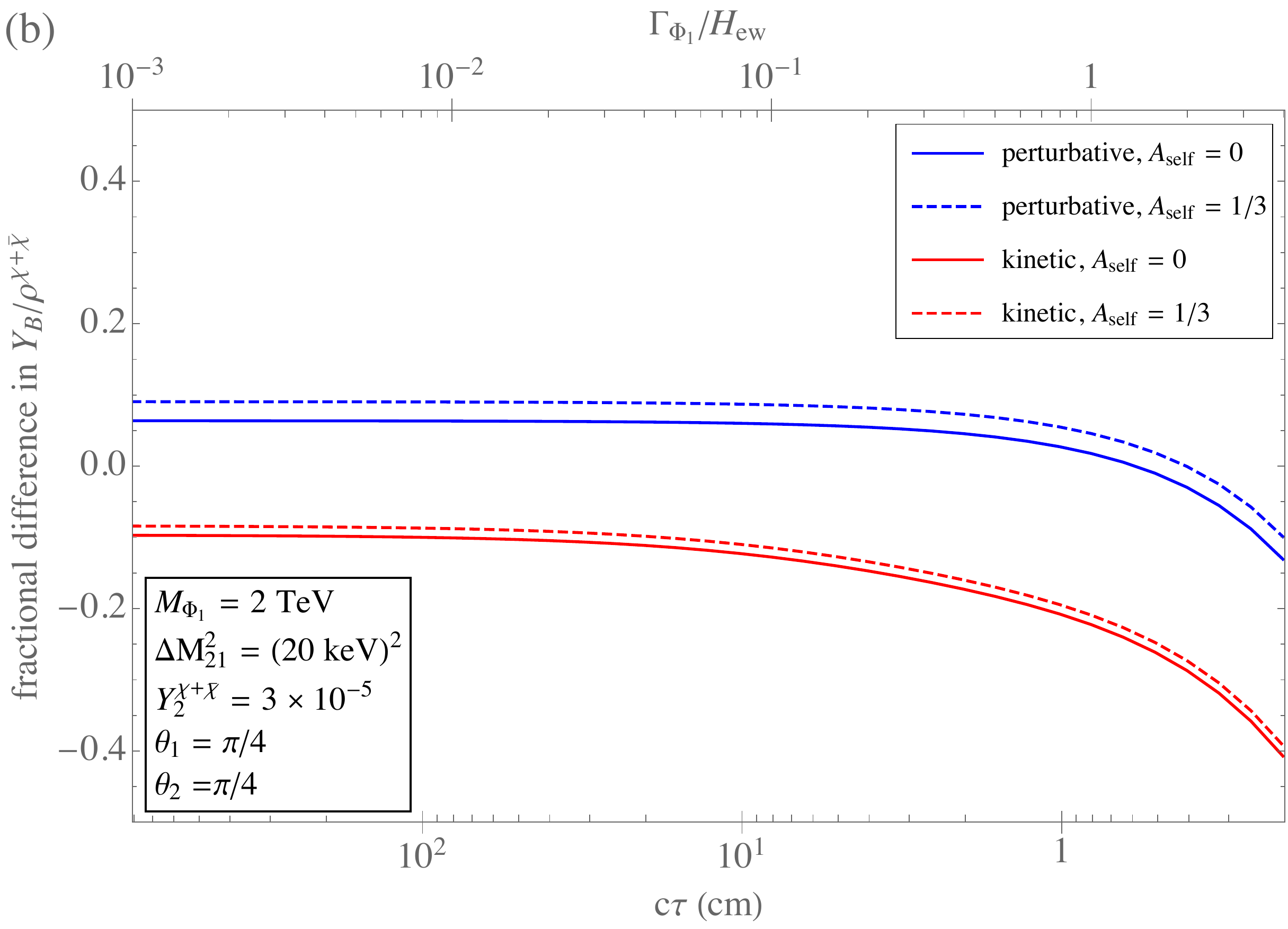}
   \caption{
Comparison of the ``mimimal'' perturbative calculation of Sec.~\ref{sec:calculation} with more refined estimates of $\YB$ and $\rhotot$.  
Here we decouple $\Phi_2$ and set the combined $\chi+{\overline \chi}$ abundance left behind from $\Phi_2$ decays to be $Y_2^{\chi+{\overline \chi}} = 3 \times 10^{-5}$, and we  take ${\mathcal J} = 1$.  The blue contours show the effect of modifying the perturbative calculation to incorporate thermal masses and the full $z$ dependence in the survival function $S_{\Phi_1}(z)$.  The red contours compare the minimal calculation with numerical solution of the kinetic equations presented in Sec.~\ref{sec:kinetic_two_scalar}; in the numerical calculation we use quantum statistics, include thermal masses,  and adopt a thermal ansatz for the $\chi$ momentum distribution.   We show the fractional difference in $\YB$ in (a) and 
the fractional difference in $\YB/\rhotot$ in (b).  For the parameters chosen, $\rhotot = \rho_\text{cdm}$ is realized for $c\tau \simeq 6$~cm and  $(\YB)_\text{max} > (\YB)_\text{obs}$ is realized for $c\tau \lsim 130$~cm.  
   }
   \label{fig:full_survival_fig}
   \end{figure*}

To make the plots of Fig.~\ref{fig:full_survival_fig}, we decouple $\Phi_2$ and  choose a value for the $\chi$ abundance left behind from $\Phi_2$ decays. 
Using  $(\YB)_{{\overline M}\rightarrow M}$ to denote the asymmetry neglecting thermal masses  and taking $S_{\Phi_i}(z) \rightarrow S_{\Phi_i}(0)$, and 
using $(\YB)_{{\overline M}}$ to denote the asymmetry  with thermal masses and the full $z$-dependence in $S_{\Phi_1}(z)$,  the blue contours of Fig.~\ref{fig:full_survival_fig}(a) show the  fractional difference
\be
2 \times
\frac{(\YB)_{{\overline M} \rightarrow M} - (\YB)_{{\overline M}}}{(\YB)_{{\overline M} \rightarrow M} + (\YB)_{{\overline M}}}.
\ee
For the inputs chosen in Fig.~\ref{fig:full_survival_fig}, neglecting thermal masses overestimates $\YB$ by roughly $25\%$.  For smaller $\Phi_1$ lifetimes, $\Gamma_{\Phi_1}/H_\text{ew} \gsim 1$, the $S_{\Phi_1}(z) \rightarrow S_{\Phi_1}(0)$ approximation significantly overestimates washout of the asymmetry by $\Phi$ decays, partially compensating for the effect of neglecting thermal masses.  
Fig.~\ref{fig:full_survival_fig}(b) shows that the fractional difference is somewhat smaller for the ratio $\YB/\rhotot$ than for $\YB$ alone.  
In general, we find that the minimal  perturbative calculation of Sec~\ref{sec:calculation} typically agrees with a perturbative calculation incorporating thermal masses at the $50\%$ level or better, for those parameter regions that are viable for baryogenesis and DM.

\subsection{A thermal ansatz for the $\chi$ momentum distribution}
\label{sec:thermal_ansatz_appendix}
In Sec.~\ref{sec:kinetic_two_scalar}, we will compare our perturbative calculations of $\YB$ and $\rhotot$ to numerical solutions of quantum kinetic equations derived using a  thermal ansatz for the $\chi$ and ${\overline \chi}$ momentum distributions.   
Here we  implement the thermal ansatz in the perturbative context to isolate its impact.  We start with  Eq.~(\ref{eq:chi_abundance}) from our perturbative calculation, where we found the energy spectrum of $\chi$ particles produced by $\Phi_j$ decays at $z'$ to be proportional to 
\be
e^{-y} e^{-M_{\Phi_j}^2 z'/(4 T_\text{ew}^2 y)},
\ee
where $y= E_{\chi}/T$. We replace this spectrum with a Maxwellian one,
 \be
e^{-y}\; e^{-M_{\Phi_j}^2 z'/(4 T_\text{ew}^2 y)}  \rightarrow   \frac{M_{\Phi_j} z'}{2T_\text{ew}} 
  {\mathcal K}_1\! \left( \frac{M_{\Phi_j} z'}{T_\text{ew}}  \right)    y^2 \,e^{-y},
       \quad\quad\quad
  \label{eqn:MB_rep}
 \ee
 where the $y$-independent factors are determined by the requirement that the integrals over $y$ be the same.  We also replace the $\chi$ energy dependence in the oscillation factor with a thermal average,
 \be
   \sin\left[ \beta_\text{osc} \left( \frac{z^3-z'^3}{y} \right)\right] \rightarrow
     \sin\left[ \beta_\text{osc} \left\langle \frac{T}{E_\chi}\right\rangle  \left( z^3-z'^3 \right)\right],
     \quad\quad\;\;\;
     \label{eqn:osc_rep}
 \ee
 with 
 \be
  \left\langle \frac{T}{E_\chi}\right\rangle = 
 \begin{cases}
\frac{\pi^2}{ 18 \zeta(3)} \simeq 0.456 &  \text{(FD)},\\
 1/2 &\text{(MB)}.
 \end{cases}
\label{eqn:y_rep}
 \ee

After making the replacements in Eqs.~(\ref{eqn:MB_rep}--\ref{eqn:osc_rep}), we carry out the $y$ integration in Eq.~(\ref{eqn:final_I_fn}) to obtain
  \begin{multline}
\YB   = 
\frac{45 g_\Phi^2}{8\pi^6 g_*}
\frac{\mathcal{K}_B}{\mathcal{K}_\Phi}
\mathcal{J}
 \left( 
 \frac{n_\text{eq}^\chi (T)}{T^3}
 \right)^{-1}
  \left( \frac{M_{\Phi_1}}{T_\text{ew}}\right)^2
    \left( \frac{M_{\Phi_2}}{T_\text{ew}}\right)^2\\
    \times
 \left( 
 \frac{\Gamma_{\Phi_1}}{H_\text{ew}}
 \right)
  \left( 
 \frac{\Gamma_{\Phi_2}}{H_\text{ew}}
 \right)
\left(
\;
I_{12}
-
\;
 I_{21}
\right),
\label{eqn:YBtwoPhi_ansatz_1}
\end{multline}
where the equilibrium $\chi$ abundance factor
is for a single mass eigenstate and helicity,
  \be
 \left( 
 \frac{n_\text{eq}^\chi (T)}{T^3}
 \right)^{-1} = 
 \begin{cases}
(4 \pi^2)/(3 \zeta(3)) \simeq  10.9 &  \text{(FD)}\\
\pi^2 &\text{(MB)},\;\;\;
 \end{cases}
\label{eqn:n_rep}
 \ee
and where in this version of $I_{ij}$, only the integrations over  $\Phi$ production and decay times remain:
\begin{multline}
I_{ij} = 
  \int_0^1 \!dz \;S_{\Phi_i}(z)\; z^3   \; {\mathcal K}_1\! \left( \frac{M_{\Phi_i}}{T_\text{ew}} z \right) \\
\times   \int_0^{z} \!dz'\;{z'}^3  {\mathcal K}_1 \! \left( \frac{M_{\Phi_j}}{T_\text{ew}} z' \right)
   \sin\left[ \beta_\text{osc} \left\langle \frac{T}{E_\chi}\right\rangle  \left( z^3-z'^3 \right)\right],
   \label{eqn:YBtwoPhi_ansatz_2}
\end{multline}
with  $S_{\Phi_i}(z)$ given in Eq.~(\ref{eqn:survival}).  Strictly speaking, Eq.~(\ref{eqn:YBtwoPhi_ansatz_1}) applies for the case of Maxwell-Boltzmann statistics, but we provide Fermi-Dirac expressions for certain quantities for reference.
We can also obtain Eqs.~(\ref{eqn:YBtwoPhi_ansatz_1}) and (\ref{eqn:YBtwoPhi_ansatz_2}) by perturbatively solving the kinetic equations presented in the following section, once we neglect thermal masses and adopt Maxwell-Boltzmann statistics. 

For a particular set of inputs, Fig.~\ref{fig:tautau_appendix}(b) compares the results of the ``direct'' calculation of Sec.~\ref{sec:calculation} with those based on Eqs.~(\ref{eqn:YBtwoPhi_ansatz_1}) and (\ref{eqn:YBtwoPhi_ansatz_2}).  As with  thermal mass effects,  the thermal ansatz only modestly impacts the preferred parameter space.

\begin{figure}
\includegraphics[width=3in]{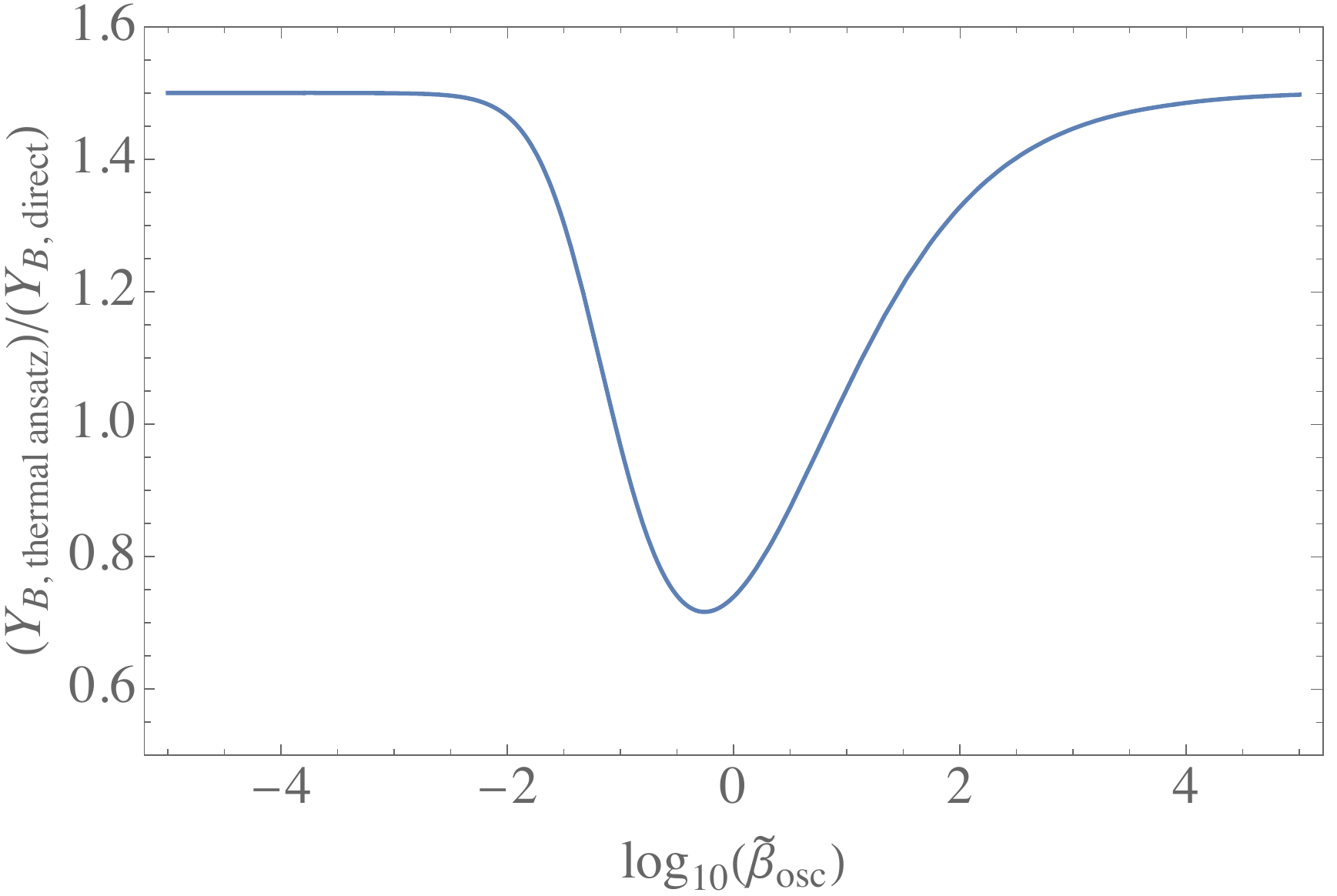}
\caption{Comparison of the approximate $Y_B$ expressions of Eqs.~(\ref{eqn:thermal_ansatz_simple}) and (\ref{eqn:YB_large_Mphi_alt}), obtained with and without the thermal ansatz for the $\chi$ momentum distribution.  
}
\label{fig:thermal_ansatz_direct_simple_compare}
\end{figure}

When
 \be
M_{\Phi_2}  \gg M_{\Phi_1} \gg T_{\rm ew}
\ee
and
\be
\frac{\Delta M_{21}^2 M_0}{M_{\Phi_2} ^3} \ll 1
 \ee
apply, we can follow the same steps that led to Eq.~(\ref{eqn:YB_large_Mphi_alt}) to approximate the thermal-ansatz results of Eqs.~(\ref{eqn:YBtwoPhi_ansatz_1}) and (\ref{eqn:YBtwoPhi_ansatz_2}) by
\begin{multline}
\YB   
\simeq
\frac{g_\Phi \mathcal{K}_B \mathcal{J}}{4\pi^2 \mathcal{K}_\Phi}
\Ytwotot
 \left( 
 \frac{n_\text{eq}^\chi (T)}{T^3}
 \right)^{-1}
  \left( \frac{T_\text{ew}}{M_{\Phi_1}}\right)^2 
   \left( 
 \frac{\Gamma_{\Phi_1}}{H_\text{ew}}
 \right)\\
  \times
 S_{\Phi_1}(0) 
  \int_0^{\infty} \!dx \;x^3  {\mathcal K}_1 \! (x)
   \sin\left[\frac{ \tilde{\beta}_\text{osc}}{8} \left\langle \frac{T}{E_\chi}\right\rangle  x^3\right].
   \label{eqn:thermal_ansatz_simple}
\end{multline}
The ratio of the  thermal-ansatz-based $\YB$ of Eq.~(\ref{eqn:thermal_ansatz_simple}) and the ``direct'' $\YB$  of Eq.~(\ref{eqn:YB_large_Mphi_alt})  depends on the single dimensionless parameter  $\tilde{\beta}_\text{osc} = 4  \Delta M_{21}^2 M_0/3 M_{\Phi_1}^3$.  Figure~\ref{fig:thermal_ansatz_direct_simple_compare} shows that the numerical discrepancy between the two expressions is less than $\sim 50\%$.  

\subsection{Numerical solution of kinetic equations}
\label{sec:kinetic_two_scalar}
To write down the kinetic equations for the two-scalar model we follow Refs.~\cite{Hambye:2017elz,Abada:2018oly}.  
We avoid having to track momentum-dependent quantities by adopting a thermal ansatz for the momentum dependence of the  $\chi$  number-density matrix,
 \be
 n_{IJ}^\chi (\bp) = \frac{n_{IJ}^\chi}{n^\chi_\text{eq}} f_\text{eq}^\chi (\bp),
 \ee
 and similarly  for ${\overline \chi}$.  Doing so leaves us with the simpler task of solving momentum-integrated kinetic equations. The $\chi$ and ${\overline \chi}$ number densities are given by the traces of  $n^\chi_{IJ}$ and $n^{\overline \chi}_{IJ}$, respectively.

 We neglect $\chi$ masses except in  oscillations. An unbroken $U(1)_{\chi - \Phi}$ symmetry under which only $\chi$ and $\Phi_i$ are charged (oppositely) then simplifies the analysis.   
 Here we are motivated by DM considerations, which lead us to focus on parameter regions with $M_{\chi} \sim 10 -100$ keV.  

Our main goal is to to compare our perturbative calculation from Section~(\ref{sec:calculation}) with numerical integration of kinetic equations that incorporate washout effects and quantum statistics.  
To that end we make further simplifying approximations that might be abandoned in future work.  
We adopt the flavor-universal  $Q$ masses given in Eqs.~(\ref{eqn:thermalQmass1}--\ref{eqn:thermalQmass3}), thereby ignoring top-Yukawa-related effects.   The  $1 \leftrightarrow 2$ processes $\Phi_i \leftrightarrow Q \chi$  are therefore kinematically allowed at all temperatures.  We neglect $2 \leftrightarrow 2$ processes, which we expect to be subdominant, as we found thermal-mass effects to be.   Finally, we  assume that quark flavor mixing is sufficiently rapid to ensure that the $Q$ chemical potentials are flavor universal.

With these assumptions, the reaction densities entering into the kinetic equations can be summarized by the matrix expression
\begin{equation}
\left[\gamma_{X,i}\right]_{IJ} 
 =  
 g_\Phi
 \!
 \left(  {F^i}^\dagger {F^i} \right)_{\!IJ}
 \!
\left( {\overline M}_{\Phi_i}^2  \!-\!  {\overline M}_{Q}^2  \right)
\!\!
\int\!\!
d\Pi_{\Phi_i}
\mathcal{F}^{\Phi}_{X, i} (\bk, \bp,\bq),
\label{eqn:RDs2}
\end{equation}
where the equilibrium distribution functions  enter through
\begin{eqnarray}
\mathcal{F}_{0,i}(\bk, \bp,\bq) 
=
\mathcal{F}_{\Phi1,i}(\bk, \bp,\bq) 
& = &
\left[1-f_\text{eq}^{Q} (\bp)  \right] f_\text{eq}^{\Phi_i} (\bq)
\nonumber
\\
\mathcal{F}_{Q1,i}(\bk, \bp,\bq) 
& = &
f_\text{eq}^{Q} (\bp)  f_\text{eq}^{\Phi_i} (\bq)
\nonumber
\\
\mathcal{F}_{Q2,i}(\bk, \bp,\bq) 
&=&
f_\text{eq}^{\chi} (\bk)f_\text{eq}^{Q} (\bp)  \nonumber
\\
\mathcal{F}_{\Phi2,i}(\bk, \bp,\bq) 
&=&
f_\text{eq}^{\chi} (\bk)f_\text{eq}^{\Phi_i} (\bq).
\label{eqn:distributionintegrandfac}
\end{eqnarray}
In Eq.~(\ref{eqn:RDs2}), the indices $i=1,2$ and $I,J = 1,2$ reference $\Phi$ and $\chi$ flavors, respectively,  while $X$ indicates whether the associated contributions to $d n^\chi/dt$ and $d n^{\overline \chi}/dt$ survive in the absence of  asymmetries (``0''), are driven by a $Q_\alpha - \overline{Q}_\alpha$ asymmetry (``$Q 1$'' and ``$Q 2$''), or are driven by  a $\Phi_i - \Phi_i^*$ asymmetry (``$\Phi 1$'' and ``$\Phi 2$''). 

The phase space factor in Eq.~(\ref{eqn:RDs2})  is
\begin{multline}
d\Pi_{\Phi_i}  = 
\frac{d^3 \bk}{(2\pi)^3}
\frac{1}{2E_\chi(\bk)}
 \frac{d^3 \bp}{(2\pi)^3}
 \frac{1}{2 E_Q(\bp)} \\
 \times
\frac{d^3 \bq}{(2\pi)^3}
\frac{1}{2 E_{\Phi_i}(\bq)}
(2\pi)^4 \delta^4(q-p-k),
\end{multline}
with $E_\chi (\bk) = |\bk|$, $E_Q (\bp) = \sqrt{|\bp|^2+\overline{M}_Q^2}$, and $E_{\Phi_i} (\bq) = \sqrt{|\bq|^2+\overline{M}_{\Phi_i}^2}$.  Carrying out all integrations besides those over $E_{\Phi_i}$ and $E_\chi$ gives
{
\begin{multline}
\left[\gamma_{X,i}\right]_{IJ} 
 =  \frac{
 g_\Phi
 \!
 \left(  {F^i}^\dagger {F^i} \right)_{\!IJ}
 \!
\left( {\overline M}_{\Phi_i}^2  \!-\!  {\overline M}_{Q}^2  \right)}
{32 \pi^3}\\
\times
\!\!
\int_{\overline{M}_{\Phi_i}}^\infty\!\!
dE_{\Phi_i}
\int_{E_\chi^-}^{E_\chi^+}\!\!
dE_{\chi}\;
\mathcal{F}^{\Phi}_{X, i} (\bk, \bp,\bq),
\label{eqn:RDs3}
\end{multline}
}
where 
\begin{equation}
E_\chi^\pm = \frac{\overline{M}_{\Phi_i}^2 - \overline{M}_Q^2}{2 \overline{M}_{\Phi_i} }
\left(
\frac{E_{\Phi_i}}{\overline{M}_{\Phi_i}}
\pm
\sqrt{
\left( \frac{E_{\Phi_i}}{\overline{M}_{\Phi_i}} \right)^2 -1
}
\right).
\;
\end{equation}
In Eq.~(\ref{eqn:RDs3}) it is to be understood that $\mathcal{F}^{\Phi}_{X, i} (\bk, \bp,\bq)$ depends on its arguments only through the associated energies, with $E_Q = E_{\Phi_i}-E_\chi$.  For example,
\be
\mathcal{F}^{\Phi}_{0, i} (\bk, \bp,\bq)
\rightarrow
\left(
\frac{e^{(E_{\Phi_i}-E_\chi)/T}}{e^{(E_{\Phi_i}-E_\chi)/T}+1}
\right)
\left(
\frac{1}{e^{E_{\Phi_i}/T}-1}
\right).\quad\quad\quad
\ee

We present the kinetic equations in terms of dimensionless interaction-picture quantities 
\be
Y^{\chi} =   \frac{ U^\dagger n^\chi U}{s}
\quad\quad
Y^{{\overline \chi}}(z) = \frac{U^\dagger n^{\overline \chi} U}{s}
\ee
and
\be\label{eq:gamma_tilde}
\tilde{\gamma} =\frac{U^\dagger {\gamma} U}{Y^\chi_\text{eq} s H}
\quad\quad
\tilde{\gamma}^* = \frac{U^\dagger {\gamma^*} U}{Y^\chi_\text{eq} s H},
\ee
where 
 $s=2\pi^2 g_* T^3/45$ is the entropy density, $H = T^2/M_0$ is the Hubble parameter,  $Y^{\chi}_\text{eq}$ is the equilibrium abundance for  a single mass eigenstate, and oscillation effects are encoded in the diagonal matrix
\be
U(z)_{IJ} &=& \exp\left[-i
\left\langle \frac{T}{E_\chi}\right\rangle
\frac{M_I^2 M_0}{6\, T_\text{ew}^3}
z^3
\right]\delta_{IJ}.
\label{eqn:U}
\ee
The thermal average $\langle T/E_\chi\rangle$ is given in Eq.~(\ref{eqn:y_rep}). Note that for $\tilde{\gamma}^* $ we complex-conjugate the couplings appearing in $\gamma$ but not the $U$ matrices.  
For the case of two $\chi$ mass eigenstates, we can apply an inconsequential overall phase to rewrite Eqn~(\ref{eqn:U}) as
\be
U(z) =  \text{diag}  \left\{
1, \exp\left\{-i \beta_\text{osc}  \left\langle \frac{T}{E_\chi}\right\rangle z^3\right\}  \right\},
\label{eqn:U2}
\ee
where we define $\beta_\text{osc} = \Delta M_{21}^2 M_0/(6 T_\text{ew}^3)$ as before.

For $\Phi_i$ and Q  we similarly define $Y = n/s$, along with the asymmetries
\be
\delta Y^Q & = &Y^Q - Y^{\overline Q} \\
\delta Y^{\Phi_i} & = & Y^{\Phi_i} - Y^{\Phi_i^*}.
\ee
We define these quantities and their equilibrium counterparts $Y^{\Phi_i}_\text{eq}$ and $Y^Q_\text{eq}$ to include a sum over gauge degrees of freedom (but not, for $Q$,  a sum over flavor degrees of freedom), while $Y^{\chi}_\text{eq}$ is the equilibrium abundance for  a single mass eigenstate (and a single helicity: $\chi$ or ${\overline \chi}$, not both). 

As is typically done, we linearize in $Q$ chemical potentials, 
\be
\frac{f_{}^{Q} (\bp,\pm\mu_Q) }{f_\text{eq}^{Q} (\bp) } \simeq 
 1 \pm 
\frac{\delta Y^{Q}}{2 Y_\text{eq}^{Q}}.
\label{eqn:Q}
\ee
However, a $\Phi_i-\Phi_i^*$  asymmetry can leave  $\Phi_i$ or $\Phi_i^*$ particles  around after  $\Phi_i -\Phi_i^*$ annihilations have effectively completed, in which case $\mu_{\Phi_i} \ll T$ is not satisfied and it is not appropriate to linearize in $\mu_{\Phi_i}$. 
We assume that ${\Phi_i} -\Phi_i^*$ annihilations keep these particles  in chemical equilibrium  even for temperatures $T \ll M_{\Phi_i}$.   

In the $T\ll M_{\Phi_i}$ regime Maxwell-Boltzmann statistics should apply, giving
\begin{equation}
\frac{f^{\Phi_i}(\bq, \pm \mu_{\Phi_i})}{ f_\text{eq}^{\Phi_i}(\bq)}  = e^{\pm \mu_{\Phi_i}/T }, 
\end{equation}
which leads to
\begin{equation}
\frac{f^{\Phi_i}(\bq, \pm \mu_{\Phi_i})}{ f_\text{eq}^{\Phi_i}(\bq)} 
=
 \sqrt{1+ \left( \frac{\delta f^{\Phi_i}(\bq) }{2  f_\text{eq}^{\Phi_i}(\bq)} \right)^2}
 \pm
 \frac{\delta f^{\Phi_i}(\bq) }{2  f_\text{eq}^{\Phi_i}(\bq)},\quad\quad
  \label{eqn:PhiMB}
\end{equation}
where $\delta f^{\Phi_i}(\bq) \equiv   f^{\Phi_i}(\bq,\mu_{\Phi_i})-f^{\Phi_i}(\bq,-\mu_{\Phi_i})$.
For the scenarios we study, Eqn.~(\ref{eqn:PhiMB}) is a good approximation even when Maxwell-Boltzmann statistics does {\em not} apply.  For $T \gtrsim M^{\Phi_i}$, it is safe to assume $\mu_{\Phi_i} \ll T$ and $\frac{\delta f^{\Phi_i}(\bq) }{2  f_\text{eq}^{\Phi_i}(\bq)}\ll 1$, and  Eq.~(\ref{eqn:PhiMB}) approximately reproduces what one gets by linearizing the full quantum-statistics distribution in $\mu_{\Phi_i}$, 
\be
f^{\Phi_i}(\bq, \pm \mu_{\Phi_i}) =  f_\text{eq}^{\Phi_i}(\bq) \pm \frac{\delta f^{\Phi_i}(\bq) }{2  }. 
\ee
Following our treatment of $f^{Q}$, we neglect momentum dependence  in the ratio $\frac{\delta f^{\Phi_i}(\bq) }{2  f_\text{eq}^{\Phi_i}(\bq)}$, giving
\be
\frac{f^{\Phi_i}(\bq,\pm \mu_{\Phi_i})}{f_\text{eq}^{\Phi_i}(\bq)} =  1+ \mathcal{G}\left(\pm\frac{\delta Y^{\Phi_i} }{2  Y_\text{eq}^{\Phi_i}}\right), 
 \label{eqn:fPhi}
\ee
where we define the function
\be
\mathcal{G}(x) = \sqrt{1+x^2} + x-1.
\ee

Having established our notation, we now give the kinetic equations describing the evolution of the $Y^{\chi}$ and $Y^{\overline \chi}$ matrices:
\begin{multline}
\frac{d\;Y^\chi_{IJ}}{d\ln z} 
=
\sum_i
\Bigg(-
\frac{1}{2}
\left\{
 {{\tilde \gamma}}^{}_{0,i}
\;
,
\; Y^\chi-Y^\chi_\text{eq}
\right\}
\\
+ \frac{\delta Y^{Q}}{2 Y^{Q}_\text{eq}}
\left[
 {\tilde \gamma}^{}_{Q1,i}
 Y^\chi_\text{eq}
+\frac{1}{2}
\left\{
{\tilde \gamma}^{}_{Q2,i}
\;
,Y^\chi
\right\}
\right]\\
+\mathcal{G} \left( -\frac{\delta Y^{\Phi_i}}{2 Y^{\Phi_i}_\text{eq}} \right)
\left[
{\tilde \gamma}^{}_{\Phi1,i}
Y^\chi_\text{eq}
-\frac{1}{2}
\left\{
{\tilde \gamma}^{}_{\Phi2,i}
\;
,Y^\chi
\right\}
\right]
\Bigg)_{IJ}
\label{eqn:chiY}
\end{multline}
and
\begin{multline}
\frac{d\;Y^{\overline \chi}_{IJ}}{d\ln z} 
=
\sum_i
\Bigg(-
\frac{1}{2}
\left\{
 {{{\tilde \gamma}}^{*}_{0,i}}
\;
,
\; Y^{\overline \chi}-Y^{\chi}_\text{eq}
\right\}
\\
- \frac{\delta Y^{Q}}{2 Y^{Q}_\text{eq}}
\left[
{{\tilde \gamma}^{*}_{Q1,i}}
Y^\chi_\text{eq}
+\frac{1}{2}
\left\{
{{\tilde \gamma}^{*}_{Q2,i}}
\;
,Y^{\overline \chi}
\right\}
\right]\\
+\mathcal{G} \left( \frac{\delta Y^{\Phi_i}}{2 Y^{\Phi_i}_\text{eq}} \right)
\left[
{{\tilde \gamma}^{*}_{\Phi1,i}}
Y^\chi_\text{eq}
-\frac{1}{2}
\left\{
{{\tilde \gamma}^{*}_{\Phi2,i}}
\;
,Y^{\overline \chi}
\right\}
\right] 
\Bigg)_{IJ}.
\label{eqn:chibarY}
\end{multline}
In any interaction involving $\Phi_i^{(*)}$, the $U(1)_{\chi - \Phi}$ symmetry requires that the changes in the $\Phi_i$, $\Phi_i^*$, $\chi$, and ${\overline \chi}$ populations are related by
\be
\Delta N^{\Phi_i} - \Delta N^{\Phi^*_i} = \Delta N^{\chi} - \Delta N^{\overline{\chi}},
\ee
which means that the evolution of $\delta Y^{\Phi_i}$ can be determined by
\be
 \frac{d\; \delta Y^{\Phi_i}}{d\ln z} 
= \text{Tr} \left[
\frac{d\;Y^\chi_{IJ}}{d\ln z} 
-
\frac{d\;Y^{\overline \chi}_{IJ}}{d\ln z} 
\right]_i,
\ee
where on the right-hand-side we only include contributions from interactions involving $\Phi_i^{(*)}$.   We therefore get
\begin{multline}
 \frac{d \;\delta Y^{\Phi_i}}{d\ln z} 
=
-
\text{Tr} 
\left[ 
 {{\tilde \gamma}}^{}_{0,i} Y^\chi
 -
 { {{\tilde \gamma}}^{*}_{0,i} }Y^{\overline \chi}
\right]
+
Y^\chi_\text{eq}\; 
\frac{\delta Y^{Q}}{Y^{Q}_\text{eq}} \text{Tr} 
\left[ 
 {{\tilde \gamma}}^{}_{Q1,i}
\right]\\
+ \frac{\delta Y^{Q}}{2 Y^{Q}_\text{eq}} \text{Tr} 
\left[
 {{\tilde \gamma}}^{}_{Q2,i}
  Y^\chi
  +
 {{{\tilde \gamma}}^{*}_{Q2,i}}
 Y^{\overline \chi}
\right]
-
Y^\chi_\text{eq}\;
\frac{\delta Y^{\Phi_i}}{Y^{\Phi_i}_\text{eq}}
\text{Tr} 
\left[
 {{\tilde \gamma}}^{}_{\Phi1,i}
 \right]\\
-
\mathcal{G} \left( - \frac{ \delta Y^{\Phi_i}}{ 2Y^{\Phi_i}_\text{eq}}
\right)
\text{Tr} 
\left[
 {{\tilde \gamma}}^{}_{\Phi2,i} Y^\chi
 \right]
+
\mathcal{G} \left(  \frac{ \delta Y^{\Phi_i}}{ 2Y^{\Phi_i}_\text{eq}}
\right)
\text{Tr} 
\left[
{ {{\tilde \gamma}}^{*}_{\Phi2,i}}  Y^{\overline \chi}
\right].
\label{eqn:YUniversal}
\end{multline}
 As we did in Sec.~(\ref{sec:calculation}), we neglect $\Phi_1 \Phi_2^* \leftrightarrow \Phi_1^* \Phi_2$ scattering.  The viability of the model does not depend on this simplification.

In Sec.~(\ref{sec:calculation}) we expressed the $\Phi$ and baryon number asymmetries in terms of the $B-L$ asymmetry stored in Standard Model particles.  Here we also need to do that for the $Q$ asymmetry: 
\begin{eqnarray}
\delta Y^{\Phi_1}+\delta Y^{\Phi_2}  & = &  {\mathcal K}_\Phi \YBmL \\
\YB & = &  {\mathcal K}_B \YBmL \\
\delta Y^{Q} & = &  {\mathcal K}_Q \YBmL, 
\end{eqnarray}
with $\mathcal{K}_B=(-54/79,\,-63/79,\,-45/79)$ and $\mathcal{K}_Q=(25/158,\, 31/79, \,40/79)$ for $Q=(Q_{\rm L}, \,u_{\rm R}, \,d_{\rm R})$, and with $\mathcal{K}_\Phi = -3$.
We replace $\delta Y^Q \rightarrow ({\mathcal K}_Q/ {\mathcal K}_\Phi)(\delta Y^{\Phi_1}+\delta Y^{\Phi_2})$ in  Eqs.~(\ref{eqn:chiY}),~(\ref{eqn:chibarY}), and~(\ref{eqn:YUniversal})  and numerically solve them to determine the final baryon asymmetry as  $\YB = ({\mathcal K}_B/ {\mathcal K}_\Phi)(\delta Y^{\Phi_1}+\delta Y^{\Phi_2})$, evaluated at sphaleron decoupling.

The red contours of  Fig.~\ref{fig:full_survival_fig} compare the minimal perturbative calculation of Sec.~(\ref{sec:calculation}) with numerical solution of the kinetic equations just introduced, incorporating thermal masses  and quantum statistics. For the chosen parameters, the fractional differences tend to be smaller than when comparing with the refined perturbative calculation (blue contours),  because the thermal ansatz  increases $\YB$ somewhat, partially compensating for the effect of including thermal masses, which decreases the asymmetry.

To conclude this discussion, we note that we can reproduce Eqs.~(\ref{eqn:YBtwoPhi_ansatz_1}) and (\ref{eqn:YBtwoPhi_ansatz_2}) as an approximate, perturbative solution to Eqs.~(\ref{eqn:chiY}),~(\ref{eqn:chibarY}), and~(\ref{eqn:YUniversal}).  We first use Eqs.~(\ref{eqn:chiY}) and (\ref{eqn:chibarY}) at to obtain leading order (order-$F^2$) expressions for $Y^\chi_{IJ}(z)$ and $Y^{\overline \chi}_{IJ}(z)$.  We use these expressions in Eq.~(\ref{eqn:YUniversal}) to determine the leading-order expression for  $\delta Y^{\Phi_i}$.  More precisely, we solve exactly the differential equation obtained from Eq.~(\ref{eqn:YUniversal}) by replacing $Y^{\chi}$ and $Y^{\overline \chi}$ with their order-$F^2$ expressions and neglecting all reaction densities besides $\tilde{\gamma}_{0,i}$ and $\tilde{\gamma}_{\Phi1,i}$, which is equal to $\tilde{\gamma}_{0,i}$ and takes into account $\Phi_i$ decays.  We adopt Maxwell-Boltzmann statistics by taking
\begin{equation}
\mathcal{F}_{0,i}(\bk, \bp,\bq) 
=
\mathcal{F}_{\Phi1,i}(\bk, \bp,\bq) 
\rightarrow
 f_\text{eq}^{\Phi_i} (\bq) \rightarrow e^{-E_{\Phi_i}(\bq)/T}
\end{equation}
in Eq.~(\ref{eqn:RDs3}), giving
\be
\left[\gamma_{0,i}\right]_{IJ}  =&&  \left[\gamma_{\Phi1,i}\right]_{IJ} =\frac{g_\Phi (F^\dagger F)_{IJ} {\overline M}_{\Phi_i}^3 T_\text{ew}}{32 \pi^3 z}\nonumber \\
&&\times  \left(1 - {\overline M}_{\Phi_i}^2 /{\overline M}_Q^2  \right)^2  {\mathcal K}_1\! \left( \frac{{\overline M}_{\Phi_i} z}{T_\text{ew}}  \right). 
\ee
If we further neglect thermal masses, this procedure finally reproduces Eqs.~(\ref{eqn:YBtwoPhi_ansatz_1}) and (\ref{eqn:YBtwoPhi_ansatz_2}).
 
\section{Constrained maximization of $Y_B$ in the decoupled-$\Phi_2$ regime.}
\label{sec:app_parameters}

In this appendix we describe  how we obtain the contours of Fig.~\ref{fig:max_ranges_various_angles}.    We use a routine that finds the maximum $Y_B$  for given $(M_{\Phi_1},\Gamma_{\Phi_1})$ values, consistent with the observed DM abundance and  with the relevant additional constraint (fixed $\Ytwotot$, or fixed $\theta_1$ and $\theta_2$). We start by setting the mixing angles and phases $\rho_i$ and $\phi_i$ to optimal values, so that the $CP$-violating factor in Eq.~\eqref{eq:jarlskog_invariant} becomes ${\mathcal J} = \sin 2\theta_1 \sin 2\theta_2$.  With $M_{\Phi_1}$ and $\Gamma_{\Phi_1}$ fixed, $Y_B$ then depends on four quantities: $\theta_1$, $\theta_2$, $\Ytwotot$, and $\Mchitwo$.  

For the blue, $\Ytwotot = 4 \times 10^{-3}$ contour of Fig.~\ref{fig:max_ranges_various_angles}, we start by turning Eq.~\eqref{eqn:DMconstraint2} into an equality and using it to solve for $\theta_2$ in terms of $\Mchitwo$ and $\theta_1$, with $\Yonetot$ determined by Eq.~\eqref{eqn:chi_Y_from_Phi_decay}. The baryon asymmetry of Eq.~\eqref{eqn:YB_large_Mphi_alt} then depends on the remaining two free quantities, $\theta_1$ and $\Mchitwo$, through the factor
\be
\sin 2\theta_1 \sin 2\theta_2  \; \tilde{I}_{12}(\tilde{\beta}_\text{osc}),
\ee
which we maximize numerically to get optimal values of $\theta_1$ and $\Mchitwo$, subject to the constraint $\Mchitwo > 10 \text{ keV}$.  In this way we determine the maximum baryon asymmetry for the given $(M_{\Phi_1},\Gamma_{\Phi_1})$ point.  For points on the blue contours of Fig.~\ref{fig:max_ranges_various_angles}, this maximum $Y_B$ equals the observed baryon asymmetry.

The green and red contours of Fig.~\ref{fig:max_ranges_various_angles} have fixed values of $\theta_1$ and $\theta_2$, so only  $\Ytwotot$ and $\Mchitwo$ need to be optimized.  In this case we use Eq.~\eqref{eqn:DMconstraint2} to solve for $\Ytwotot$ in terms of the other parameters, and then we numerically maximize
\be
\Ytwotot \; \tilde{I}_{12}(\tilde{\beta}_\text{osc})
\ee
with respect to $\Mchitwo$.

We can understand the $\Ytwotot = 4 \times 10^{-3}$ contour of Fig.~\ref{fig:max_ranges_various_angles} qualitatively by considering two separate regimes in turn. 
We first work under the assumption that the  $\Phi_1$ couplings are too small for $\Phi_1$ decays to contribute significantly to the DM energy density, which therefore must originate almost entirely from $\Phi_2$ decays.  
In that case, the parameter $\theta_1$, which determines the relative coupling of $\Phi_1^{(*)}$ to $\chi_1$ vs.~$\chi_2$, effectively drops out of the DM constraint.  It only enters into the baryon asymmetry calculation via the $CP$-violating factor ${\mathcal J}$. In this case, the baryon asymmetry is maximized when the $\sin 2\theta_1$ factor in Eq.~\eqref{eq:jarlskog_invariant} is maximal. After we also choose optimal values for $\rho_i$ and  $\phi_i$ we are left with ${\mathcal J} = \sin 2\theta_2$.  

\begin{figure}
\includegraphics[width=3in]{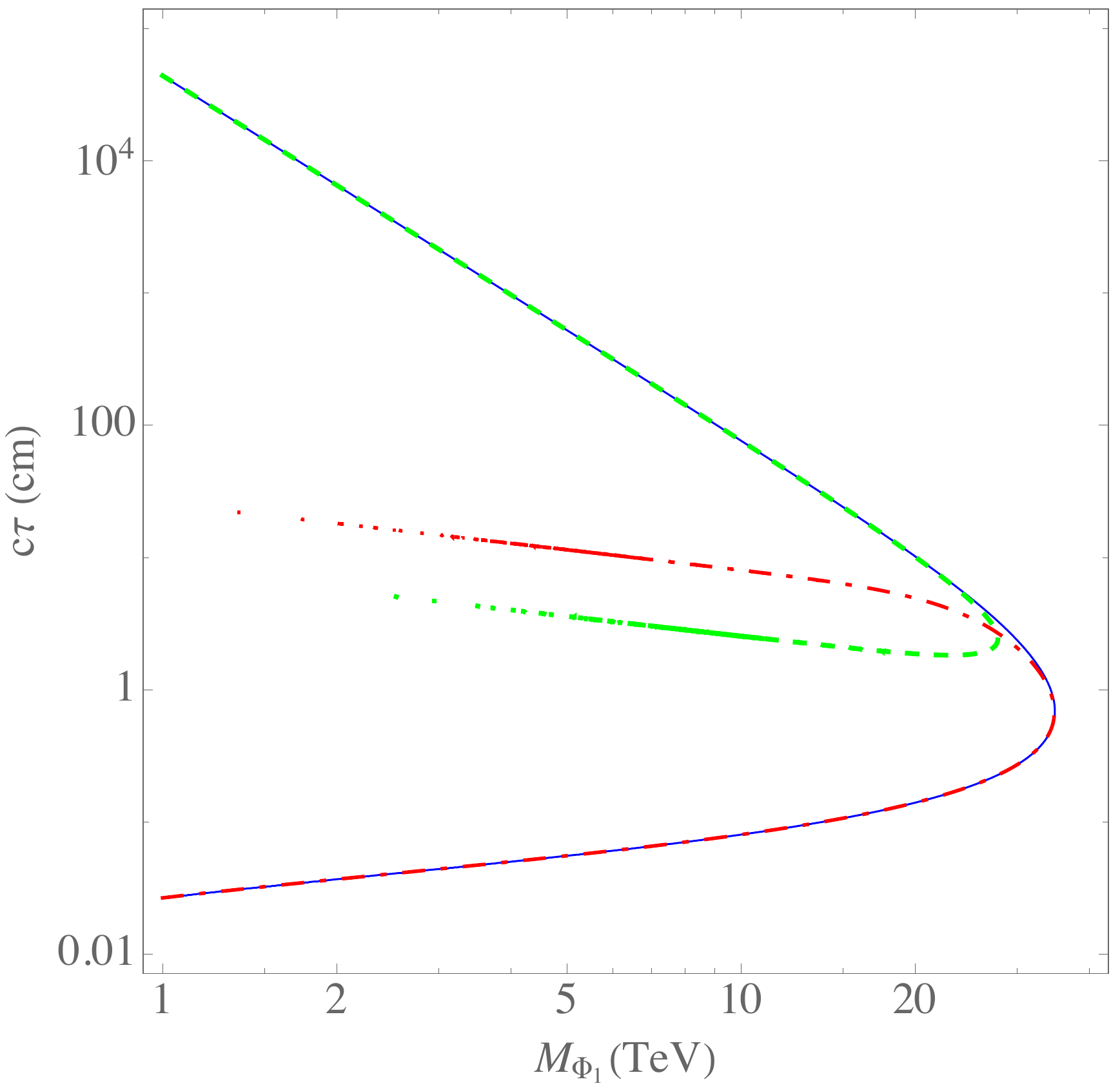}
\caption{Contours of $(Y_B)_\text{max} = (Y_B)_\text{obs}$ for $\Ytwotot = 4\times 10^{-3}$.  For the blue, solid contour we impose the DM constraint of Eq.~\eqref{eqn:DMconstraint2} and require $\Mchitwo>10$ keV (this reproduces the contour from Fig.~\ref{fig:max_ranges_various_angles}). As described in the text, the other contours impose additional constraints.
Green, dashed: we set $\theta_1=\pi/4$ and use Eq.~(\ref{eqn:large_ctau_Mchi2}) to determine $\Mchitwo$.  Red, dot-dashed: we use Eqs.~(\ref{eqn:small_ctau_angles}) and ~(\ref{eqn:small_ctau_Mchi2}) to determine $\theta_1$, $\theta_2$, and $\Mchitwo$. 
 }
 \label{fig:optimal_param_plot}
\end{figure}

Having fixed $\theta_1 = \pi/4$,  maximizing $Y_B$ for arbitrary $(M_{\Phi_1},\Gamma_{\Phi_1})$ is straightforward. We saturate Eq.~\eqref{eqn:DMconstraint2} to determine $\theta_2$ in terms of $\Mchitwo$, and then all that remains is to determine an optimal value of $\Mchitwo$. Here we use our working assumption that the DM density comes predominantly from $\Phi_2^{(*)}$ decays, which means that Eq.~\eqref{eqn:DMconstraint2} gives
\be
\theta_2 \simeq \sqrt{\frac{\rho_\text{cdm}/s}{\Mchitwo\Ytwotot}}.
\ee
The small angle approximation  for $\theta_2$ is justified given that we consider $\Ytwotot = 4\times 10^{-3}$ and $\Mchitwo > 10$ keV.  In this approximation, the dependence of $Y_B$ on $\Mchitwo$ is then contained in the factors
\be
\tilde{\beta}_\text{osc}^{-1/4} \tilde{I}_{12}(\tilde{\beta}_\text{osc}),
\ee
which is maximized for
\be
\label{eqn:large_ctau_Mchi2}
\Mchitwo=(16.2 \text{ keV}) \times \left(\frac{M_{\Phi_1}}{\text{TeV}} \right)^{3/2}.
\ee
With all parameters besides $(M_{\Phi_1},\Gamma_{\Phi_1})$ finally determined, we show the points that give the observed baryon asymmetry in the green contour in Fig.~\ref{fig:optimal_param_plot}.  

Now consider the opposite regime, in which the couplings of $\Phi_1$ are so large that the DM constraint requires $\theta_1 \ll 1$.  In that case, the small angle approximation applies for both $\theta_1$ and $\theta_2$, and it is not difficult to show that the maximum baryon asymmetry is realized when  $\Phi_1$ and  $\Phi_2$ decays contribute equally to the DM energy density. So, we take
\be
\label{eqn:small_ctau_angles}
\sin^2\theta_1 = \frac{\rho_\text{cdm}/s}{2 \Mchitwo\Yonetot}\quad
\sin^2\theta_2 = \frac{\rho_\text{cdm}/s}{2 \Mchitwo\Ytwotot},
\ee
and because the baryon asymmetry depends on the factor $\sin 2\theta_1 \sin 2\theta_2$, which scales approximately as  $1/\Mchitwo$, we obtain an optimal value of $\Mchitwo$  by maximizing 
\be
\tilde{\beta}_\text{osc}^{-1/2} \tilde{I}_{12}(\tilde{\beta}_\text{osc}),
\ee
which leads to
\be
\label{eqn:small_ctau_Mchi2}
\Mchitwo=(12.9 \text{ keV}) \times \left(\frac{M_{\Phi_1}}{\text{TeV}} \right)^{3/2}.
\ee
For these inputs, the points that give the observed baryon asymmetry lie on the red contour in Fig.~\ref{fig:optimal_param_plot}.    Taken together, the two contours we obtain by considering the opposite extremes $\theta_1 = \pi/4$ and $\theta_1 \ll 1$ 
reproduce almost all of the $(\YB)_\text{max} = (\YB)_\text{obs}$ contour for $\Ytwotot = 4 \times 10^{-3}$. 

\section{Mixing angles and phases in the single-scalar model}\label{app:singlescalar_param}
The single-scalar model is characterized by a single matrix, $F_{\alpha I}$, which gives the Yukawa couplings between each SM quark flavor $\alpha$ and $\chi$ mass eigenstate $I$. The matrix $F^\dagger F$ enters into the expression for the baryon asymmetry, and it can be parametrized as follows (by analogy with the two-scalar case in Sec.~\ref{sec:calculation}):
\be
\cos\theta_1 &=& \sqrt{\frac{(F^\dagger F)_{11}}{\mathrm{Tr}F^\dagger F}},\\
\cos\theta_2 &=& \frac{1}{\sin\theta_1}\sqrt{\frac{(F^\dagger F)_{22}}{\mathrm{Tr}F^\dagger F}},\\
\cos\rho_1 &=& \frac{|(F^\dagger F)_{12}|}{\sqrt{(F^\dagger F)_{11}(F^\dagger F)_{22}}},\\
\cos\rho_2 &=& \frac{|(F^\dagger F)_{23}|}{\sqrt{(F^\dagger F)_{22}(F^\dagger F)_{33}}},\\
\cos\rho_3 &=& \frac{|(F^\dagger F)_{13}|}{\sqrt{(F^\dagger F)_{11}(F^\dagger F)_{33}}},\\
\phi_1 &=& \arg(F^\dagger F)_{12},\\
\phi_2 &=& \arg(F^\dagger F)_{23},\\
\phi_3 &=& \arg(F^\dagger F)_{31}.
\ee
In this case, we can compute the Jarlskog-like invariant in Eq.~\ref{eq:singlescalar_Jarlskog} as
\be
\mathcal{J} &=& \cos\rho_1\cos\rho_2\cos\rho_3\cos^2\theta_1\sin^4\theta_1\label{eq:order6_Jarlskog}\\
&&{}\quad\sin^2(2\theta_2)\sin(\phi_1+\phi_2+\phi_3).
\ee

However, unlike in the case of two scalars, these are not completely independent parameters:~the reason is that there may not always exist a matrix $F$ corresponding to that set of parameters. For example, it is possible to have $\cos\rho_i=1$ for all $\rho_i$, but in this case the sum of the phases $\phi_1+\phi_2+\phi_3=0$! Thus, $\mathcal{J}$ is not optimized by requiring maximal mixing angles, since in that case the effect of the phases vanishes.

Rather than do a systematic study of the mixing angles and phases, we instead construct only the optimal value of $\mathcal{J}$, which allows us to map out the largest possible space of baryogenesis for the other parameters (such as particle masses and decay widths). Since $\mathcal{J}$ is independent of basis, we can construct the optimal $\mathcal{J}$ through a judicious choice of basis. 

First, we think of $F_{\alpha I}$ as a collection of 3 three-vectors in active quark flavor space, $F_{\alpha 1}$, $F_{\alpha 2}$, and $F_{\alpha 3}$. $(F^\dagger F)_{IJ}$, which appears in the expression for $\mathcal{J}$, can therefore be seen as a dot product of pairs of these vectors. $\mathcal{J}$ is also independent of the overall magnitudes of the three-vectors and how the magnitudes are distributed amongst the three vectors, and so we take each to have unit norm. We can always choose a flavor basis where the coupling of $\chi_1$ is exclusively to a single flavor (which we take to be $\alpha=3$), in which case $F_{\alpha 1} = (0,0,1)$. Similarly, the freedom to choose a basis and re-phase the quark fields allows us to write $F_{\alpha 2} = (0,\cos\varphi,\sin\varphi)$. Finally, there is no advantage to having $F_{13}\neq0$, since its contribution to any dot product is necessarily zero, and the optimal $CP$-violation comes from a maximal relative phase between $F_{23}$ and $F_{33}$ (any phase that is the same between the two entries contributes only to the overall normalization factor and is irrelevant), while the imaginary part of the dot product is maximized if $|F_{22}|=|F_{23}|=|F_{32}|=|F_{33}| = 1/\sqrt{2}$. The Yukawa texture for our benchmark case is thus 
\be
F_{\alpha I} &=& \left(\begin{array}{ccc}
0 & 0 & 0 \\
0 & \frac{1}{\sqrt 2} & \frac{i}{\sqrt 2} \\
1 & \frac{1}{\sqrt 2} & \frac{1}{\sqrt 2}
\end{array}
\right).
\ee
  It is straightforward to check that this set of Yukawa couplings corresponds to $\rho_1=\rho_2=\rho_3=\phi_1+\phi_2+\phi_3=\theta_2= \pi/4$, and $\cos\theta_1=1/\sqrt{3}$. The corresponding value of $\mathcal{J}=1/27$, which we use in Sec.~\ref{sec:single_scalar}. 

Because this texture of Yukawa couplings  has a zero eigenvalue of $F^\dagger F$, it has special properties with respect to the equilibration of the $\chi_I$ states. We therefore consider a second benchmark for the single-scalar study.  We modify the $F_{11}$ coupling to give three non-zero eigenvalues of $F^\dagger F$. The modified benchmark Yukawa texture is
\be
F_{\alpha I} &=& \left(\begin{array}{ccc}
1 & 0 & 0 \\
0 & \frac{1}{\sqrt 2} & \frac{i}{\sqrt 2} \\
1 & \frac{1}{\sqrt 2} & \frac{1}{\sqrt 2}
\end{array}
\right).
\ee

\bibliography{biblio}

\end{document}